\documentclass[12pt]{article}
\pdfoutput=1

\usepackage{putex}
\usepackage{graphicx}
\usepackage[font={small}]{caption}
\usepackage{amsmath, amsbsy, mathtools}
\usepackage{array}
\usepackage{diagbox}
\usepackage{subcaption}
\usepackage{epstopdf}
\usepackage{enumerate}
\usepackage{cite}
\usepackage{youngtab}
\usepackage{tensor}
\usepackage{slashed}
\usepackage[aligntableaux=center]{ytableau}
\usepackage[utf8]{inputenc}
\usepackage[
      colorlinks=true,
      linkcolor=blue,
      urlcolor=blue,
      filecolor=black,
      citecolor=red,
      ]{hyperref}
\usepackage{dsfont}
\usepackage{bm}
\usepackage{bbm}

\usepackage{booktabs}
\usepackage{comment}
\usepackage{placeins}
\usepackage{blkarray}
\usepackage{pgfplots} 
\pgfplotsset{compat=1.18}
\usepackage{braket}
\usepackage[normalem]{ulem}
\usepackage[T2A, T1]{fontenc}

\newcommand {\be} {\begin {equation}}
\newcommand {\ee} {\end {equation}}

\newcommand {\bes} {\begin {equation*}}
\newcommand {\ees} {\end {equation*}}

\newcommand{\cO}{{\mathcal O}}

\newcommand{\beq}{\begin{equation}}
\newcommand{\eeq}{\end{equation}}

\def\ie{\begin{equation}\begin{aligned}}
\def\fe{\end{aligned}\end{equation}}

\numberwithin{equation}{section}

\def\<{\langle}
\def\>{\rangle}

\let\U\relax
\DeclareMathOperator{\U}{U}

\def\V{\mathcal{V}}

\newcommand{\qsol}{\langle Q_\text{sol}\rangle}

\usepackage{tikz}
\usetikzlibrary{calc,matrix,positioning}
\usetikzlibrary{decorations.markings, arrows.meta}
\tikzstyle{tensor}=[circle,draw=black!50,thick, minimum size = 40pt]

\tikzstyle{square}=[fill=white, draw=black, shape=rectangle, rounded corners, minimum width=1.cm, minimum height=1.cm, inner sep=0pt, outer sep=0pt,thick]
\tikzstyle{circle}=[fill=white, draw=black, shape=circle, minimum width=1.cm, minimum height=1.cm, inner sep=0pt, outer sep=0pt, thick]
\tikzstyle{shade}=[dotted,fill=gray!30,rounded corners=5mm]

\tikzset{
	diagram/.style={
		baseline={([yshift=-.5ex]current bounding box.center)},
		every node/.style={scale=0.7},
		thick
	}
}
\tikzset{
	arrow mid/.style={
		postaction={decorate},
		decoration={
			markings,
			mark=at position 0.5 with {
				\pgftransformxshift{3pt}
				\arrow{Stealth}
			}
		}
	}
}

\begin{document}

\preprint{PUPT-2667\\MIT-CTP/6160}

\institution{MIT}{Center for Theoretical Physics -- a Leinweber Institute, Massachusetts Institute of Technology, Cambridge, MA 02139, USA}
\institution{PU}{Joseph Henry Laboratories and Leinweber Forum for Theoretical Physics, Princeton University, Princeton, NJ 08544, USA}
\institution{PCTS}{Princeton Center for Theoretical Science, Princeton University, Princeton, NJ 08544, USA}

\title{String breaking in a (1+1)-dimensional \\ 
$e^+e^-$ collider}

\authors{Ross Dempsey,\worksat{\MIT} Anna-Maria E.~Gl\"uck,\worksat{\PU} Silviu S.~Pufu,\worksat{\PU,\PCTS}\\[10pt] and Benjamin T.~S\o{}gaard\worksat{\PU}}

\abstract{In this paper, we develop a framework for studying solitons in $(1+1)$-dimensional gauge theories using tensor networks, and we apply it to simulate $e^+e^-$ collisions in the massive Schwinger model. We work at $\theta = \pi$ and at a bare fermion mass exceeding the critical value; in this regime, there are two degenerate vacua, and the $e^+$ and $e^-$ particles are realized as soliton states interpolating between them. We carefully explain how to find and characterize such soliton states using infinite tensor network methods, and then exhibit real-time simulations of their collisions. These collisions involve the formation of an energetic flux string, and we demonstrate that our final states are superpositions of elastic and string-breaking trajectories.} 

\date{November 2026}

\maketitle

\tableofcontents

\section{Introduction}

The massive Schwinger model \cite{Schwinger:1962tp,Coleman:1976uz}, describing electrodynamics in $1+1$ dimensions, was the first tractable quantum field theory to exhibit confinement. Its action is
\begin{equation}\label{eq:schwinger_action}
    S = \int d^2x\,\left\lbrack -\frac{1}{4g^2} F_{\mu\nu} F^{\mu\nu} - \frac{\theta}{4\pi} \epsilon^{\mu\nu}F_{\mu\nu} + \bar{\psi} \left(i \gamma^\mu (\partial_\mu + i A_\mu) - m\right)\psi \right\rbrack\,,
\end{equation}
where $\eta_{\mu\nu} = \diag(1,-1)$ and $\epsilon^{01} = 1$. Here, $\psi$ is a Dirac spinor field of bare mass $m$, describing particles of electric charge $\pm 1$ coupled to the $\U(1)$ gauge field $A_\mu$ with coupling $g$. The (dimensionless) electric field is $E \equiv F_{01}/g^2$; it is quantized in $\mathbb{Z}+\frac{\theta}{2\pi}$.  With these conventions, $\psi^\dagger$ creates positrons (denoted $e^+$, with charge $+1$) and $\psi$ creates electrons ($e^-$, with charge $-1$). Schwinger solved the model at $m = 0$ (where the dynamics becomes independent of $\theta$) \cite{Schwinger:1962tp}, and found that the massless model has a single stable particle: a bosonic $e^+e^-$ bound state of mass $\frac{g}{\sqrt{\pi}}$. The electron and positron themselves do not exist as asymptotic states. 

At $\theta = \pi$ and for sufficiently large mass $m$, however, finite-energy excitations corresponding to the electron and positron become possible, as pointed out by Coleman~\cite{Coleman:1976uz}. In $1+1$ dimensions, the $\theta$-angle corresponds to a background electric field $\theta/2\pi$, which can be thought of as sourced by a pair of fractional charges at $\pm\infty$. Pair production of integer-charged dynamical fermions can only change this field by integer amounts. For $|\theta| < \pi$ the background field therefore cannot be screened, while at $\theta = \pi$ the configurations with field $+\frac{1}{2}$ and $-\frac{1}{2}$ have the same energy. Indeed, for $m \gg g$, where a semiclassical analysis applies, one finds two degenerate vacua at $\theta = \pi$, in which the electric field equals $+\frac{1}{2}$ or $-\frac{1}{2}$ everywhere in space. These vacua are exchanged by charge conjugation $C$. There is then a finite-energy $e^-$ excitation for which the electric field is $+\frac{1}{2}$ to the left and $-\frac{1}{2}$ to the right, and similarly for $e^+$ with the two vacua interchanged. The electron and positron are thus realized as solitons (domain walls) interpolating between the two degenerate vacua. As $m/g$ is decreased, the degenerate vacua and the corresponding solitons persist as long as $C$ is spontaneously broken, which is the case for $m > m_c$, with $m_c/g \approx 0.3335$ \cite{Byrnes:2002nv,ArguelloCruz:2024xzi,Fujii:2024reh,Okuda:2026hbg}. 

Studying solitons on the lattice is considerably more challenging than studying particles. Particle states (i.e., local excitations above a single vacuum) can be approximated by excited states of a Hamiltonian on a finite lattice, and thus studied via exact diagonalization on small lattices \cite{Chakraborty:2020uhf,Honda:2021aum,Dempsey:2022nys,Sriganesh:1999ws} or using the matrix product state (MPS) ansatz \cite{Banuls:2013jaa,VanAcoleyen:2014suo,Buyens:2015dkc,Buyens:2017crb,Buyens:2016hhu,Banuls:2016gid,Funcke:2019zna,Dempsey:2023gib,Cuomo:2026kmy,Fujii:2024reh,Itou:2024psm,Honda:2022edn} on larger lattices. However, soliton states are most naturally formulated on infinite lattices, for which we can sharply define the vacua at left and right infinity and decompose the Hilbert space accordingly.

For this reason, lattice studies of solitons in gauge theories rely crucially on the uniform matrix product state (uMPS) ansatz \cite{Zauner-Stauber:2017eqw,Vanderstraeten:2019voi}, which allows for approximating the ground states of lattice Hamiltonians directly in the thermodynamic limit. Uniform matrix product states have been applied to lattice gauge theories before \cite{Buyens:2013yza,Buyens:2015tea,Buyens:2016hhu,Buyens:2017crb,Zapp:2017fcr,Fujii:2024reh}, but the electric energy term in the Hamiltonian presents an extra difficulty: either the link variables corresponding to the gauge field are kept explicitly, which standard matrix product operator methods do not naturally accommodate, or they are eliminated using the Gauss law, which produces long-range interactions. This problem was recently addressed using a new tensor network construction, the link-enhanced matrix product operator (LEMPO) acting on symmetric matrix product states \cite{Dempsey:2025wia}. A LEMPO can represent the Hamiltonian including its electric energy term in a local and efficient manner that allows for infinite tensor network methods.

In this paper, we study solitons in the Schwinger model at $\theta = \pi$ using LEMPOs on an infinite Hamiltonian lattice. We construct solitons and antisolitons on the lattice as domain walls interpolating from a lattice vacuum $\ket{v_-}$ with negative electric field on its left to the other vacuum $\ket{v_+}$ on its right; schematically
\begin{equation}\label{eq:sol_antisol_schematic_1}
    \tikz[baseline=-0.75cm]{
        \node[anchor=east] at (-4,0) {$e^+$};
        \draw[ultra thick,dashed] (-3.5,0) -- node[below] {\footnotesize $\ket{v_-}$} (0,0);
        \draw[ultra thick,dotted] (0,0) -- node[below] {\footnotesize $\ket{v_+}$} (3.5,0);
        \draw[thick,red!50,fill=red!30] (0,0) circle (.3) node[white] {$+$};
        \node[anchor=east] at (-4,-1.5) {$e^-$};
        \draw[ultra thick,dotted] (-3.5,-1.5) -- node[below] {\footnotesize $\ket{v_+}$} (0,-1.5);
        \draw[ultra thick,dashed] (0,-1.5) -- node[below] {\footnotesize $\ket{v_-}$} (3.5,-1.5);
        \draw[thick,blue!50,fill=blue!30] (0,-1.5) circle (.3) node[white] {$-$};
    }\,.
\end{equation}
We compute their masses and the expectation values of their electric charges (which differ from $\pm 1$ due to partial screening, as we describe in Section~\ref{sec:model}) and extrapolate these values to the continuum limit of vanishing lattice spacing $ag$. In this limit, our results interpolate between analytic results for the weak-coupling behavior at $m \gg g$ and a scaling behavior at $m - m_c \ll g$ that is governed by the Ising conformal field theory (CFT)\@.

We then also construct two-soliton states in which a soliton interpolates from a vacuum $\ket{v_-}$ on its left to the other vacuum $\ket{v_+}$ on its right, and then an anti-soliton to the right of the soliton interpolates back to $\ket{v_-}$. Schematically, the configuration is \begin{equation}\label{eq:sol_antisol_schematic_2}
    \tikz[baseline=0cm]{
        \draw[ultra thick,dashed] (-5.5,0) -- node[below] {\footnotesize $\ket{v_-}$} (-3.5,0);
        \draw[ultra thick,dashed] (5.5,0) -- node[below] {\footnotesize $\ket{v_-}$} (3.5,0);
        \draw[ultra thick,dotted] (-3.5,0) -- node[below] {\footnotesize $\ket{v_+}$} (3.5,0);
        \draw[thick,red!50,fill=red!30] (-3.5,0) circle (.3) node[white] {$+$};
        \draw[thick,blue!50,fill=blue!30] (3.5,0) circle (.3) node[white] {$-$};
    }\,.
\end{equation}
By building a wavepacket out of these states, and using the lattice Hamiltonian to simulate time evolution, we can collide the soliton and anti-soliton. These lattice $e^+e^-$ collisions clearly demonstrate the half-asymptotic nature of these particles, first pointed out by Coleman \cite{Coleman:1976uz,Coleman:1975pw}: when the two particles pass each other, the space between them has a finite energy density above that of $\ket{v_\pm}$. That region can be interpreted as an energetic flux string, which gives rise to a constant attractive force. Indeed, at weak coupling and low center-of-mass energy $\sqrt{s}$, the particles follow the classical trajectories of relativistic massive particles subject to a constant force. After the particles rebound and pass each other again, the space between them is again described by the $\ket{v_+}$ vacuum: 
\begin{equation}\label{eq:psifinal_el}
\tikz[baseline=0cm]{
        \draw[ultra thick,dashed] (-5.5,0) -- node[below] {\footnotesize $\ket{v_-}$} (-3.5,0);
        \draw[ultra thick,dashed] (5.5,0) -- node[below] {\footnotesize $\ket{v_-}$} (3.5,0);
        \draw[ultra thick,dotted] (-3.5,0) -- node[below] {\footnotesize $\ket{v_+}$} (3.5,0);
        \draw[thick,red!50,fill=red!30] (-3.5,0) circle (.3) node[white] {$+$};
        \draw[thick,blue!50,fill=blue!30] (3.5,0) circle (.3) node[white] {$-$};
        \draw[ultra thick,-latex] (3.25,.5) -- (3.75,.5);
        \draw[ultra thick,-latex] (-3.25,.5) -- (-3.75,.5);
    }\,.
\end{equation}

As we increase $\sqrt{s}$, we expect another possibility to become significant: the flux string can break, forming an additional soliton-anti-soliton pair. In this case, the final state consists of an unbound soliton-anti-soliton pair moving to the left and another unbound soliton-anti-soliton pair moving to the right. The space between the two pairs is then described by the $\ket{v_-}$ vacuum:
\begin{equation}\label{eq:psifinal_inel}
\tikz[baseline=0cm]{
        \draw[ultra thick,dashed] (-5.5,0) -- node[below] {\footnotesize $\ket{v_-}$} (-3.5,0);
        \draw[ultra thick,dashed] (5.5,0) -- node[below] {\footnotesize $\ket{v_-}$} (3.5,0);
        \draw[ultra thick,dotted] (-3.5,0) -- node[below] {\footnotesize $\ket{v_+}$} (-2,0);
        \draw[ultra thick,dotted] (3.5,0) -- node[below] {\footnotesize $\ket{v_+}$} (2,0);
        \draw[ultra thick,dashed] (-2,0) -- node[below] {\footnotesize $\ket{v_-}$} (2,0);
        \draw[thick,red!50,fill=red!30] (-3.5,0) circle (.3) node[white] {$+$};
        \draw[thick,blue!50,fill=blue!30] (-2,0) circle (.3) node[white] {$-$};
        \draw[thick,red!50,fill=red!30] (2,0) circle (.3) node[white] {$+$};
        \draw[thick,blue!50,fill=blue!30] (3.5,0) circle (.3) node[white] {$-$};
        \draw[ultra thick,-latex] (3.25,.5) -- (3.75,.5);
        \draw[ultra thick,-latex] (1.75,.5) -- (2.25,.5);
        \draw[ultra thick,-latex] (-1.75,.5) -- (-2.25,.5);
        \draw[ultra thick,-latex] (-3.25,.5) -- (-3.75,.5);
    }\,.
\end{equation}

A key result of this paper is a test of this expected picture. We show that the distribution of observables in the region between the post-collision excitations is well-described using a linear combination of their distributions in $\ket{v_+}$ and $\ket{v_-}$. The weights of the combination provide an estimate of the string-breaking probability, and this estimate agrees well with a value obtained by directly projecting the final state onto the space of two-particle states. We can also use the entanglement between the central electric field and the outgoing excitations to project onto either string-unbroken or string-broken final states, clearly demonstrating that our real-time simulation evolves into a superposition of both outcomes.

Finally, we study solitons near the Ising critical point at $m=m_c$. We compute the charge-mass relation of the solitons in this regime by using the integrability of the $\epsilon$-deformation of the 2D Ising CFT~\cite{Wu:1975mw,McCoy:1977er,DiFrancesco:1997nk}. We also argue using effective field theory that near $m_c$, string breaking becomes significant only at center-of-mass energies of order $g$, far above the kinematic threshold $4M_\text{sol}$ (where $M_\text{sol}$ is the mass of a soliton or anti-soliton). 

Real-time scattering of the $e^+$ and $e^-$ solitons in the Schwinger model was studied previously in \cite{Belyansky:2023rgh}, also using infinite tensor network methods to prepare the scattering states. The underlying lattice model used in \cite{Belyansky:2023rgh} is very different, however: the authors of \cite{Belyansky:2023rgh} use abelian bosonization to map the continuum Schwinger model to a massive sine-Gordon model, which they then discretize. Because the sine-Gordon coupling runs with the scale, one would have to account for this running in order to extract a continuum limit. In contrast, we use the fermionic Kogut-Susskind formulation \cite{Kogut:1974ag,Banks:1975gq}, which is super-renormalizable, so we can take a continuum limit without any non-trivial renormalization calculation. Our setup also does not rely on abelian bosonization, so it extends readily to non-abelian gauge groups, which LEMPOs naturally accommodate~\cite{Dempsey:2025wia}. 

More broadly, real-time dynamics in the Schwinger model and related (1+1)-dimensional systems have been studied extensively with numerical methods. In most studies of string breaking, the string is stretched either between static charges \cite{Hebenstreit:2013baa,Kuhn:2015zqa,Buyens:2015tea,Grieninger:2026bdq,Gupta:2026tcg} or between dynamical charges in a prepared initial state \cite{Pichler:2015yqa,Verdel:2019chj,Lee:2023urk,Gupta:2026tcg}. Scattering studies have mostly focused on collisions of confined bound states \cite{Pichler:2015yqa,Rigobello:2021fxw,Papaefstathiou:2024zsu}, while studies of jet production have modeled the outgoing charges as external sources receding at the speed of light \cite{Hebenstreit:2014rha,Florio:2023dke,Florio:2024aix,Janik:2025bbz}. String-breaking dynamics have also recently been observed on analog and digital quantum simulators \cite{De:2024smi,Liu:2024lut,Gonzalez-Cuadra:2024xul,Cochran:2024rwe,Alexandrou:2025vaj}, and quantum simulations of hadron scattering are beginning to appear \cite{Farrell:2024fit,Davoudi:2025rdv}. Our results provide classical benchmarks for the collision experiments such platforms are beginning to target.

The rest of the paper is organized as follows. In Section~\ref{sec:model}, we describe the continuum theory in more detail, and in particular the phase diagram and the presence of solitonic states in the $C$-broken region; we also give the details of the lattice model we use. In Section~\ref{sec:method}, we explain the tensor network methods we use to construct the degenerate lattice vacua, to find and characterize the soliton states between them, and to prepare initial states for soliton collisions. In Section~\ref{sec:results}, we show our real-time soliton collision simulations and analysis of them. In Section~\ref{sec:critical}, we discuss the soliton properties and collisions near the 2d Ising critical point at $m = m_c$. We conclude in Section~\ref{sec:conclusion}. Several technical details are relegated to appendices.

\section{The Schwinger model}\label{sec:model}
In this section we will describe key aspects of the Schwinger model that will be relevant for our study of $e^\pm$ states and their collisions. In Section~\ref{sec:model_continuum}, we review the phase diagram of the Schwinger model, and we discuss solitons in the weak- and strong-coupling limits. Then, in Section~\ref{sec:model_lattice}, we review the lattice Hamiltonian for the Schwinger model.

\subsection{Continuum theory}\label{sec:model_continuum}

We will give a brief review of the phase diagram of the one-flavor Schwinger model with action \eqref{eq:schwinger_action}. We gauge-fix $A_0 = 0$ and work mostly in a Hamiltonian picture (except in Section~\ref{sec:critical_scattering}). The dimensionless electric field is 
\begin{equation}
    E = \frac{1}{g^2}F_{01}\,,
\end{equation}
and the momentum conjugate to $A_1$ is $\Pi = E-\frac{\theta}{2\pi}$. The Hamiltonian is given by
\begin{equation}\label{eq:continuum_H}
    H = \int dx_1\left\lbrack \frac{g^2}{2} E^2 - i \psi^\dagger(x_1) \gamma^5 (\partial_1 + i A_1) \psi(x_1) + m \psi^\dagger(x_1) \gamma^0 \psi(x_1)\right\rbrack\,
\end{equation}
where $\gamma^5 = \gamma^0 \gamma^1$. Physical states obey the Gauss law
\begin{equation}
    \partial_{x_1} E = \psi^\dagger\psi\,.
\end{equation}
The Gauss law is supplemented by the condition that $\Pi$ has integer eigenvalues, so the electric field is valued in $\mathbb{Z}+\frac{\theta}{2\pi}$; this condition is the only place where $\theta$ will appear in the model.

The dimensionless parameters of the model are $m/g$ and $\theta$. Note that $\theta\sim \theta+2\pi$, so it suffices to consider only $\theta \in [0,2\pi)$. Additionally, by making an axial transformation $\psi \mapsto e^{i\frac{\pi}{2}\gamma^5}\psi$, one can show that the theory at parameters $(m/g,\theta)$ is equivalent to that at $(-m/g,\theta + \pi)$, so we may also restrict to $m \ge 0$. (See, for instance, Appendix 2 of \cite{Dempsey:2022nys}.) Charge conjugation $C$ and parity $P$ act on the fields as
\begin{equation}\label{eq:CP_continuum}
\begin{aligned}
    C&:\quad \psi(t,x)\mapsto \mathsf{C}\,\bar\psi^T(t,x)\,,\qquad \; A_\mu(t,x)\mapsto -A_\mu(t,x)\,,\\
    P&:\quad \psi(t,x)\mapsto \gamma^0\psi(t,-x)\,,\qquad
    A_0(t,x)\mapsto A_0(t,-x)\,,\quad A_1(t,x)\mapsto -A_1(t,-x)\,,
\end{aligned}
\end{equation}
where $\mathsf{C}$ is such that $\mathsf{C}^{-1}\gamma^\mu\mathsf{C}=-(\gamma^\mu)^T$. Both transformations map $E\mapsto -E$, and hence map the condition $E\in\mathbb{Z}+\frac{\theta}{2\pi}$ to $E\in\mathbb{Z}-\frac{\theta}{2\pi}$; they are therefore symmetries only at $\theta=0$ or $\theta=\pi$.

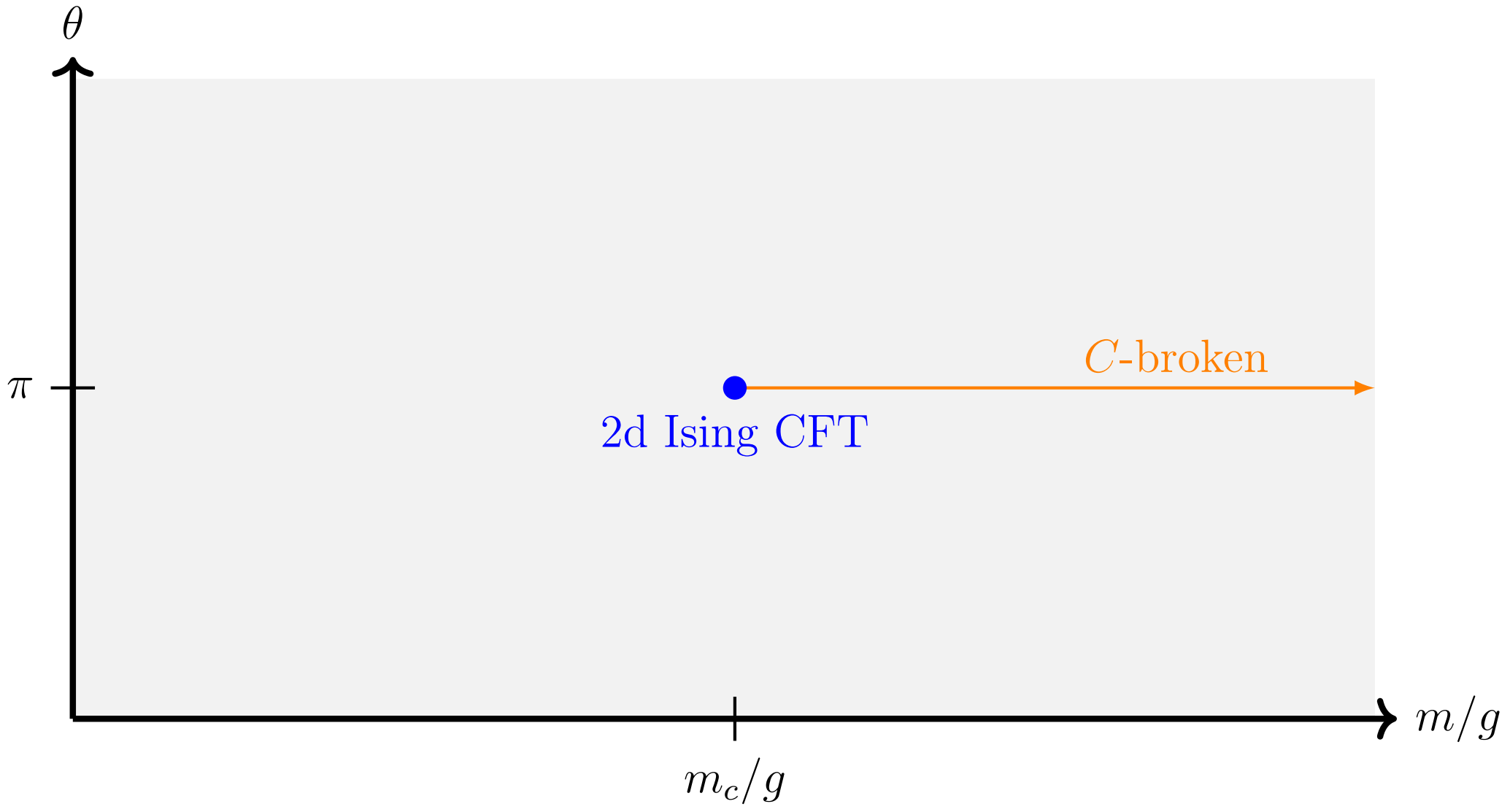
\begin{figure}
    \centering
    \begin{tikzpicture}[scale=2]
        \fill[gray!10] (0,0) rectangle (5.9,2.9);
        
        \draw[ultra thick,->] (0,0) -- (0,3) node[above] {$\theta$};
        \draw[ultra thick,->] (0,0) -- (6,0) node[right] {$m/g$};

        \draw[thick] (3,.1) -- (3,-.1) node[below] {$m_c/g$};
        \draw[thick] (.1,1.5) -- (-.1,1.5) node[left] {$\pi$};

        \draw[orange,thick,-latex] (3,1.5) -- (5.9,1.5);
        \draw[fill,blue] (3,1.5) circle (.05);

        \node[blue,below] at (3,1.45) {2d Ising CFT};
        \node[orange,above] at (5,1.5) {$C$-broken};
    \end{tikzpicture}
    \caption{The phase diagram of the one-flavor Schwinger model for $m/g>0$ and $0\leq \theta < 2\pi$. For $\theta = \pi$ and $m>m_c$, where $m_c/g \approx 0.3335$ \cite{Byrnes:2002nv,ArguelloCruz:2024xzi,Fujii:2024reh,Okuda:2026hbg}, charge conjugation symmetry is spontaneously broken and there are two degenerate vacua. This line of first-order phase transitions ends in a second-order critical point at $\theta = \pi$ and $m = m_c$, which can be described by the 2d Ising CFT.}
    \label{fig:phase}
\end{figure}

The phase diagram is given in Figure~\ref{fig:phase}. At $\theta = \pi$ for $m \gg g$, there is a first-order phase transition as $\theta$ crosses $\pi$: for $\theta = \pi + \epsilon$, with $0 < \epsilon \ll 1$, the electric field in the ground state will be just above $-\frac{1}{2}$, whereas for $\theta = \pi-\epsilon$ it will be just below $+\frac{1}{2}$. At precisely $\theta = \pi$, these are two degenerate vacuum states, which we denote by $\ket{v_\pm}$ (with the sign indicating the sign of the electric field). These vacua are exchanged by charge conjugation:
\begin{equation} \label{eq:C_vac}
    C\ket{v_\pm} = \ket{v_\mp}\,.
\end{equation}
That is, charge conjugation symmetry is spontaneously broken in this region. Parity symmetry also relates the two vacua:
\begin{equation} \label{eq:P_vac}
    P\ket{v_\pm} = \ket{v_\mp}\,.
\end{equation}
Each vacuum is thus invariant under $CP$.

At a critical mass $m_c$, there is a second-order transition in the 2d Ising universality class at which these two vacua merge into a single vacuum and charge conjugation symmetry is restored (as is parity symmetry). The critical mass is known to a precision of about $10^{-5}$: two separate estimates $m_c/g = 0.333556(5)$ \cite{Fujii:2024reh} and $m_c/g = 0.333561(4)$ \cite{ArguelloCruz:2024xzi} agree within their uncertainties (see also \cite{Okuda:2026hbg}).

In this paper, we focus on the $C$-broken region, in which there are positively charged soliton states interpolating from $\ket{v_-}$ on the left to $\ket{v_+}$ on the right, and negatively charged anti-soliton states which do the opposite. These are dressed positrons and electrons, respectively. In the weak-coupling limit $m/g\to\infty$, the solitons correspond to the partons in the Lagrangian,\footnote{More precisely, a soliton is the excitation created by acting with $\psi^\dagger(x)$ dressed with a half-infinite Wilson line to make it gauge invariant.} with mass $M_\text{sol} = m$ and definite charge $Q_\text{sol} = \pm 1$. However, as we increase the coupling, the soliton mass is adjusted away from $m$, and its charge is no longer definite. We therefore consider the expectation value of the soliton charge, which by the Gauss law is $Q_\text{sol} = E(+\infty) - E(-\infty)$.  Then
\begin{equation}\label{eq:qsol_def}
\begin{split}
    \qsol = \braket{v_+|E|v_+} - \braket{v_-|E|v_-} = 2 \braket{v_+|E|v_+}\,,
\end{split}
\end{equation}
where the first equality uses cluster decomposition, and the second equality uses $C E C = -E$. The expectation value of the anti-soliton charge is $-\qsol $.

The weak-coupling expansions are\footnote{The expansion of $M_\text{sol}$ can be found in Appendix~C of \cite{Cuomo:2026kmy}. For $\qsol$, we first note that, with $m/g\to\infty$, we have a pure gauge theory, and at $\theta = \pi$ there are two vacua with $\langle F_{01}\rangle = \pm g^2/2$. The leading effect of the large-mass fermion is then to adjust the coupling $g^2 \mapsto g_\text{eff}^2 = g^2\left(1-\frac{g^2}{6\pi m^2}\right)$, and using \eqref{eq:qsol_def}, we find \eqref{eq:weak_coupling}. See again \cite{Cuomo:2026kmy} for a very similar calculation with two flavors.}
\begin{equation}\label{eq:weak_coupling}
    M_\text{sol} = m - \frac{g^2}{2\pi m} + \ldots, \qquad  \qsol = 1 - \frac{g^2}{6\pi m^2} + \ldots
\end{equation}
The decrease of $ \qsol$ as we increase the coupling reflects the fact that, if we insert $\mp\tfrac12$ probe charges at $\pm\infty$ in order to set $\theta = \pi$, those charges will be partially screened by dynamical fermions.

As $m \to m_c$, this screening becomes near-complete, and so $ \qsol \to 0$. Likewise, as we approach the gapless critical point, we will have $M_\text{sol} \to 0$. Near the critical mass $m_c$, the theory follows the critical exponents of the 2d Ising CFT\@. In particular, the scaling of the soliton charge and the soliton mass as we approach $m\to m_c$ from above are given by
\begin{equation}\label{eq:scaling}
    M_\text{sol} \propto m - m_c\,, \qquad \qsol \propto (m - m_c)^{1/8}\,.
\end{equation}
In Section~\ref{sec:critical_charge_mass}, we will explore this scaling behavior in more detail. 

\subsection{Lattice theory}\label{sec:model_lattice}

To study the Schwinger model on the lattice, we will use the standard Kogut-Susskind discretization \cite{Kogut:1974ag,Banks:1975gq} of \eqref{eq:continuum_H}:
\begin{equation}
    H = \frac{g^2 a}{2} \sum_n \left(L_n + \frac{\theta}{2\pi}\right)^2 - \frac{i}{2a} \sum_n \left(c^\dagger_n U_n c_{n+1} - c_{n+1}^\dagger U_n^\dagger c_n\right) + m_\text{lat} \sum_n (-1)^n c^\dagger_n c_n\,,
    \label{eq:lattice_H}
\end{equation}
with $m_\text{lat} = m - \frac{g^2 a}{8}$ to improve convergence to the continuum limit \cite{Dempsey:2022nys}. Here, $c_n$ is a complex fermion on each site, with $\{ c_n, c_m^\dagger\} = \delta_{nm}$; $U_n$ is a Wilson line connecting adjacent sites; and $L_n+\theta/2\pi$ is the lattice analogue of $E$, with $[L_n, U_m] = \delta_{nm} U_m$ and $L_n$ integer-valued. The Hamiltonian \eqref{eq:lattice_H} acts on the Hilbert space of gauge-invariant states, namely those satisfying the Gauss law
\begin{equation}
    L_n - L_{n-1} = Q_n, \qquad Q_n \equiv c^\dagger_n c_n - \begin{cases} 0 & n\text{ even}\,,\\ 1 & n\text{ odd}\,.\end{cases} \label{eq:Gauss}
\end{equation}
For concreteness, in this section we will work on a finite chain with sites $n=1,\ldots,N$; in Section~\ref{sec:method} we will explain how to move to an infinite lattice.

In (1+1)d, the gauge field is not dynamical. Indeed, we can use a gauge transformation to fix all $U_n \to \mathds{1}$, and then solve the Gauss law \eqref{eq:Gauss} to write the electric field variables $L_n$ in terms of on-site charges. For the second step, we also need boundary conditions
\begin{equation}\label{eq:L_bdry}
    L_0 = 0 \qquad\text{and}\qquad L_N = 0\,.
\end{equation}
Using \eqref{eq:Gauss} and the boundary conditions \eqref{eq:L_bdry}, we can eliminate $L_n$ in terms of the charge operators to obtain
\begin{equation}
    L_n = \sum_{j=1}^n Q_j \,, \label{eq:Ln_solved}
\end{equation}
which we can substitute into \eqref{eq:lattice_H} to give an explicit Hamiltonian solely in terms of $c_n$. The second condition in \eqref{eq:L_bdry} leads to the constraint $\sum_{j=1}^N Q_j = 0$.

We can furthermore perform a Jordan-Wigner transform to write the fermion operators in terms of Pauli matrices $\{\sigma_n^x, \sigma_n^y, \sigma_n^z\}$ and their linear combinations $\sigma_n^\pm = \frac{\sigma_n^x\pm i\sigma_n^y}{2}$ as 
\begin{equation}
    c_n = \left(\prod_{j=1}^{n-1} (-i \sigma_j^{z})\right) \sigma_n^{-}, \qquad  c_n^\dagger = \left(\prod_{j=1}^{n-1} (i \sigma_j^{z})\right) \sigma_n^{+}\,.
\end{equation}
After performing this transformation, the Hamiltonian \eqref{eq:lattice_H} becomes
\begin{equation}
    H = \sum_n\left[ \frac{g^2 a}{2} \left(L_n + \frac{\theta}{2\pi}\right)^2 + \frac{1}{2a} \left(\sigma_n^+ \sigma_{n+1}^- + \sigma_n^- \sigma_{n+1}^+ \right) + \frac{m_\text{lat}}{2} (-1)^n \sigma_n^z\right]\,,
    \label{eq:lattice_H_fixed}
\end{equation}
and 
\begin{equation}
    Q_n = \frac{1}{2}(\sigma_n^z + (-1)^n), \qquad\text{so that}\qquad L_n = \sum_{j=1}^n \frac{1}{2}(\sigma_j^z + (-1)^j) \,. \label{eq:Ln_solved_jw}
\end{equation}

Since we will be studying soliton excitations between vacua that are exchanged by the spontaneously-broken charge conjugation symmetry, we will need the charge conjugation operator $\mathcal{C}$ on the lattice, which involves a one-site translation. It is given by
\begin{equation}\label{eq:C}
    \mathcal{C}L_n\mathcal{C}^{-1} = -L_{n+1} - \frac{\theta}{\pi}, \qquad \mathcal{C} U_n\mathcal{C}^{-1} = U_{n+1}^\dagger, \qquad \mathcal{C}c_n\mathcal{C}^{-1} = c_{n+1}^\dagger.
\end{equation}
One can check that $\mathcal{C} H \mathcal{C}^{-1} = H$, and also that conjugation by $\mathcal{C}$ preserves \eqref{eq:Gauss} whenever $\theta = 0$ or $\pi$ modulo $2\pi$. 

It will also be useful to us to have the parity operator $\mathcal{P}$ on the lattice. We use a site-centered parity, which acts as
\begin{equation}
    \mathcal{P} L_n \mathcal{P}^{-1} = -L_{-n-1} - \frac{\theta}{\pi}, \qquad \mathcal{P} U_n \mathcal{P}^{-1} = U^\dagger_{-n-1}, \qquad \mathcal{P} c_n \mathcal{P}^{-1} = (-1)^n c_{-n}.
\end{equation}
This transformation also commutes with $H$ and preserves the Gauss law for $\theta = 0$ or $\pi$ modulo $2\pi$.

In Appendix~\ref{app:symmetries}, we describe in more detail how to implement $\mathcal{C}$ and $\mathcal{P}$ in the tensor network setup we use in the remainder of the paper.

\section{Preparing soliton wavepacket states}\label{sec:method}

The goal of this paper, which we carry out in Section~\ref{sec:results}, is to perform real-time simulations of colliding solitons in the lattice Schwinger model. In this section, we describe the ingredients we need to prepare an initial state.  

In Section~\ref{sec:method_lempo}, we start with a review of the tensor network methods used in our simulations, namely matrix product states (MPSs) \cite{Fannes:1990ur,Ostlund:1995zz,Perez-Garcia:2006nqo,Cirac:2020obd}, matrix product operators \cite{Verstraete:2008xyc,Schollwoeck:2010uqf}, and link-enhanced matrix product operators (LEMPOs) \cite{Dempsey:2025wia}. In this first section, we focus on tensor networks on finite lattices. However, as mentioned in the introduction, we need infinite lattices to sharply define solitons between the degenerate Schwinger vacua $\ket{v_\pm}$ at $\theta = \pi$ and $m>m_c$. Therefore, we have to generalize the methods introduced in Section~\ref{sec:method_lempo} to work with uniform matrix product states (uMPSs) on infinite lattices in the following sections. 

Our first step, which we describe in Section~\ref{sec:method_boundary}, is to build uMPSs that approximate the degenerate Schwinger vacua $\ket{v_\pm}$. We characterize these states using local expectation values. Then, in Section~\ref{sec:method_envs}, we explain how to treat the extensive energy of a uMPS using environments. This step is crucial to later obtaining the finite soliton energy in the background of a diverging groundstate energy on an infinite lattice. Our next step, in Section~\ref{sec:method_qp}, is then to construct lattice states approximating the continuum solitons of Section~\ref{sec:model_continuum}, using the quasiparticle ansatz \cite{Haegeman:2013xcv,Haegeman:2013vud,Zauner-Stauber:2018gqr,Vanderstraeten:2019voi}. Determining the energy and momentum of these states is subtle, because both are ambiguous on the solitonic Hilbert space.\footnote{Previous papers have studied solitons using infinite tensor network methods \cite{Haegeman:2013vud,Zauner-Stauber:2018gqr}, including in the Schwinger model \cite{Belyansky:2023rgh} (albeit starting from a different continuum formulation). The details of the methods vary and are not always made explicit. In this section, we aim to clearly explain the potential difficulties and give a general prescription for addressing them.} We describe different ways to fix these ambiguities, including using the discrete symmetries of the model (see also Appendix~\ref{app:symmetries}), and compute the soliton dispersion relation.

After having successfully characterized solitons and their properties on infinite lattices, in Section~\ref{sec:method_wavepacket}, we finally describe how we build two-soliton wavepacket states, which we use as the initial states for our collisions in Section~\ref{sec:results}. We also define a basis of two-soliton, nearly plane-wave states and use them to numerically calculate the momentum-space wavefunctions of our initial states.

\subsection{Finite tensor network ingredients}\label{sec:method_lempo}

Throughout this paper, we use symmetric matrix product states as an ansatz for gauge-invariant states on the lattice. First, let us define an ordinary matrix product state on a tensor product Hilbert space $\mathcal{H} = \bigotimes_{n=1}^N \mathcal{H}_n$, with bases $\ket{s_{i_n}}$, $i_n=1,\ldots,\dim \mathcal{H}_n$. Using three-index tensors $A_1,\ldots,A_N$ and boundary vectors $v_L$ and $v_R$, we can define a state
\begin{equation}\label{eq:finite_mps}
\begin{split}
    \ket{\{A_n\},v_L,v_R} &= \sum_{\{i_n\},\{\alpha_n\}} v_L^{\alpha_0} ((A_1)_{i_1}^{\alpha_0,\alpha_1} \ket{s_{i_1}}) \otimes ((A_2)_{i_2}^{\alpha_1,\alpha_2} \ket{s_{i_2}}) \otimes \cdots \otimes ((A_N)_{i_N}^{\alpha_{N-1},\alpha_N} \ket{s_{i_N}}) v_R^{\alpha_N} \\
    &= \begin{tikzpicture}[diagram]
			\node[style=circle] at (-3.5,0) (vL) {$v_L$};
			\node[style=square] at (-2.5,0) (A0) {$A_1$};
            \node at (-1.25,0) {$\cdots$};
            \node at (-1.25,-1) {$\cdots$};
			\node[style=square] at (0,0) (A1) {$A_{n-1}$};
			\node[style=square] at (1.5,0) (A2) {$A_n$};
			\node[style=square] at (3,0) (A3) {$A_{n+1}$};
            \node at (4.25,0) {$\cdots$};
            \node at (4.25,-1) {$\cdots$};
			\node[style=square] at (5.5,0) (AN) {$A_N$};
			\node[style=circle] at (6.5,0) (vR) {$v_R$};
			\node at (-2.5,-1) {$\ket{s_{i_{1}}}$};
			\node at (0,-1) {$\ket{s_{i_{n-1}}}$};
			\node at (1.5,-1) {$\ket{s_{i_n}}$};
			\node at (3,-1) {$\ket{s_{i_{n+1}}}$};
			\node at (5.5,-1) {$\ket{s_{i_{N}}}$};
            \draw (vL) -- (A0);
			\draw (A1) -- (A2) -- (A3);
            \draw (AN) -- (vR);
			\draw (A0) -- ++(0,-.8);
			\draw (A1) -- ++(0,-.8);
			\draw (A2) -- ++(0,-.8);
			\draw (A3) -- ++(0,-.8);
			\draw (AN) -- ++(0,-.8);
			\draw (A0) -- ++(0.75,0);
			\draw (A1) -- ++(-0.75,0);
			\draw (A3) -- ++(0.75,0);
			\draw (AN) -- ++(-0.75,0);
		\end{tikzpicture}\,.
\end{split}
\end{equation}
The indices $\alpha_n$ correspond to virtual spaces $\mathcal{V}_n$; the dimension of $\mathcal{V}_n$ is called the bond dimension of the $n$th link.

For a lattice gauge theory, we do not work on the full tensor product Hilbert space, but only on the subspace of states obeying a Gauss law of the form \eqref{eq:Gauss}, where $Q_n$ is the charge operator on site $n$. We restrict to this subspace by working with a symmetric matrix product state \cite{Singh:2009cd,Buyens:2013yza}. To build such a state, we define a $\U(1)$-grading\footnote{See \cite{Dempsey:2025wia} for the non-abelian case.} of each virtual space $\mathcal{V}_n$ by defining integer-valued operators $\mathcal{L}_n:\mathcal{V}_n\to \mathcal{V}_n$, and then impose the local $\U(1)$-invariance condition
\begin{equation}\label{eq:gauss_symmps}
    \mathcal{L}_n^{\alpha_n,\beta_n} (A_n)_{i_n}^{\alpha_{n-1},\beta_n} - \mathcal{L}_{n-1}^{\alpha_{n-1},\beta_{n-1}} (A_n)_{i_n}^{\beta_{n-1},\alpha_n} = (Q_n)_{i_n,j_n}(A_n)_{j_n}^{\alpha_{n-1},\alpha_n} \,,
\end{equation}
where $Q_n$ assigns $\U(1)$-charges to $\mathcal{H}_n$.  As pointed out in \cite{Dempsey:2025wia}, the condition \eqref{eq:gauss_symmps} implies that we can act on a state with $L_n$ by inserting $\mathcal{L}_n$ on a virtual link in \eqref{eq:finite_mps}:
\begin{equation}\label{eq:finite_mps_Ln}
\begin{split}
    L_n\ket{\{A_n\},v_L,v_R} = \begin{tikzpicture}[diagram]
			\node[style=circle] at (-3.5,0) (vL) {$v_L$};
			\node[style=square] at (-2.5,0) (A0) {$A_1$};
            \node at (-1.25,0) {$\cdots$};
            \node at (-1.25,-1) {$\cdots$};
			\node[style=square] at (0,0) (A1) {$A_{n-1}$};
			\node[style=square] at (1.5,0) (A2) {$A_n$};
			\node[style=circle] at (2.5,0) (L) {$\mathcal{L}_n$};
			\node[style=square] at (3.5,0) (A3) {$A_{n+1}$};
            \node at (4.75,0) {$\cdots$};
            \node at (4.75,-1) {$\cdots$};
			\node[style=square] at (6,0) (AN) {$A_N$};
			\node[style=circle] at (7,0) (vR) {$v_R$};
			\node at (-2.5,-1) {$\ket{s_{i_{1}}}$};
			\node at (0,-1) {$\ket{s_{i_{n-1}}}$};
			\node at (1.5,-1) {$\ket{s_{i_n}}$};
			\node at (3.5,-1) {$\ket{s_{i_{n+1}}}$};
			\node at (6,-1) {$\ket{s_{i_{N}}}$};
            \draw (vL) -- (A0);
			\draw (A1) -- (A2) --(L) -- (A3);
            \draw (AN) -- (vR);
			\draw (A0) -- ++(0,-.8);
			\draw (A1) -- ++(0,-.8);
			\draw (A2) -- ++(0,-.8);
			\draw (A3) -- ++(0,-.8);
			\draw (AN) -- ++(0,-.8);
			\draw (A0) -- ++(0.75,0);
			\draw (A1) -- ++(-0.75,0);
			\draw (A3) -- ++(0.75,0);
			\draw (AN) -- ++(-0.75,0);
		\end{tikzpicture}\,.
\end{split}
\end{equation}
Indeed, for MPS tensors obeying \eqref{eq:gauss_symmps}, the state $\ket{\{A_n\},v_L,v_R}$ obeys the Gauss law \eqref{eq:Gauss}.

For the remainder of this paper, all matrix product states will be $\U(1)$-symmetric matrix product states obeying \eqref{eq:gauss_symmps}. The norm of such a state is given by the contraction
\begin{equation}
    \braket{\{A_n\},v_L,v_R|\{A_n\},v_L,v_R} = \begin{tikzpicture}[diagram,yscale=0.8]
			\node[style=circle] at (-3.5,0) (vL) {$v_L$};
			\node[style=circle] at (-3.5,-2) (vLp) {$\bar{v}_L$};
			\node[style=square] at (-2.5,0) (A0) {$A_1$};
			\node[style=square] at (-2.5,-2) (A0p) {$\bar{A}_1$};
            \node at (-1.25,0) {$\cdots$};
            \node at (-1.25,-1) {$\cdots$};
            \node at (-1.25,-2) {$\cdots$};
			\node[style=square] at (0,0) (A1) {$A_{n-1}$};
			\node[style=square] at (1.5,0) (A2) {$A_n$};
			\node[style=square] at (3,0) (A3) {$A_{n+1}$};
			\node[style=square] at (0,-2) (A1p) {$\bar{A}_{n-1}$};
			\node[style=square] at (1.5,-2) (A2p) {$\bar{A}_n$};
			\node[style=square] at (3,-2) (A3p) {$\bar{A}_{n+1}$};
            \node at (4.25,0) {$\cdots$};
            \node at (4.25,-1) {$\cdots$};
            \node at (4.25,-2) {$\cdots$};
			\node[style=square] at (5.5,0) (AN) {$A_N$};
			\node[style=square] at (5.5,-2) (ANp) {$\bar{A}_N$};
			\node[style=circle] at (6.5,0) (vR) {$v_R$};
			\node[style=circle] at (6.5,-2) (vRp) {$\bar{v}_R$};
            \draw (vL) -- (A0);
            \draw (vLp) -- (A0p);
			\draw (A1) -- (A2) -- (A3);
			\draw (A1p) -- (A2p) -- (A3p);
            \draw (AN) -- (vR);
            \draw (ANp) -- (vRp);
			\draw (A0) -- (A0p);
            \draw (A1) -- (A1p);
            \draw (A2) -- (A2p);
            \draw (A3) -- (A3p);
            \draw (AN) -- (ANp);
			\draw (A0) -- ++(0.75,0);
			\draw (A1) -- ++(-0.75,0);
			\draw (A3) -- ++(0.75,0);
			\draw (AN) -- ++(-0.75,0);
			\draw (A0p) -- ++(0.75,0);
			\draw (A1p) -- ++(-0.75,0);
			\draw (A3p) -- ++(0.75,0);
			\draw (ANp) -- ++(-0.75,0);
		\end{tikzpicture}\,;\label{eq:finite_mps_norm}
\end{equation}
we will always assume our states are normalized. To calculate \eqref{eq:finite_mps_norm} in practice, we do the vertical contractions first, forming transfer matrices
\begin{equation}\label{eq:T}
    T(A_n, \bar{A}_n) \equiv \begin{tikzpicture}[diagram]
        \node[square] at (0,0) (An) {$A_n$};
        \node[square] at (0,-1.5) (Anp) {$\bar{A}_n$};
        \draw (An) -- (Anp);
        \draw (An) -- ++(.75,0);
        \draw (Anp) -- ++(.75,0);
        \draw (An) -- ++(-.75,0);
        \draw (Anp) -- ++(-.75,0);
    \end{tikzpicture}\,,
\end{equation}
and then multiply the transfer matrices:
\begin{equation}\label{eq:finite_mps_norm_T}
    \braket{\{A_n\},v_L,v_R|\{A_n\},v_L,v_R} = (v_L \otimes \bar{v}_L)^T \left(\prod_{j=1}^N T(A_j, \bar{A}_j)\right) (v_R \otimes \bar{v}_R) \,.
\end{equation}

To calculate an expectation value of some local operator $\mathcal{O}_n:\mathcal{H}_n\to \mathcal{H}_n$, we repeat the contraction \eqref{eq:finite_mps_norm} with ${\cal O}_n$ inserted on the $n$th physical leg:
\begin{equation}
    \braket{\{A_n\},v_L,v_R|\mathcal{O}_n|\{A_n\},v_L,v_R} = \begin{tikzpicture}[diagram]
			\node[style=circle] at (-2.5,0) (vL) {$v_L$};
			\node[style=circle] at (-2.5,-2) (vLp) {$\bar{v}_L$};
			\node[style=square] at (-1.5,0) (A0) {$A_1$};
			\node[style=square] at (-1.5,-2) (A0p) {$\bar{A}_1$};
            \node at (-.25,0) {$\cdots$};
            \node at (-.25,-1) {$\cdots$};
            \node at (-.25,-2) {$\cdots$};
			\node[style=square] at (1,0) (A2) {$A_n$};
            \node[style=circle] at (1,-1) (On) {$\mathcal{O}_n$};
			\node[style=square] at (1,-2) (A2p) {$\bar{A}_n$};
            \node at (2.25,0) {$\cdots$};
            \node at (2.25,-1) {$\cdots$};
            \node at (2.25,-2) {$\cdots$};
			\node[style=square] at (3.5,0) (AN) {$A_N$};
			\node[style=square] at (3.5,-2) (ANp) {$\bar{A}_N$};
			\node[style=circle] at (4.5,0) (vR) {$v_R$};
			\node[style=circle] at (4.5,-2) (vRp) {$\bar{v}_R$};
            \draw (vL) -- (A0);
            \draw (vLp) -- (A0p);
            \draw (AN) -- (vR);
            \draw (ANp) -- (vRp);
			\draw (A0) -- (A0p);
            \draw (A2) -- (On) -- (A2p);
            \draw (AN) -- (ANp);
			\draw (A0) -- ++(0.75,0);
			\draw (A0p) -- ++(0.75,0);
			\draw (A2) -- ++(-0.75,0);
			\draw (A2p) -- ++(-0.75,0);
			\draw (A2) -- ++(0.75,0);
			\draw (A2p) -- ++(0.75,0);
			\draw (AN) -- ++(-0.75,0);
			\draw (ANp) -- ++(-0.75,0);
		\end{tikzpicture}\,.\label{eq:finite_On_expval}
\end{equation}
Furthermore, using \eqref{eq:finite_mps_Ln}, we can calculate the expectation value of the electric field $L_n$ by inserting $\mathcal{L}_n$ on a virtual leg:
\begin{equation}
    \braket{\{A_n\},v_L,v_R|L_n|\{A_n\},v_L,v_R} = \begin{tikzpicture}[diagram,yscale=.8]
			\node[style=circle] at (-2.5,0) (vL) {$v_L$};
			\node[style=circle] at (-2.5,-2) (vLp) {$\bar{v}_L$};
			\node[style=square] at (-1.5,0) (A0) {$A_1$};
			\node[style=square] at (-1.5,-2) (A0p) {$\bar{A}_1$};
            \node at (-.25,0) {$\cdots$};
            \node at (-.25,-1) {$\cdots$};
            \node at (-.25,-2) {$\cdots$};
			\node[style=square] at (1,0) (A2) {$A_n$};
			\node[style=circle] at (2,0) (Ln) {$\mathcal{L}_n$};
			\node[style=square] at (3,0) (A3) {$A_{n+1}$};
			\node[style=square] at (1,-2) (A2p) {$\bar{A}_n$};
			\node[style=square] at (3,-2) (A3p) {$\bar{A}_{n+1}$};
            \node at (4.25,0) {$\cdots$};
            \node at (4.25,-1) {$\cdots$};
            \node at (4.25,-2) {$\cdots$};
			\node[style=square] at (5.5,0) (AN) {$A_N$};
			\node[style=square] at (5.5,-2) (ANp) {$\bar{A}_N$};
			\node[style=circle] at (6.5,0) (vR) {$v_R$};
			\node[style=circle] at (6.5,-2) (vRp) {$\bar{v}_R$};
            \draw (vL) -- (A0);
            \draw (vLp) -- (A0p);
			\draw (A2) -- (Ln) -- (A3);
			\draw (A2p) -- (A3p);
            \draw (AN) -- (vR);
            \draw (ANp) -- (vRp);
			\draw (A0) -- (A0p);
            \draw (A2) -- (A2p);
            \draw (A3) -- (A3p);
            \draw (AN) -- (ANp);
			\draw (A0) -- ++(0.75,0);
			\draw (A2) -- ++(-0.75,0);
			\draw (A2p) -- ++(-0.75,0);
			\draw (A3) -- ++(0.75,0);
			\draw (AN) -- ++(-0.75,0);
			\draw (A0p) -- ++(0.75,0);
			\draw (A3p) -- ++(0.75,0);
			\draw (ANp) -- ++(-0.75,0);
		\end{tikzpicture}\,.\label{eq:finite_Ln_expval}
\end{equation}
We can again make use of transfer matrices to evaluate these contractions.

In addition to local operators, we will be interested in the Hamiltonian. Outside the context of gauge theories, it has long been understood how to represent a Hamiltonian as a matrix product operator \cite{Verstraete:2008xyc,Verstraete:2008cex}, given by a contraction of rank-4 tensors $W_n$:
\begin{equation} \label{eq:finite_MPO}
\begin{split}
    H = \begin{tikzpicture}[diagram]
			\node[style=circle] at (-3.5,0) (dL) {$\delta_f$};
			\node[style=circle] at (-2.5,0) (A0) {$W_1$};
			\node at (-2.5,-1) {$\ket{s_{i_1}}$};
			\node at (-2.5,1) {$\bra{s_{j_1}}$};
            \node at (-1.25,0) {$\cdots$};
            \node at (-1.25,-1) {$\cdots$};
            \node at (-1.25,1) {$\cdots$};
			\node[style=circle] at (0,0) (A1) {$W_{n-1}$};
			\node[style=circle] at (1.5,0) (A2) {$W_n$};
			\node[style=circle] at (3,0) (A3) {$W_{n+1}$};
			\node at (0,-1) {$\ket{s_{i_{n-1}}}$};
			\node at (1.5,-1) {$\ket{s_{i_n}}$};
			\node at (3,-1) {$\ket{s_{i_{n+1}}}$};
			\node at (0,1) {$\bra{s_{j_{n-1}}}$};
			\node at (1.5,1) {$\bra{s_{j_n}}$};
			\node at (3,1) {$\bra{s_{j_{n+1}}}$};
            \node at (4.25,0) {$\cdots$};
            \node at (4.25,-1) {$\cdots$};
            \node at (4.25,1) {$\cdots$};
			\node[style=circle] at (5.5,0) (AN) {$W_N$};
			\node at (5.5,-1) {$\ket{s_{i_N}}$};
			\node at (5.5,1) {$\bra{s_{j_N}}$};
			\node[style=circle] at (6.5,0) (dR) {$\delta_i$};
            \draw (dL) -- (A0);
			\draw (A1) -- (A2) -- (A3);
            \draw (AN) -- (dR);
			\draw (A0) -- ++(0,-.8);
			\draw (A1) -- ++(0,-.8);
			\draw (A2) -- ++(0,-.8);
			\draw (A3) -- ++(0,-.8);
			\draw (AN) -- ++(0,-.8);
			\draw (A0) -- ++(0,.8);
			\draw (A1) -- ++(0,.8);
			\draw (A2) -- ++(0,.8);
			\draw (A3) -- ++(0,.8);
			\draw (AN) -- ++(0,.8);
			\draw (A0) -- ++(0.75,0);
			\draw (A1) -- ++(-0.75,0);
			\draw (A3) -- ++(0.75,0);
			\draw (AN) -- ++(-0.75,0);
		\end{tikzpicture}\,.
\end{split}
\end{equation}
Like in \eqref{eq:finite_mps}, vertical lines denote contractions with bases for the on-site Hilbert spaces, while the horizontal lines are contractions in virtual spaces. Unlike the virtual spaces $\V_n$ of the matrix product state tensors $A_n$, the virtual spaces of $W_n$ have some fixed, small dimension, chosen to be just large enough to represent a given Hamiltonian. For instance, if $H = \sum_n \mathcal{O}_n \mathcal{O}_{n+1}$, where each $\mathcal{O}_n$ is an on-site operator, we would take
\begin{equation}
    W_n = \begin{pmatrix}
        \mathds{1} & \mathcal{O}_n & \\
        & & \mathcal{O}_n \\
        & & \mathds{1}
    \end{pmatrix}\,,
\end{equation}
where the matrix indices are for the dimension-three virtual spaces, entries of the matrix are operators on $\mathcal{H}_n$, and the boundary vectors are $\delta_i = (0,0,1)$ and $\delta_f = (1,0,0)$. One can check that contracting these $W_n$ tensors as in \eqref{eq:finite_MPO} gives the desired Hamiltonian. More generally, we shall always take $\delta_i = (0, \ldots, 0, 1)$ and $\delta_f = (1, 0, \ldots, 0)$, and we assume that $W_n$ is an upper triangular matrix, with identity operators in the upper left and lower right corners, and zeroes in the other diagonal entries \cite{Verstraete:2008xyc}:
\begin{equation}
W_n =
\begin{tikzpicture}[baseline=(m.center)]
\matrix (m) [
    matrix of math nodes,
    ampersand replacement=\&,
    left delimiter={(},
    right delimiter={)},
    column sep=.7em,
    row sep=.6em,
    nodes={
        minimum width=1.4em,
        minimum height=1.2em,
        anchor=center
    }
]
{
  \mathds{1}           \& \bullet              \& \bullet              \& \bullet    \\
  \phantom{\mathds{1}} \&               \& \bullet              \& \bullet    \\
  \phantom{\mathds{1}} \& \phantom{\mathds{1}} \&               \& \bullet    \\
  \phantom{\mathds{1}} \& \phantom{\mathds{1}} \& \phantom{\mathds{1}} \& \mathds{1} \\
};

\node[anchor=east, xshift=-0.8em]
    at (m.west |- m-1-1) {$f$};
\path (m.west |- m-2-1) --
      (m.west |- m-3-1)
    node[midway, anchor=east, xshift=-0.8em] {$\vdots$};
\node[anchor=east, xshift=-0.8em]
    at (m.west |- m-4-1) {$i$};

\node[anchor=south, yshift=0.2em]
    at (m-1-1 |- m.north) {$f$};
\path (m-1-2 |- m.north) --
      (m-1-3 |- m.north)
    node[midway, anchor=south, yshift=0.2em] {$\cdots$};
\node[anchor=south, yshift=0.2em]
    at (m-1-4 |- m.north) {$i$};
\draw[thick]
    (m-1-1.south west) --
    (m-1-1.north west) --
    (m-1-4.north east) --
    (m-4-4.south east) --
    (m-4-4.south west) -- cycle;
\end{tikzpicture} \,.
\label{eq:W}
\end{equation}
As explained in \cite{Verstraete:2008xyc}, with these choices in place, the contraction \eqref{eq:finite_MPO} can be interpreted as executing a finite state machine that builds a sum of finite-range Hamiltonian terms over the lattice.

To evaluate the expectation value of the Hamiltonian in some state, we contract the MPO \eqref{eq:finite_MPO} with two MPSs:
\begin{equation}
    \braket{\{A_n\},v_L,v_R|H|\{A_n\},v_L,v_R} = \begin{tikzpicture}[diagram,yscale=0.8]
			\node[style=circle] at (-3.5,0) (vL) {$v_L$};
			\node[style=circle] at (-3.5,-1.5) (dL) {$\delta_f$};
			\node[style=circle] at (-3.5,-3) (vLp) {$\bar{v}_L$};
			\node[style=square] at (-2.5,0) (A0) {$A_1$};
			\node[style=circle] at (-2.5,-1.5) (W0) {$W_1$};
			\node[style=square] at (-2.5,-3) (A0p) {$\bar{A}_1$};
            \node at (-1.25,0) {$\cdots$};
            \node at (-1.25,-1.5) {$\cdots$};
            \node at (-1.25,-3) {$\cdots$};
			\node[style=square] at (0,0) (A1) {$A_{n-1}$};
			\node[style=square] at (1.5,0) (A2) {$A_n$};
			\node[style=square] at (3,0) (A3) {$A_{n+1}$};
			\node[style=circle] at (0,-1.5) (W1) {$W_{n-1}$};
			\node[style=circle] at (1.5,-1.5) (W2) {$W_n$};
			\node[style=circle] at (3,-1.5) (W3) {$W_{n+1}$};
			\node[style=square] at (0,-3) (A1p) {$\bar{A}_{n-1}$};
			\node[style=square] at (1.5,-3) (A2p) {$\bar{A}_n$};
			\node[style=square] at (3,-3) (A3p) {$\bar{A}_{n+1}$};
            \node at (4.25,0) {$\cdots$};
            \node at (4.25,-1.5) {$\cdots$};
            \node at (4.25,-3) {$\cdots$};
			\node[style=square] at (5.5,0) (AN) {$A_N$};
			\node[style=circle] at (5.5,-1.5) (WN) {$W_N$};
			\node[style=square] at (5.5,-3) (ANp) {$\bar{A}_N$};
			\node[style=circle] at (6.5,0) (vR) {$v_R$};
			\node[style=circle] at (6.5,-1.5) (dR) {$\delta_i$};
			\node[style=circle] at (6.5,-3) (vRp) {$\bar{v}_R$};
            \draw (vL) -- (A0);
            \draw (dL) -- (W0);
            \draw (vLp) -- (A0p);
			\draw (A1) -- (A2) -- (A3);
			\draw (W1) -- (W2) -- (W3);
			\draw (A1p) -- (A2p) -- (A3p);
            \draw (AN) -- (vR);
            \draw (WN) -- (dR);
            \draw (ANp) -- (vRp);
			\draw (A0) -- (W0) -- (A0p);
            \draw (A1) -- (W1) -- (A1p);
            \draw (A2) -- (W2) -- (A2p);
            \draw (A3) -- (W3) -- (A3p);
            \draw (AN) -- (WN) -- (ANp);
			\draw (A0) -- ++(0.75,0);
			\draw (A1) -- ++(-0.75,0);
			\draw (A3) -- ++(0.75,0);
			\draw (AN) -- ++(-0.75,0);
			\draw (W0) -- ++(0.75,0);
			\draw (W1) -- ++(-0.75,0);
			\draw (W3) -- ++(0.75,0);
			\draw (WN) -- ++(-0.75,0);
			\draw (A0p) -- ++(0.75,0);
			\draw (A1p) -- ++(-0.75,0);
			\draw (A3p) -- ++(0.75,0);
			\draw (ANp) -- ++(-0.75,0);
		\end{tikzpicture}\,.\label{eq:finite_mpo_expval}
\end{equation}
To carry out this contraction, we define the MPO transfer matrix $T(A_n, W_n, \bar{A}_n)$, in analogy with the MPS transfer matrix $T(A_n, \bar{A}_n)$ in \eqref{eq:T}:
\begin{equation}
    T(A_n, W_n, \bar{A}_n) = \begin{tikzpicture}[diagram,yscale=.8]
			\node[style=square] at (0,0) (A) {$A_n$};
			\node[style=circle] at (0,-1.5) (W) {$W_n$};
			\node[style=square] at (0,-3) (Ap) {$\bar{A}_n$};
			\draw (A) -- (W) -- (Ap);
			\draw (A) -- ++(-0.75,0);
			\draw (A) -- ++(0.75,0);
			\draw (W) -- ++(-0.75,0);
			\draw (W) -- ++(0.75,0);
			\draw (Ap) -- ++(-0.75,0);
			\draw (Ap) -- ++(0.75,0);
		\end{tikzpicture}\,;\label{eq:mpo_transfer_matrix}
\end{equation}
this procedure gives
\begin{equation}
    \braket{\{A_n\},v_L,v_R|H|\{A_n\},v_L,v_R} = (v_L \otimes \delta_f \otimes \bar{v}_L)^T \left(\prod_{j=1}^N T(A_j, W_j, \bar{A}_j)\right) (v_R \otimes \delta_i \otimes \bar{v}_R).
\end{equation}

The construction \eqref{eq:finite_MPO} allows for operators acting on the spaces $\mathcal{H}_n$, but the Hamiltonian of a lattice gauge theory also includes operators acting on links. As we exhibited in \eqref{eq:finite_mps_Ln}, a symmetric matrix product state also admits such operators. To incorporate them into the Hamiltonian, we generalize a matrix product operator to a link-enhanced matrix product operator, or LEMPO \cite{Dempsey:2025wia}. The expectation value of a LEMPO $H$, defined using the tensors $W_n$ discussed above as well as link versions $W_n^{(L)}$, is given by
\begin{equation}
    \braket{\{A_n\},v_L,v_R|H|\{A_n\},v_L,v_R} = \begin{tikzpicture}[diagram,yscale=.8]
			\node[style=circle] at (-2.5,0) (vL) {$v_L$};
			\node[style=circle] at (-2.5,-1.5) (dL) {$\delta_f$};
			\node[style=circle] at (-2.5,-3) (vLp) {$\bar{v}_L$};
            \node at (-1.25,0) {$\cdots$};
            \node at (-1.25,-1.5) {$\cdots$};
            \node at (-1.25,-3) {$\cdots$};
			\node[style=square] at (0,0) (A1) {$A_n$};
			\node[style=circle] at (0,-1.5) (W1) {$W_n$};
			\node[style=square] at (0,-3) (A1p) {$\bar{A}_n$};
			\node[style=circle] at (1.2,0) (L1) {$W_n^{(L)}$};
			\node[style=square] at (2.4,0) (A2) {$A_{n+1}$};
			\node[style=circle] at (2.4,-1.5) (W2) {$W_{n+1}$};
			\node[style=square] at (2.4,-3) (A2p) {$\bar{A}_{n+1}$};
			\node[style=circle] at (3.6,0) (L2) {$W_{n+1}^{(L)}$};
            \node at (4.9,0) {$\cdots$};
            \node at (4.9,-1.5) {$\cdots$};
            \node at (4.9,-3) {$\cdots$};
			\node[style=circle] at (6.15,0) (vR) {$v_R$};
			\node[style=circle] at (6.15,-1.5) (dR) {$\delta_i$};
			\node[style=circle] at (6.15,-3) (vRp) {$\bar{v}_R$};
			\draw (A1) -- (L1) -- (A2);
			\draw (W1) -- (L1) -- (W2);
			\draw (A1p) -- (A2p);
			\draw (A2) -- (L2);
			\draw (W2) -- (L2);
            \draw (A1) -- (W1) -- (A1p);
            \draw (A2) -- (W2) -- (A2p);
			\draw (vL) -- ++(0.75,0);
			\draw (dL) -- ++(0.75,0);
			\draw (vLp) -- ++(0.75,0);
			\draw (A1) -- ++(-0.75,0);
			\draw (W1) -- ++(-0.75,0);
			\draw (A1p) -- ++(-0.75,0);
			\draw (L2) -- ++(0.8,0);
			\draw (L2) -- (4.4,-1.2);
			\draw (A2p) -- ++(2,0);
			\draw (vR) -- ++(-0.75,0);
			\draw (dR) -- ++(-0.75,0);
			\draw (vRp) -- ++(-0.75,0);
		\end{tikzpicture}\,.\label{eq:finite_lempo_expval}
\end{equation}
To evaluate this contraction, we modify the transfer matrix \eqref{eq:mpo_transfer_matrix} to include the contraction with $W_n^{(L)}$:
\begin{equation}
    T(A_n, W_n, W_n^{(L)}, \bar{A}_n) = \begin{tikzpicture}[diagram,yscale=.8]
			\node[style=square] at (0,0) (A) {$A_n$};
			\node[style=circle] at (0,-1.5) (W) {$W_n$};
			\node[style=square] at (0,-3) (Ap) {$\bar{A}_n$};
			\node[style=circle] at (1.2,0) (L) {$W_n^{(L)}$};
			\draw (A) -- (W) -- (Ap);
			\draw (A) -- (L);
			\draw (W) -- (L);
			\draw (A) -- ++(-0.75,0);
			\draw (W) -- ++(-0.75,0);
			\draw (Ap) -- ++(-0.75,0);
			\draw (L) -- (2.1,0);
			\draw (L) -- (2.1,-1.2);
			\draw (Ap) -- (2.1,-3);
		\end{tikzpicture}\,.\label{eq:lempo_link_transfer_matrix}
\end{equation}
We replace the condition \eqref{eq:W} on the form of $W_n$ with the requirement that each $W_n$ and $W_n^{(L)}$ is upper triangular, and that the product
\begin{equation}
    W_n W^{(L)}_n = \begin{tikzpicture}[diagram]
			\node[style=circle] at (0,0) (W) {$W_n$};
			\node[style=circle] at (1.5,1) (L) {$W^{(L)}_n$};
			\draw (W) -- (L);
			\draw (W) -- ++(-0.75,0);
			\draw (L) -- ++(0.75,-1);
			\draw (W) -- ++(0,.8) node[above] {$\mathcal{H}_n$};
			\draw (W) -- ++(0,-.8) node[below] {$\mathcal{H}_n$};
			\draw (L) -- ++(-0.75,0) node[above] {$\mathcal{V}_n$};
			\draw (L) -- ++(0.75,0) node[above] {$\mathcal{V}_n$};
		\end{tikzpicture}\,,\label{eq:combined_lempo_tensor}
\end{equation}
written as a matrix of operators acting on $\mathcal{H}_n \otimes \mathcal{V}_n$, has the particular upper-triangular form \eqref{eq:W}. More details on the form of $W_n^{(L)}$ can be found in \cite{Dempsey:2025wia}, and we give the explicit tensors $W_n$ and $W_n^{(L)}$ for the lattice Schwinger Hamiltonian \eqref{eq:lattice_H_fixed} below.

Let us now apply these constructions to the lattice Schwinger model. In the spin formulation \eqref{eq:lattice_H_fixed} of the lattice Schwinger model, each $\mathcal{H}_n$ is the two-dimensional space of the spin on site $n$. As we described after \eqref{eq:Ln_solved}, the Gauss law and the boundary conditions $L_0 = L_N = 0$ require us to impose the half-filling constraint $\sum_{j=1}^N \sigma^z_j = 0$, which the symmetric matrix product states \eqref{eq:gauss_symmps} implement with $Q_n$ given by \eqref{eq:Ln_solved_jw}.

For the spin Hamiltonian \eqref{eq:lattice_H_fixed}, the site tensors are
\begin{equation}\label{eq:schwinger_MPO}
	W_n = \left(\begin{array}{c|c|c|c}
		\mathds{1}_n & \sigma^+_n & \sigma^-_n & \frac{(-1)^n m_\mathrm{lat}}{2}  \sigma^z_n \\
		\hline
		&  &  & \frac{1}{2a}\sigma^-_n \\
		\hline
		&  &  & \frac{1}{2a}\sigma^+_n \\
		\hline
		&  &  & \mathds{1}_n
	\end{array} \right) \,.
\end{equation}
The operators $\sigma^\pm_n$ and $\sigma^z_n$ act in the usual way on a two-dimensional physical Hilbert space, but due to the staggering of charge in \eqref{eq:Gauss}, these spaces have charges $\{0, 1\}$ and $\{-1, 0\}$ on even and odd sites, respectively. The link tensors, which implement the electric field energy term, are
\begin{equation}\label{eq:schwinger_link_MPO}
	W_n^{(L)} = \left(\begin{array}{c|c|c|c}
		I_n &  &  & \frac{g^2a}{2} \left(\mathcal{L}_n+\frac{\theta}{2\pi}\right)^2 \\
		\hline
		& I_n &  &  \\
		\hline
		&  & I_n &  \\
		\hline
		&  &  & I_n
	\end{array}\right)\,,
\end{equation}
where $I_n$ is the identity on $\mathcal{V}_n$. The link tensor passes the hopping channels through unchanged, and one can check that the product $W_n W_n^{(L)}$ takes the form \eqref{eq:W}. The simple form of $W_n^{(L)}$ reflects the fact that the gauge kinetic term acts on a single link; in a non-abelian theory, $\left(\mathcal{L}_n + \frac{\theta}{2\pi}\right)^2$ is replaced by the quadratic Casimir \cite{Dempsey:2025wia,Dempsey:2026hlf}.

\subsection{Vacua on an infinite lattice}\label{sec:method_boundary}

To prepare states of colliding solitons, we first need to prepare individual solitons; and to do so, we need to prepare the degenerate vacua $\ket{v_\pm}$ of the lattice Schwinger model at $\theta = \pi$. The key step in finding these vacuum states, relative to Section~\ref{sec:method_lempo}, is to pass from a finite lattice to an infinite one.

To go to an infinite lattice while keeping the number of variational parameters finite, we impose manifest translation invariance: $A_n = A_{n+k}$, where $k$ is the size of a unit cell. Then, in all the diagrams of Section~\ref{sec:method_lempo}, we replace a contraction spanning a finite $N$-site lattice with an infinite contraction; that infinite contraction can be framed as the repeated application of a transfer matrix. A state on an infinite lattice obeying the periodicity $A_n = A_{n+k}$ is called a uniform matrix product state (uMPS) \cite{Vidal:2006ofj}. We will denote uniform matrix product states using the same notation as for finite matrix product states, $\ket{\{A_n\}, v_L, v_R}$; it will be clear from context when we mean for the MPS tensors to obey the uniformity condition.

We can start by normalizing a uMPS. Let
\begin{equation}
    T_{1\cdots k} = \prod_{j=1}^k T(A_j, \bar{A}_j) \,,
\end{equation}
with the factors defined in \eqref{eq:T}. Then, taking the lattice size infinite in \eqref{eq:finite_mps_norm_T}, we find
\begin{equation}\label{eq:umps_norm}
    \braket{\{A_n\},v_L,v_R|\{A_n\},v_L,v_R} = \lim_{N\to\infty} \left\lbrack (v_L \otimes \bar{v}_L)^T T_{1\cdots k}^N (v_R \otimes \bar{v}_R)\right\rbrack\,.
\end{equation}
In order for the limit to be well-defined, the largest-magnitude eigenvalue of $T_{1\cdots k}$ must be 1; we normalize the MPS tensors so that this condition holds, and we assume throughout that there are no other eigenvalues of unit magnitude.\footnote{One can show that cluster decomposition on the lattice only holds when $A$ is injective, meaning the transfer matrix $T_{1\cdots k}$ has a unique dominant eigenvalue \cite{Fannes:1990ur,Perez-Garcia:2006nqo}. When we have a unique ground state, it will be approximated by an injective uMPS; when we have degenerate ground states, enforcing cluster decomposition picks out the basis we want, and each cluster decomposing vacuum will be approximated by an injective uMPS.} We denote the left- and right-dominant eigenvectors of $T_{1\cdots k}$ by $l$ and $r$, respectively, so that
\begin{equation}\label{eq:lr}
    l^T T_{1\cdots k} = l^T, \qquad T_{1\cdots k} r = r\,.
\end{equation}
We normalize these eigenvectors by $(l, r) = 1$, where $(x, y) \equiv x^T y$ will be the inner product on transfer matrix spaces going forward. Then,  \eqref{eq:umps_norm} reduces to
\begin{equation}\label{eq:normA_3}
    \braket{\{A_n\},v_L,v_R|\{A_n\},v_L,v_R} = (v_L \otimes \bar{v}_L, r) (l, v_R \otimes \bar{v}_R)\,.
\end{equation}
In order to normalize the state, we require that this product is 1. For convenience, we will furthermore require $(v_L\otimes \bar{v}_L, r) = (l, v_R \otimes \bar{v}_R) = 1$ separately.

Once we have a normalized uniform matrix product state, we can calculate local expectation values by passing to the infinite version of \eqref{eq:finite_Ln_expval}; we will use a link operator (acting on $\mathcal{V}_n$) as an example, although the same idea holds for a site operator (acting on $\mathcal{H}_n$). Let us define
\begin{equation}
    \tilde{T}_{1\cdots k}(L_n) = \begin{tikzpicture}[diagram,yscale=.8]
			\node[style=square] at (-2.5,0) (A0) {$A_1$};
			\node[style=square] at (-2.5,-2) (A0p) {$\bar{A}_1$};
            \node at (-1.25,0) {$\cdots$};
            \node at (-1.25,-1) {$\cdots$};
            \node at (-1.25,-2) {$\cdots$};
			\node[style=square] at (0,0) (A1) {$A_n$};
			\node[style=square] at (0,-2) (A1p) {$\bar{A}_n$};
			\node[style=circle] at (1,0) (L) {$\mathcal{L}_n$};
			\node[style=square] at (2,0) (A2) {$A_{n+1}$};
			\node[style=square] at (2,-2) (A2p) {$\bar{A}_{n+1}$};
            \node at (3.25,0) {$\cdots$};
            \node at (3.25,-1) {$\cdots$};
            \node at (3.25,-2) {$\cdots$};
			\node[style=square] at (4.5,0) (AN) {$A_k$};
			\node[style=square] at (4.5,-2) (ANp) {$\bar{A}_k$};
			\draw (A1) -- (L) -- (A2);
			\draw (A1p) -- (A2p);
			\draw (A0) -- (A0p);
            \draw (A1) -- (A1p);
            \draw (A2) -- (A2p);
            \draw (AN) -- (ANp);
			\draw (A0) -- ++(-0.75,0);
			\draw (A0p) -- ++(-0.75,0);
			\draw (A0) -- ++(0.75,0);
			\draw (A0p) -- ++(0.75,0);
			\draw (A1) -- ++(-0.75,0);
			\draw (A1p) -- ++(-0.75,0);
			\draw (A2) -- ++(0.75,0);
			\draw (A2p) -- ++(0.75,0);
			\draw (AN) -- ++(-0.75,0);
			\draw (ANp) -- ++(-0.75,0);
			\draw (AN) -- ++(0.75,0);
			\draw (ANp) -- ++(0.75,0);
		\end{tikzpicture}\,.\label{eq:umps_transfer_link_insertion}
\end{equation}
We then have
\begin{equation}\label{eq:Osandwich_limit}
    \braket{\{A_n\},v_L,v_R|L_n|\{A_n\},v_L,v_R} = \lim_{N\to\infty} (v_L \otimes \bar{v}_L)^T \,T_{1\cdots k}^N \;\tilde{T}_{1\cdots k}(L_n)\; T_{1\cdots k}^N\, (v_R\otimes \bar{v}_R)\,.
\end{equation}
Using the eigendecomposition of $T_{1\cdots k}$ to evaluate the limit, we find
\begin{equation}\label{eq:exp_On}
    \braket{\{A_n\},v_L,v_R|L_n|\{A_n\},v_L,v_R} = (l, \tilde{T}_{1\cdots k}(L_n) r)\,.
\end{equation}
Note that the boundary vectors $v_L$ and $v_R$ drop out of the right-hand side (once we use the normalization condition given after \eqref{eq:normA_3}). Indeed, in an infinite system, we would not expect the boundary conditions to affect the expectation value of a local operator.

To determine a vacuum state, we need to minimize the energy of a uMPS\@. Of course, this is not actually well-defined: the energy is extensive and thus is infinite for a uMPS\@. The finite quantity we can minimize is the {\em energy density}, defined as the total energy divided by the number of unit cells. The most straightforward way to calculate the energy density, for a Hamiltonian like \eqref{eq:lattice_H_fixed} that is a sum of repeating local terms, is to define a Hamiltonian density operator $h_n$ and calculate its expectation value using the approach of \eqref{eq:exp_On}. (In practice, we find uMPS approximations to vacuum states using the VUMPS algorithm \cite{Zauner-Stauber:2017eqw,Vanderstraeten:2019voi}, which works directly with the LEMPO and the environments we will introduce in Section~\ref{sec:method_envs}; see Appendix~\ref{app:lempo} for how VUMPS is adapted to LEMPOs.)

Let us now turn to the lattice Schwinger model, for which the unit cell has $k = 2$ sites. At $\theta = \pi$ and $m > m_c$, we need uMPS approximations to the two degenerate vacua $\ket{v_\pm}$. Specifically, we determine one vacuum, $\ket{v_-}$, and then construct $\ket{v_+}$ by acting with $\mathcal{C}$, as described in Appendix~\ref{app:symmetries}\@. For the moment, let us set aside the manner in which we obtain the states $\ket{v_\pm}$, and focus on studying them using expectation values.

\begin{figure}
    \centering
    \includegraphics[width=0.6\linewidth]{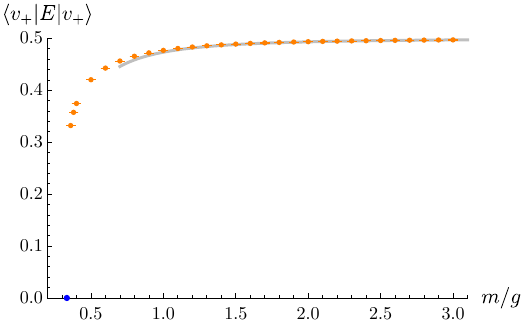}
    \caption{We plot the expectation value $\braket{v_+|E|v_+}$ as a function of $m/g$. The expectation value turns on rapidly above $m = m_c$, reflecting the 1/8 power-law scaling of \eqref{eq:scaling}, and it approaches 1/2 at $m\gg g$. We calculate the expectation value in uMPS approximations to $\ket{v_+}$ at several finite lattice spacings $ag$, and then extrapolate the result as $ag\to 0$.}
    \label{fig:eplus}
\end{figure}

First, we can calculate $\braket{v_+|E|v_+}$, which appeared in \eqref{eq:qsol_def}. To approximate this expectation value on the lattice, we use the identification
\begin{equation}\label{eq:eeff_lattice}
    E(na) \mapsto \frac{L_n + L_{n+1}}{2} + \frac{\theta}{2\pi}\,.
\end{equation}
We expect that for $m\gg g$ the expectation value is approximately $1/2$, and as $m$ approaches $m_c$ from above it approaches 0. And indeed, in Figure~\ref{fig:eplus}, we see this behavior. The rapid onset of this expectation value as $m$ increases above $m_c$ reflects the 1/8 power-law scaling in \eqref{eq:scaling}.

\begin{figure}
    \centering
    \includegraphics[width=0.75\linewidth]{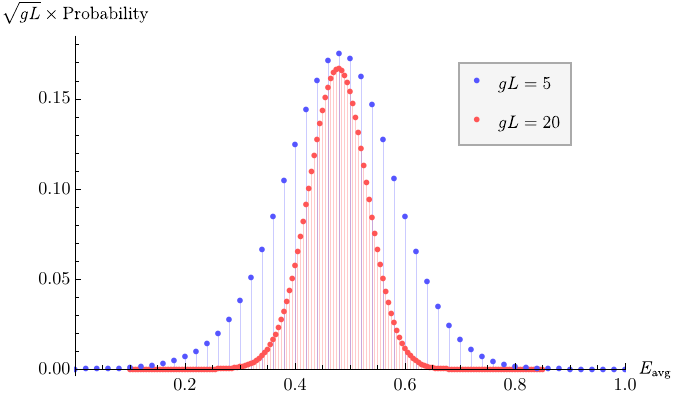}
    \caption{The probabilities for measuring different values of $E_\text{avg}$ on intervals of lengths $gL = 5$ and 20, computed at $m/g = 1$ and $ag = 0.1$ using the method described in Appendix~\ref{app:avg_e_field}\@. We rescale the probabilities by $\sqrt{gL}$ so that the peaks of the distributions are similar; when we increase $gL$ from 5 to 20, the standard deviation is roughly halved, consistent with the scaling in \eqref{eq:sigma_eavg}.}
    \label{fig:vacuum_histogram}
\end{figure}

We can also measure the electric field averaged over some region of fixed physical width. This is arguably more physical: a measurement apparatus probes infrared observables rather than the details of the ultraviolet definition of a theory. In the continuum, we can define
\begin{equation}
    E_\text{avg}(x_1,x_2) = \frac{1}{x_2-x_1} \int_{x_1}^{x_2} dx\, E(x)\,,
\end{equation}
which is straightforwardly discretized using \eqref{eq:eeff_lattice}. In a translation-invariant state, its variance depends only on $L \equiv x_2 - x_1$ and is given by
\begin{equation}\label{eq:sigma_eavg}
    \sigma^2({E}_\text{avg}) = \frac{2}{L^2} \int_0^L dx\,(L-x) \langle E(0) E(x)\rangle_c\,.
\end{equation}
When we have a finite correlation length (in units of $g^{-1}$), the correlator in the integral is bounded by an exponential and so we find $\sigma(E_\text{avg}) \sim (gL)^{-1/2}$ at large $gL$. In Figure~\ref{fig:vacuum_histogram}, we plot histograms illustrating this variance suppression on the lattice at $m/g = 1$; the method for calculating such histograms is detailed in Appendix~\ref{app:avg_e_field}. We will study the probability distribution of $E_\text{avg}$ again in Section~\ref{sec:results}.

\subsection{Hamiltonian and environments}\label{sec:method_envs}

In Section~\ref{sec:method_boundary}, we explained that we could find uMPS approximations to vacuum states by minimizing their energy density, calculated as the expectation value of a local Hamiltonian density operator. However, our next step will be to calculate the dispersion relation for the soliton states, which entails finding the energy of solitonic momentum eigenstates.\footnote{\label{fn:normalizability}Such eigenstates are not normalizable, so by their energy, we mean the expectation value of the Hamiltonian divided by the norm of the eigenstate.} Since the energy of a soliton exceeds that of the vacuum by a finite amount, the energy density of a soliton state of definite momentum is identical to that of the vacuum (or of the degenerate vacua). Thus, as a step towards calculating the finite energy of a soliton state, we now show how to handle the extensive, divergent energy of a uMPS.

Put another way, we need to understand the $N\to\infty$ limit of \eqref{eq:finite_lempo_expval}. To this end, we define
\begin{equation}
\begin{split}
    \mathbb{T} &= T(A_1, W_1, W_1^{(L)}, \bar{A}_1) \cdots T(A_k, W_k, W_k^{(L)}, \bar{A}_k) \\
    &= \begin{tikzpicture}[diagram,yscale=.8]
			\node[style=square] at (0,0) (A1) {$A_1$};
			\node[style=circle] at (0,-1.5) (W1) {$W_1$};
			\node[style=square] at (0,-3) (A1p) {$\bar{A}_1$};
			\node[style=circle] at (1.5,0) (L1) {$W^{(L)}_1$};
            \node at (2.75,0) {$\cdots$};
            \node at (2.75,-1.5) {$\cdots$};
            \node at (2.75,-3) {$\cdots$};
			\node[style=square] at (4,0) (Ak) {$A_k$};
			\node[style=circle] at (4,-1.5) (Wk) {$W_k$};
			\node[style=square] at (4,-3) (Akp) {$\bar{A}_k$};
			\node[style=circle] at (5.5,0) (Lk) {$W^{(L)}_k$};
			\draw (A1) -- (W1) -- (A1p);
			\draw (Ak) -- (Wk) -- (Akp);
			\draw (A1) -- (L1);
			\draw (W1) -- (L1);
			\draw (Ak) -- (Lk);
			\draw (Wk) -- (Lk);
			\draw (A1) -- ++(-0.75,0);
			\draw (W1) -- ++(-0.75,0);
			\draw (A1p) -- ++(-0.75,0);
			\draw (L1) -- (2.25,0);
			\draw (L1) -- (2.25,-1.5);
			\draw (A1p) -- (2.25,-3);
			\draw (Ak) -- ++(-0.75,0);
			\draw (Wk) -- ++(-0.75,0);
			\draw (Akp) -- ++(-0.75,0);
			\draw (Lk) -- (6.5,0);
			\draw (Lk) -- (6.5,-1.5);
			\draw (Akp) -- (6.5,-3);
		\end{tikzpicture}\,.
\end{split}
\label{eq:lempo_unit_cell_transfer_matrix}
\end{equation}
We then see from \eqref{eq:finite_lempo_expval} that, formally,
\begin{equation}\label{eq:uMPS_H}
\begin{split}
    \braket{\{A_n\},v_L,v_R|H|\{A_n\},v_L,v_R} = \lim_{N\to\infty} (\mathbbm{v}_L, \mathbb{T}^N \mathbbm{v}_R)\\
    \text{with}\qquad \mathbbm{v}_L \equiv v_L \otimes \delta_f \otimes \bar{v}_L\,, \quad \mathbbm{v}_R \equiv v_R \otimes \delta_i \otimes \bar{v}_R\,.
\end{split}
\end{equation}

We had a very similar expression in \eqref{eq:umps_norm}. In that case, we used an eigendecomposition of $T_{1\cdots k}$ to evaluate the infinite product. Since $T_{1\cdots k}$ had a unique dominant eigenvalue, the boundary vectors $v_L$ and $v_R$ survived only through their inner products with the fixed point vectors $l$ and $r$, and we normalized those inner products to 1. That style of argument must fail here, because the expectation value of the Hamiltonian should behave like
\begin{equation}\label{eq:aha_form}
    \braket{\{A_n\},v_L,v_R|H|\{A_n\},v_L,v_R} = N \epsilon + E_\text{bdry}
\end{equation}
on a lattice of $N$ unit cells. Here, $\epsilon$ is the energy per unit cell, which can also be computed by evaluating the expectation value of the Hamiltonian density, and $E_\text{bdry}$ is a finite contribution that, as we will see, depends on details of the boundary vectors $v_L$ and $v_R$.

Indeed, there is a key difference here: $\mathbb{T}$ has a dominant eigenvalue of 1, but with an algebraic multiplicity of 2. This property is a consequence of the two identity entries on the diagonal of each $W_n W^{(L)}_n$ (see \eqref{eq:combined_lempo_tensor}), which takes the form \eqref{eq:W}.\footnote{Since $\mathbb{T}$ is then a product of upper triangular factors, its characteristic polynomial factorizes as
 $$
    \det(\mathbb{T}-\lambda \mathbbm{I}) = \prod_{a=1}^D \det\left(\mathbb{T}_{aa} - \lambda \mathbbm{I}\right),
    \qquad
    \mathbb{T}_{aa} = \prod_{j=1}^{k} T\big(A_j, (W_j)_{aa}, (W^{(L)}_j)_{aa}, \bar{A}_j\big)\,.
$$
Here $(W_j)_{aa}$ and $(W^{(L)}_j)_{aa}$ are the diagonal entries of the LEMPO tensors, $D$ is the virtual space dimension of the LEMPO, and $T\big(A_j, (W_j)_{aa}, (W^{(L)}_j)_{aa}, \bar{A}_j\big)$ is the LEMPO transfer matrix defined in \eqref{eq:lempo_link_transfer_matrix}. For $a = 1$ and $a = D$, the diagonal entries are the identity, so $\mathbb{T}_{11} = \mathbb{T}_{DD} = T_{1\cdots k}$, whose dominant eigenvalue is 1; for $1 < a < D$, they vanish, so $\mathbb{T}_{aa} = 0$. Since $T_{1\cdots k}$ has a unique dominant eigenvalue, the algebraic multiplicity of the eigenvalue 1 of $\mathbb{T}$ is exactly 2.} Generically, the multiplicity does not imply the existence of two eigenvectors with eigenvalue 1; it only implies that, in the Jordan form of $\mathbb{T}$, there is a $2\times 2$ Jordan block with 1s on the diagonal.

Using this fact, we will show that \eqref{eq:uMPS_H} leads to \eqref{eq:aha_form}. In \eqref{eq:umps_norm}, the $N\to\infty$ limit projected away all the information about $v_R$ except for its overlap with the dominant eigenvector $r$ of $T_{1\cdots k}$. In this case, however, the two-dimensional projection of $\mathbbm{v}_R$ onto the generalized eigenspace of $\mathbb{T}$ with eigenvalue $1$ survives the limit, and so more of the boundary information is relevant to us. 

To see how, we need a basis for the generalized eigenspace. We will use a vector $\mathbb{E}_R$ called the right environment (standard terminology) and a vector $\mathbb{A}_R$ called the right accumulator (less so), satisfying the generalized eigenvalue equations
\begin{equation}\label{eq:E_and_A}
    \mathbb{T} \mathbb{E}_R = \mathbb{E}_R + \epsilon \mathbb{A}_R \,,  \qquad \mathbb{T} \mathbb{A}_R = \mathbb{A}_R \,,
\end{equation}
with $\epsilon$ being a non-vanishing constant that (so far) we are free to choose.  These are the standard generalized eigenvalue equations for a matrix with a $2 \times 2$ Jordan block corresponding to eigenvalue $1$.\footnote{In linear algebra, we would say that $(\epsilon \mathbb{A}_R , \mathbb{E}_R)$ form a Jordan chain.} The interpretation is that, when we build the contraction \eqref{eq:finite_lempo_expval} by adding an additional unit cell to the left, the energy of that unit cell is recorded in the coefficient of the accumulator. We can similarly define the left environment $\mathbb{E}_L$ and left accumulator $\mathbb{A}_L$ via the solutions to the generalized eigenvalue equations
\begin{equation}\label{eq:E_and_A_left}
    \mathbb{E}_L^T \mathbb{T}  = \mathbb{E}_L^T + \epsilon \mathbb{A}_L^T \,, \qquad \mathbb{A}_L^T \mathbb{T}  = \mathbb{A}_L^T \,,
\end{equation}
which tell us what happens when we extend \eqref{eq:finite_lempo_expval} by adding an additional unit cell to the right.

Using the constrained form \eqref{eq:W} of the product \eqref{eq:combined_lempo_tensor}, we can construct the accumulators in terms of the boundary vectors $\delta_i$ and $\delta_f$ for the LEMPO and of the left and right fixed points $l$ and $r$ of $T_{1\cdots k}$:
\begin{equation}\label{eq:A_def}
	\begin{tikzpicture}[diagram]
		\draw (-1.5,0) -- (0,0) (-1.5,-1) -- (0,-1) (-1.5,-2) -- (0,-2);
		\draw[rounded corners,fill=white] (-.5,-2.3) rectangle node[scale=1.2] {$\mathbb{A}_R$} (.5,.3);
	\end{tikzpicture}=
	\begin{tikzpicture}[diagram]
		\node[style=circle] at (0,-1) (d) {$\delta_f$};
		\node[style=circle] at (1,-1) (r) {$r$};
		\draw (-1,-1) -- (d);
		\draw[rounded corners=8pt] (-1,0) -- (1,0) -- (r) -- (1,-2) -- (-1,-2);
	\end{tikzpicture}\;,\qquad
	\begin{tikzpicture}[diagram]
		\draw (0,0) -- (1.5,0) (0,-1) -- (1.5,-1) (0,-2) -- (1.5,-2);
		\draw[rounded corners,fill=white] (-.5,-2.3) rectangle node[scale=1.2] {$\mathbb{A}_L$} (.5,.3);
	\end{tikzpicture}=
	\begin{tikzpicture}[diagram]
		\node[style=circle] at (0,-1) (d) {$\delta_i$};
		\node[style=circle] at (-1,-1) (l) {$l$};
		\draw (d) -- (1,-1);
		\draw[rounded corners=8pt] (1,0) -- (-1,0) -- (l) -- (-1,-2) -- (1,-2);
	\end{tikzpicture}
\end{equation}
Additionally, one can check that consistency of the four inner products $(\mathbb{A}_L, \mathbb{T} \mathbb{A}_R)$, $(\mathbb{A}_L, \mathbb{T} \mathbb{E}_R)$, $(\mathbb{E}_L, \mathbb{T} \mathbb{A}_R)$, $(\mathbb{E}_L, \mathbb{T} \mathbb{E}_R)$, computed by acting with $\mathbb{T}$ either on the left or on the right, imposes the conditions
 \begin{equation} \label{eq:automatic}
     (\mathbb{A}_L, \mathbb{E}_R) = (\mathbb{E}_L, \mathbb{A}_R) \,, \qquad (\mathbb{A}_L, \mathbb{A}_R) = 0 \,.
 \end{equation}
We can then further impose the condition
\begin{equation}\label{eq:EA_products}
    (\mathbb{A}_L, \mathbb{E}_R) = 1  \,.
\end{equation}
The condition in \eqref{eq:EA_products}, combined with \eqref{eq:A_def}, fixes the normalizations of both the environments and the accumulators such that $\epsilon$ (as defined in \eqref{eq:E_and_A}) takes the value of the energy per unit cell (see also Appendix C in \cite{Zauner-Stauber:2017eqw}). Therefore, for a uMPS with extensive energy, we can calculate the energy density by solving the generalized eigenvalue equations \eqref{eq:E_and_A} for its LEMPO transfer matrix and imposing \eqref{eq:A_def} and \eqref{eq:EA_products}. The conditions \eqref{eq:E_and_A}--\eqref{eq:A_def}, and \eqref{eq:EA_products} do not fully fix the environment vectors; these equations are still satisfied if we change the environments by 
\begin{equation}\label{eq:etas}
\mathbb{E}_L \to \mathbb{E}_L + \eta_L \mathbb{A}_L\,,\qquad \mathbb{E}_R \to \mathbb{E}_R + \eta_R \mathbb{A}_R\,,
\end{equation}
for some constants $\eta_L,\eta_R$.

To evaluate $\braket{\{A_n\},v_L,v_R|H|\{A_n\},v_L,v_R}$, we first decompose the boundary vectors $\mathbbm{v}_R$ and $\mathbbm{v}_L$ into linear combinations of the (generalized) eigenvectors of $\mathbb{T}$:  
\begin{equation}
 \begin{aligned}
    \mathbbm{v}_R &= \alpha_R \mathbb{A}_R + \mathbb{E}_R + \mathbbm{v}_R' \,, \qquad
    \mathbbm{v}_L = \alpha_L \mathbb{A}_L + \mathbb{E}_L + \mathbbm{v}_L'\,, \label{eq:vDecomp}
 \end{aligned}
\end{equation}
where $\mathbbm{v}_R'$ and $\mathbbm{v}_L'$ are linear combinations of other right and left (generalized) eigenvectors (with eigenvalues less than one) of $\mathbb{T}$.\footnote{Using \eqref{eq:A_def}--\eqref{eq:EA_products}, one finds that the coefficients of $\mathbb{E}_R$ and $\mathbb{E}_L$ in \eqref{eq:vDecomp} are $(l, v_R\otimes \bar{v}_R)$ and $(v_L\otimes \bar{v}_L, r)$, respectively, which were set to 1 below \eqref{eq:normA_3}.} 
Then, for the contraction at large~$N$, up to exponentially small terms, we have
\begin{equation}\label{eq:jordan_power_1}
 \begin{aligned}
    \mathbb{T}^N \mathbbm{v}_R &\approx \mathbb{E}_R + (\alpha_R + N\epsilon) \mathbb{A}_R  \,, \qquad
   \mathbbm{v}_L^T \mathbb{T}^N  \approx \mathbb{E}_L^T + (\alpha_L + N\epsilon) \mathbb{A}_L^T  \,, 
 \end{aligned}
\end{equation}
and so (ignoring the remainder that vanishes as $N\to\infty$)
\begin{equation}\label{eq:jordan_power_2}
    \braket{\{A_n\},v_L,v_R|H|\{A_n\},v_L,v_R} = (\mathbbm{v}_L, \mathbb{T}^N \mathbbm{v}_R) = N\epsilon + (\mathbb{E}_L, \mathbb{E}_R) +\alpha_L + \alpha_R \,.
\end{equation}
We see that we have reproduced \eqref{eq:aha_form}, with $E_\text{bdry} = (\mathbb{E}_L, \mathbb{E}_R) + \alpha_L + \alpha_R$. 

In the following, it will be convenient to fix half of the ambiguity \eqref{eq:etas} by imposing the additional condition
\begin{equation}\label{eq:EE0}
    (\mathbb{E}_L, \mathbb{E}_R) = 0\,,
\end{equation}
so that $E_\text{bdry} = \alpha_L + \alpha_R$.

\subsection{Particles and solitons}\label{sec:method_qp}
 
Let us now turn from preparing vacua to preparing soliton states. For this task we use the quasiparticle ansatz~\cite{Haegeman:2011lcd, Haegeman:2013xcv}, which describes momentum eigenstates of particles or solitons. In this ansatz, we construct matrix product states by using the tensors $A_1,\ldots,A_k$ of one uMPS to the left, the tensors $\tilde{A}_1,\ldots,\tilde{A}_k$ of another uMPS to the right, and a new set of tensors $B_1,\ldots,B_k$ inserted at a unit cell $n$ between them:\footnote{Note that in \eqref{eq:Binsertion} we are using an abbreviation for the product of the $k$ tensors in a unit cell, where
\begin{equation}
 \begin{aligned}
	\begin{tikzpicture}[diagram]
		\node[style=square] at (0,0) (A) {$\{A\}$};
		\draw (A) -- ++(-0.75,0);
		\draw (A) -- ++(0.75,0);
		\draw (A) -- ++(0,-.8);
	\end{tikzpicture}
	\; &\equiv \;
	\begin{tikzpicture}[diagram]
		\node[style=square] at (0,0) (A1) {$A_1$};
		\node[style=square] at (1.25,0) (A2) {$A_2$};
		\node at (2.5,0) {$\cdots$};
		\node[style=square] at (3.75,0) (Ak) {$A_k$};
		\draw (A1) -- (A2);
		\draw (A2) -- ++(0.75,0);
		\draw (Ak) -- ++(-0.75,0);
		\draw (A1) -- ++(-0.75,0);
		\draw (Ak) -- ++(0.75,0);
		\draw (A1) -- ++(0,-.8);
		\draw (A2) -- ++(0,-.8);
		\draw (Ak) -- ++(0,-.8);
	\end{tikzpicture}\,, \qquad
%%%%%%%%%%%%%%%%%%%%%
   	\begin{tikzpicture}[diagram]
		\node[style=square] at (0,0) (A) {$\{\tilde A\}$};
		\draw (A) -- ++(-0.75,0);
		\draw (A) -- ++(0.75,0);
		\draw (A) -- ++(0,-.8);
	\end{tikzpicture}
	\; \equiv \;
	\begin{tikzpicture}[diagram]
		\node[style=square] at (0,0) (A1) {$\tilde A_1$};
		\node[style=square] at (1.25,0) (A2) {$\tilde A_2$};
		\node at (2.5,0) {$\cdots$};
		\node[style=square] at (3.75,0) (Ak) {$\tilde A_k$};
		\draw (A1) -- (A2);
		\draw (A2) -- ++(0.75,0);
		\draw (Ak) -- ++(-0.75,0);
		\draw (A1) -- ++(-0.75,0);
		\draw (Ak) -- ++(0.75,0);
		\draw (A1) -- ++(0,-.8);
		\draw (A2) -- ++(0,-.8);
		\draw (Ak) -- ++(0,-.8);
	\end{tikzpicture}\,,  \\[.5em]
    	\begin{tikzpicture}[diagram]
		\node[style=square] at (0,0) (B) {$\{B\}$};
		\draw (B) -- ++(0,-.8);
		\draw (B) -- ++(-0.75,0);
		\draw (B) -- ++(0.75,0);
	\end{tikzpicture}
	\; &\equiv \;  \sum_{j=1}^k e^{ipaj}\; \begin{tikzpicture}[diagram]
		\node[style=square] at (0,0) (A1) {$A_1$};
		\node[style=square] at (1.5,0) (A2) {$A_2$};
		\node at (2.5,0) (dots1) {$\cdots$};
		\node[style=square] at (3.5,0) (B) {$B_j$};
		\node at (4.5,0) (dots2) {$\cdots$};
		\node[style=square] at (5.5,0) (A3) {$\tilde{A}_k$};
		\draw (A1) -- (A2) -- (dots1) -- (B) -- (dots2) -- (A3);
		\draw (A1) -- ++(0,-.8);
		\draw (A2) -- ++(0,-.8);
		\draw (A3) -- ++(0,-.8);
		\draw (B) -- ++(0,-.8);
		\draw (A1) -- ++(-0.75,0);
		\draw (A3) -- ++(0.75,0);
	\end{tikzpicture}\,.
	\label{eq:unit_cell_abbreviation}
  \end{aligned}
\end{equation}}
 \begin{equation}
	\ket{n, \{B\}} =
	\begin{tikzpicture}[diagram]
		\node[style=square] at (-1.5,0) (A0) {$\{A\}$};
		\node[style=square] at (0,0) (A1) {$\{A\}$};
		\node[style=square] at (1.5,0) (A2) {$\{B\}$};
		\node[style=square] at (3,0) (A3) {$\{\tilde{A}\}$};
		\node[style=square] at (4.5,0) (A4) {$\{\tilde{A}\}$};
		\node at (-2.5,0) (dots1) {$\cdots$};
		\node at (5.5,0) (dots2) {$\cdots$};
		\node at (-2.5,-1) {$\cdots$};
		\node at (5.5,-1) {$\cdots$};
 
		\draw (A0) -- (A1) -- (A2) -- (A3) -- (A4);
		\draw (A0) -- ++(0,-.8);
		\draw (A1) -- ++(0,-.8) node[below] {$n-1$};
		\draw (A2) -- ++(0,-.8) node[below] {$n$};
		\draw (A3) -- ++(0,-.8) node[below] {$n+1$};
		\draw (A4) -- ++(0,-.8);
		\draw (A0) -- ++(-0.75,0);
		\draw (A4) -- ++(0.75,0);
	\end{tikzpicture}\,.
	\label{eq:Binsertion}
\end{equation}
By summing over the insertion point $n$ with a phase factor, we can build a state of definite momentum $p$:\footnote{In the case of particle excitations (with $\{A\} = \{\tilde{A}\}$), when $p = 0$ the ansatz \eqref{eq:Bsoliton} could give us back the vacuum state if $\{B\} = \{A\}$. To disallow this, we require \eqref{eq:Bsoliton} to be orthogonal to the vacuum, which amounts to \cite{Vanderstraeten:2019voi}
\begin{equation}\label{eq:p0_ortho}
	\begin{tikzpicture}[diagram,baseline=(L.center)]
		\node[style=square] at (0,0) (B) {$\{B\}$};
		\node[style=square] at (0,-1.5) (A) {$\{\bar A\}$};
		\node[style=circle] at (-1,-.75) (L) {$l$};
		\node[style=circle] at (1,-.75) (R) {$r$};
 
		\draw (B.south) -- (A.north);
		\draw[rounded corners=8pt]
		(L.north) -- (-1,0) -- (B.west);
		\draw[rounded corners=8pt]
		(A.west) -- (-1,-1.5) -- (L.south);
		\draw[rounded corners=8pt]
		(B.east) -- (1,0) -- (R.north);
		\draw[rounded corners=8pt]
		(R.south) -- (1,-1.5) -- (A.east);
	\end{tikzpicture}
	= 0\,.
\end{equation}
For $p \neq 0$ or for soliton states, no such condition is needed because \eqref{eq:Bsoliton} will be orthogonal to the vacuum/vacua for any $\{B\}$ (due to translation symmetry and to orthogonality of the vacua, respectively).}
\begin{equation}
	\ket{\{B\}, p} =  \sum_n e^{ip(kan)}\ket{n, \{B\}} \,,
	\label{eq:Bsoliton}
\end{equation}
where $a$ is the lattice spacing, so that $ka$ is the length of a unit cell, and $p$ is the dimensionful lattice momentum, with $p \sim p + 2\pi/(ka)$. Far to the left, the state \eqref{eq:Bsoliton} looks like $\ket{\{A\}}$, and far to the right it looks like $\ket{\{\tilde A\}}$. If $\{A\} = \{\tilde{A}\}$ are uMPS tensors for a ground state, then this ansatz describes a particle excitation over a single vacuum; if $\{A\} \neq \{\tilde{A}\}$ are distinct sets of tensors corresponding to two degenerate ground states, then this ansatz describes a soliton interpolating between these two ground states.
 
Determining the dispersion relation $\mathcal{E}_\text{sol}(p)$ of a soliton is subtle. The expectation value of $H$ in the state \eqref{eq:Bsoliton} includes a boundary contribution as in \eqref{eq:jordan_power_2}. Moreover, the momentum $p$ in \eqref{eq:Bsoliton} can be shifted by changing the phase of $\{\tilde A\}$ relative to $\{A\}$. These phases are arbitrary when we determine the uMPS states, but they change the eigenvalue of \eqref{eq:Bsoliton} under translation: if $A\rightarrow e^{i\phi} A$ and $\tilde{A} \rightarrow e^{i\tilde{\phi}}\tilde A$, then the momentum shifts as $p\rightarrow p +a^{-1}(\phi-\tilde \phi)$. (This ambiguity does not arise for particle excitations, because the same tensors $A$ are used on both sides of the $B$ tensor.) In the remainder of this subsection, we will describe how to fix both ambiguities using various methods.\footnote{Note that these ambiguities are not specific to our LEMPO construction. Indeed, they are also present in systems for which the Hamiltonian is represented by a regular MPO\@.}

The first and most general method is to appeal to Lorentz symmetry in the continuum limit. Neither the momentum ambiguity nor the energy ambiguity changes the curvature of $\mathcal{E}_\text{sol}(p)$ at its minimum. By Lorentz invariance, the curvature of the shifted lattice dispersion relation $\mathcal{E}(p;a)$ at the position $p_0$ of its minimum,
\begin{equation}
    \left.\frac{d^2 \mathcal{E}(p;a)}{dp^2}\right|_{p=p_0} \equiv \frac{1}{M_\text{curv}(a)}\,,
\end{equation}
must approach $1/M_\text{sol}$ as we take $ag\to 0$.

The remaining methods involve discrete symmetries of the lattice model. We henceforth assume we have a symmetry that exchanges the vacua, and also a symmetry involving parity that leaves each vacuum invariant. For concreteness, we phrase the discussion in terms of the symmetries of the Schwinger model at $\theta = \pi$, where these symmetries are charge conjugation $C$ and $CP$, respectively. 

To fix the momentum ambiguity, we can utilize $C$ to construct $\{\tilde A\}$ from $\{A\}$ (see \eqref{eq:C_vac}), as we did in Section~\ref{sec:method_boundary}. Since $C$ acts linearly, any rephasing of $A$ rephases $\tilde A$ identically, so $\tilde\phi = \phi$, and the momentum becomes unambiguous.\footnote{Fixing the ambiguity does not necessarily imply that this procedure also gives the minimum of the dispersion relation at $p=0$. However,  see the argument after \eqref{eq:Esol_final} for why the dispersion relation has to have an extremum at $p=0$. To make sure this extremum is a minimum, we have to choose the sign of $C$ appropriately. } Details on how we implement the action of $C$ on a uMPS in the case of the Schwinger model are given in Appendix~\ref{app:symmetries}\@.

Next, we turn to the energy. To simplify the discussion, let us first shift the Hamiltonian so that the energy density $\epsilon$ vanishes in the uMPS ground state.\footnote{Explicitly, we first calculate $\epsilon$ from \eqref{eq:E_and_A} with properly normalized environments and accumulators using $H$. Then, we subtract $(\epsilon/k)\mathds{1}$ from the top-right entry of each tensor \eqref{eq:W}. Since the corner entries of \eqref{eq:W} are $\mathds{1}$ and the rest of the first column and last row vanish, $H$ is linear in each top-right entry, and so this replacement subtracts $\epsilon$ per unit cell from the energy of both uMPSs.} We call the resulting Hamiltonian $H_{\epsilon=0}$. As a first step, we consider its expectation value in \eqref{eq:Bsoliton}, divided by the norm of \eqref{eq:Bsoliton} (as in footnote~\ref{fn:normalizability}): $\left\langle  \{B\}, p  \middle| H_{\epsilon = 0} \middle| \{B\}, p \right\rangle/
     \left\langle  \{B\}, p \middle|  \{B\}, p \right\rangle$.

Using \eqref{eq:Bsoliton}, both the numerator and denominator of this fraction can be written as double sums.  Translation invariance implies 
\begin{equation}\label{eq:qp_energy_sum}
    \frac{\braket{\{B\},p|H_{\epsilon=0}|\{B\},p}}{\left\langle  \{B\}, p \middle|  \{B\}, p \right\rangle} = \frac{\sum_{n} e^{ipka n} \braket{0,\{B\}|H_{\epsilon=0}|n,\{B\}}}{\sum_{n} e^{ipka n} \braket{0,\{B\}|n,\{B\}}}\,.
\end{equation} 
To the left of unit cell $\min(n,0)$, both the ket and the bra consist of the tensors $\{A\}$; to the right of $\max(n,0)$, both consist of $\{\tilde{A}\}$. These semi-infinite pieces can be replaced by $\mathbb{E}_L + \alpha_L \mathbb{A}_L$ and $\tilde{\mathbb{E}}_R + \tilde{\alpha}_R \tilde {\mathbb{A}}_R$ since, with $\epsilon = 0$, \eqref{eq:jordan_power_1} and its analog for the left boundary vector give
  \begin{equation}\label{eq:jordan_power_3}
   \begin{aligned}
     \tilde{\mathbb{T}}^N \tilde{\mathbbm{v}}_R &\approx \tilde{\mathbb{E}}_R + \tilde{\alpha}_R  \tilde{\mathbb{A}}_R \,, \qquad
     \mathbbm{v}_L^T  \mathbb{T}^N \approx \mathbb{E}_L^T + \alpha_L  \mathbb{A}_L^T  \,,
   \end{aligned}
 \end{equation}
as $N \to \infty$, up to exponentially small terms. Using this fact, the terms of \eqref{eq:qp_energy_sum} become finite contractions. For $n < 0$,
\begin{equation}\label{eq:qp_energy_term_diag}
    \braket{0,\{B\}|H_{\epsilon=0}|n,\{B\}} =
    \begin{tikzpicture}[diagram]
		\draw[rounded corners,fill=white] (-1.5,-2.3) rectangle node[rotate=90] {$\mathbb{E}_L + \alpha_L \mathbb{A}_L$} (-.5,.3);
		\draw[rounded corners,fill=white] (5.5,-2.3) rectangle node[rotate=-90] {$\tilde{\mathbb{E}}_R + \tilde{\alpha}_R \tilde{\mathbb{A}}_R$} (6.5,.3);
		\node[style=square] at (.5,0) (B) {$\{B\}$};
		\node[style=circle] at (.5,-1) (W1) {$\{W\}$};
		\node[style=square] at (.5,-2) (A1) {$\{\bar{A}\}$};
		\node[style=square] at (2,0) (At2) {$\{\tilde{A}\}$};
		\node[style=circle] at (2,-1) (W2) {$\{W\}$};
		\node[style=square] at (2,-2) (A2) {$\{\bar{A}\}$};
		\node at (3.25,0) {$\cdots$};
		\node at (3.25,-1) {$\cdots$};
		\node at (3.25,-2) {$\cdots$};
		\node[style=square] at (4.5,0) (At3) {$\{\tilde{A}\}$};
		\node[style=circle] at (4.5,-1) (W3) {$\{W\}$};
		\node[style=square] at (4.5,-2) (Bb) {$\{\bar{B}\}$};
		\draw (-.5,0) -- (B) -- (At2) -- ++(0.75,0);
		\draw (-.5,-1) -- (W1) -- (W2) -- ++(0.75,0);
		\draw (-.5,-2) -- (A1) -- (A2) -- ++(0.75,0);
		\draw (At3) -- ++(-0.75,0);
		\draw (W3) -- ++(-0.75,0);
		\draw (Bb) -- ++(-0.75,0);
		\draw (At3) -- (5.5,0);
		\draw (W3) -- (5.5,-1);
		\draw (Bb) -- (5.5,-2);
		\draw (B) -- (W1) -- (A1);
		\draw (At2) -- (W2) -- (A2);
		\draw (At3) -- (W3) -- (Bb);
		\node at (.5,-2.65) {$n$};
		\node at (4.5,-2.65) {$0$};
	\end{tikzpicture}\,,
\end{equation}
where $\{W\}$ is a schematic abbreviation for all the MPO tensors and link tensors appearing in a unit cell. The cases $n = 0$ and $n > 0$ are analogous.\footnote{There is a further subtlety in summing the contractions \eqref{eq:qp_energy_term_diag} in \eqref{eq:qp_energy_sum}. In particular, the powers of $\mathbb{T}$ appearing between the insertions of the $B$s become geometric series in \eqref{eq:qp_energy_sum} of the form $\sum_{n=0}^\infty e^{ipka n}\mathbb{T}^n$, which diverges for $\{A\} = \{\tilde{A}\}$ and $p= 0$ because $1$ is an eigenvalue of $\mathbb T$. However, the divergence is proportional to the contraction \eqref{eq:p0_ortho}, which we have fixed to zero, and so it does not change the quasiparticle energy.}

Because of the form \eqref{eq:W} of $W_n W_n^{(L)}$, the boundary vector $\delta_i$ in $\mathbb{A}_L$ passes through the finite contraction in \eqref{eq:qp_energy_term_diag} unchanged, and likewise for $\delta_f$ in $\tilde{\mathbb{A}}_R$. The $\alpha$-dependent terms in the numerator of \eqref{eq:qp_energy_sum} therefore sum to $(\alpha_L + \tilde{\alpha}_R)\braket{0,\{B\}|n,\{B\}}$, and so we find
\begin{equation} \label{eq:Esol}
   \frac{\left\langle  \{B\}, p  \middle| H_{\epsilon = 0} \middle| \{B\}, p \right\rangle}
     {\left\langle  \{B\}, p \middle|  \{B\}, p \right\rangle} = \mathcal{E}_0(p) + \alpha_L + \tilde\alpha_R\,,
\end{equation} 
where $\mathcal{E}_0(p)$ denotes the contraction \eqref{eq:qp_energy_term_diag} with the environments alone (i.e., with $\alpha_L$ and $\tilde{\alpha}_R$ set to zero). Even though we have not yet settled on the physical interpretation of $\mathcal{E}_0(p)$, we see that the shift $\alpha_L + \tilde{\alpha}_R$ is independent of $\{B\}$. Thus, we can determine $\{B\}$ by minimizing \eqref{eq:Esol}, making \eqref{eq:Bsoliton} into a variational approximation of the true soliton- or particle-like eigenstate of the lattice Hamiltonian.

Now we consider the distinction between particles and solitons. For particles, $\alpha_L = \tilde\alpha_{L} $ and  $\alpha_R = \tilde\alpha_{R}$ because the same vacua and boundary vectors appear on both sides.  We have $\alpha_L + \alpha_R = \langle \{A\}| H_{\epsilon = 0} | \{A\} \rangle$, as can be seen from the fact that $\langle \{A\}| H_{\epsilon = 0} | \{A\} \rangle = \lim_{N \to \infty} (\mathbb{T}^N \mathbbm{v}_L, \mathbb{T}^N \mathbbm{v}_R)$, which after using \eqref{eq:jordan_power_3} and \eqref{eq:automatic}, \eqref{eq:EA_products} and \eqref{eq:EE0}, evaluates to $\alpha_L + \alpha_R$. Consequently, the energy of a particle can be defined by 
\begin{equation}\label{eq:E_particle}
    \mathcal{E}_\mathrm{particle}(p) \equiv \mathcal{E}_0(p) = \frac{\left\langle  \{B\}, p  \middle| H_{\epsilon = 0} \middle| \{B\}, p \right\rangle}{\left\langle  \{B\}, p \middle| \{B\}, p \right\rangle} - \langle \{A\}| H_{\epsilon = 0} | \{A\} \rangle\,.
\end{equation}
For a soliton, in contrast, the shift $\alpha_L + \tilde\alpha_R$ mixes the boundary contributions of the two vacua; the individual vacuum energies only determine the combinations $\alpha_L + \alpha_R$ and $\tilde\alpha_L + \tilde\alpha_R$.

One straightforward remedy for this mixing is to calculate the sum of soliton and anti-soliton energies (with the latter obtained from anti-soliton tensors $\{\tilde{B}\}$, with $\{\tilde{A}\}$ on the left and $\{A\}$ on the right, determined by an optimization analogous to the one for $\{B\}$). This sum contains the shift $(\alpha_L + \tilde{\alpha}_R) + (\tilde{\alpha}_L + \alpha_R)$, which can be canceled by the sum of vacuum energies. Then, since $C$ exchanges a soliton of momentum $p$ with an anti-soliton of the same momentum (see Table~\ref{tab:CP}), we know that the soliton and anti-soliton energies are equal, and so we may write
\begin{equation} \label{eq:SolAntisolAverage}
\begin{split}
        \mathcal{E}_\mathrm{sol}(p) = \frac{1}{2} \Bigg(&\frac{\left\langle  \{B\}, p  \middle| H_{\epsilon = 0} \middle| \{B\}, p \right\rangle}{\left\langle  \{B\}, p \middle| \{B\}, p \right\rangle} + \frac{\left\langle  \{\tilde{B}\}, p  \middle| H_{\epsilon = 0} \middle| \{\tilde{B}\}, p \right\rangle}{\left\langle  \{\tilde{B}\}, p \middle| \{\tilde{B}\}, p \right\rangle} \\
        & - \langle \{A\}| H_{\epsilon = 0} | \{A\} \rangle - \langle \{\tilde A\}| H_{\epsilon = 0} | \{\tilde A\} \rangle\Bigg)\,.
\end{split}
\end{equation}

Another approach, which we use in this paper, is to employ $CP$ symmetry, which leaves each vacuum invariant while exchanging left and right. Because the two ends of the chain are related by $CP$, we can choose the boundary vectors of each uMPS so that they are $CP$-images of each other (and, additionally, use $CP$ to fix the remaining freedom \eqref{eq:etas} in the environments). See Appendix~\ref{app:CP_states} for details. With these choices, the boundary energy of each vacuum is split equally between its two ends,
\begin{equation}
    \alpha_L = \alpha_R \equiv \alpha\, ,\qquad \tilde \alpha_L = \tilde \alpha_R \equiv \tilde \alpha\,.
\end{equation}
The soliton shift $\alpha + \tilde{\alpha}$ is then half the sum of the vacuum energies $2\alpha$ and $2\tilde{\alpha}$, and so
\begin{equation}\label{eq:Esol_final}
     \mathcal{E}_\mathrm{sol}(p) = \frac{\left\langle  \{B\}, p  \middle| H_{\epsilon = 0} \middle| \{B\}, p \right\rangle}{\left\langle  \{B\}, p \middle| \{B\}, p \right\rangle} - \frac{1}{2} \left(\langle \{A\}| H_{\epsilon = 0} | \{A\} \rangle + \langle \{\tilde A\}| H_{\epsilon = 0} | \{\tilde A\} \rangle\right)\,.
\end{equation}
Note that the right-hand side is equal to $\mathcal{E}_0(p)$ in \eqref{eq:Esol}.
 
With both ambiguities fixed, the soliton dispersion relation is even in $p$: $C$ maps a soliton of momentum $p$ to an anti-soliton of momentum $p$, while $CP$ maps it to an anti-soliton of momentum $-p$ (see Table~\ref{tab:CP}), so $\mathcal{E}_\text{sol}(p) = \mathcal{E}_\text{sol}(-p)$. Combined with the periodicity $p\sim p+2\pi/ka$, this argument implies that at finite lattice spacing the dispersion relation has extrema at $p=0$ and $p = \pi/(ka)$.

Let us now apply these results to the lattice Schwinger model. For our application, we will take $\{A\}$ and $\{\tilde A\}$ to be uMPS tensors corresponding to $\ket{v_-}$ and $\ket{v_+}$ respectively, so that \eqref{eq:Bsoliton} describes a soliton in the lattice Schwinger model.  We can compute the electric charge of the soliton without knowing the specific tensors $\{B\}$. Indeed, as we discussed around \eqref{eq:qsol_def}, the soliton charge is given by $Q_\text{sol} = E(+\infty)-E(-\infty)$. We cannot quite use this definition along with \eqref{eq:Bsoliton}, because there is no notion of positions $\pm\infty$ in a translation-invariant state; but we could form a normalizable state by truncating the sum in \eqref{eq:Bsoliton}, and then place the electric field operators far to the right and left of the soliton region. One can then use the transfer-matrix techniques of Section~\ref{sec:method_boundary} to show that the expectation value of $E(+\infty)$ in this state reduces to the expectation value of $E(0)$ in the uMPS corresponding to $\ket{v_+}$, and likewise for $E(-\infty)$ and $\ket{v_-}$. Thus, our estimate for the soliton charge is just twice our estimate of $\braket{v_+|E|v_+}$ given in Figure~\ref{fig:eplus}.

\begin{figure}
    \centering
    \includegraphics[width=0.7\linewidth]{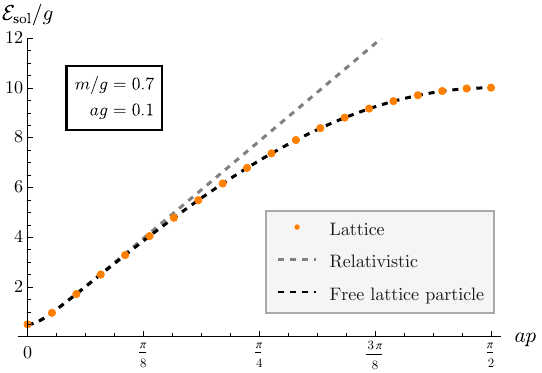}
    \caption{The dispersion relation of solitons on the lattice with $m/g = 0.7$ and $ag=0.1$. The lattice points agree very well (all within $0.1\%$ in this case) with the dispersion relation of a free particle on the lattice, \eqref{eq:free_dispersion}. For sufficiently small momentum, this dispersion relation is in turn very close to the continuum relativistic dispersion relation.}
    \label{fig:dispersion}
\end{figure}

Using \eqref{eq:Esol_final}, we can calculate the soliton dispersion relation on the lattice. As we take $ag \to 0$, it approximates the continuum relativistic dispersion relation
\begin{equation} \label{eq:Lorentzdispersion}
    \mathcal{E}_\text{sol}(p) = \sqrt{M_\text{sol}^2+p^2}
\end{equation}
increasingly well. Figure~\ref{fig:dispersion} shows an example at $m/g = 0.7$ and $ag = 0.1$. At finite lattice spacing, the data are in fact very well described by the dispersion relation of a free particle\footnote{Namely, a particle excitation of \eqref{eq:lattice_H} with $g=0$, $U_n\to \mathds{1}$, and $m_\text{lat}\to M_\text{sol}$, i.e.\ $H_\text{free} = -\frac{i}{2a}\sum_n\left(c_n^\dagger c_{n+1}-c_{n+1}^\dagger c_n\right) + M_\text{sol}\sum_n(-1)^n c_n^\dagger c_n$. The single-particle dispersion of this Hamiltonian is \eqref{eq:free_dispersion}.},
\begin{equation}\label{eq:free_dispersion}
    \mathcal{E}_\text{lat,free}(p) = \sqrt{M_\text{sol}^2 + \frac{1}{a^2}\sin^2(ap)}\,,
\end{equation}
which reduces to \eqref{eq:Lorentzdispersion} for $ap \ll 1$. This free dispersion does not occur generically; for instance, the dispersion relation of a lattice meson in the confined phase of the Schwinger model would not follow such a curve.

\begin{figure}[t]
    \centering
    \includegraphics[width=0.7\linewidth]{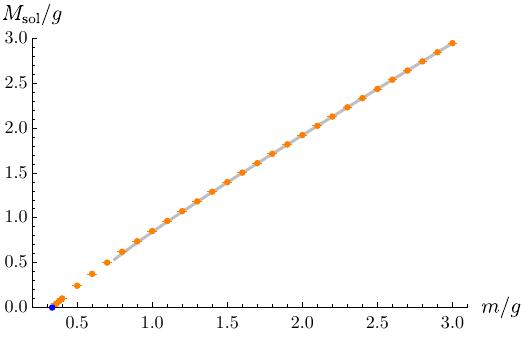}
    \caption{The mass of solitons in the Schwinger model at $\theta = \pi$ as a function of the bare fermion mass $m/g$. We indicate the critical mass $m_c$ with a blue point; $M_\text{sol}$ approaches zero linearly as $m$ approaches $m_c$, in accordance with \eqref{eq:scaling}. We also plot the weak-coupling expansion \eqref{eq:weak_coupling} in gray, showing good agreement.}
    \label{fig:m}
\end{figure}

We can now determine the soliton mass $M_\text{sol}$ for any $m > m_c$ by extrapolating the zero-momentum quasiparticle energies on the lattice to $ag \to 0$. In Figure~\ref{fig:m}, we plot $M_\text{sol}/g$ as a function of $m/g$. We see that it turns on linearly near the critical point, as expected from \eqref{eq:scaling}, and that at weak coupling it agrees with \eqref{eq:weak_coupling}.

\subsection{Soliton wavepackets}\label{sec:method_wavepacket}

The quasiparticle states \eqref{eq:Bsoliton} are momentum eigenstates, and so they are spread uniformly over the infinite lattice. To simulate collisions, we need localized states. In this section, we explain how to build such localized states from the quasiparticle tensors for the solitons of the Schwinger model. First we will work with a single soliton, and then we will consider soliton-anti-soliton pairs; in each case we characterize the states we build by their momentum-space wavefunctions.

Taking $\{A\}$ and $\{\tilde{A}\}$ to describe $\ket{v_-}$ and $\ket{v_+}$ respectively, and $\{B_+(p_0)\}$ to be the quasiparticle tensors that minimize the energy at momentum $p_0$, we define a positron wavepacket centered at position $x_0$ with width $\sigma$ by
\begin{equation}\label{eq:single_wavepacket}
    \ket{\{B_+(p_0)\}, x_0, \sigma} = \sum_{|2na - x_0| < \ell_\text{cut}} f(n) \ket{n, \{B_+(p_0)\}}\,, \qquad f(n) = e^{ip_0(2an)}\, e^{-(2an - x_0)^2/2\sigma^2}\,,
\end{equation}
and similarly for an electron, with $\{A\}$ and $\{\tilde{A}\}$ exchanged. (Note that we have substituted $k = 2$ for the Schwinger model.) The cutoff $\ell_\text{cut}$ is inessential for a one-soliton wavepacket, but it will be needed for the two-soliton states below; we take $\ell_\text{cut} = 2\sigma$, so that the truncated Gaussian tails are small.

There are two subtleties with \eqref{eq:single_wavepacket}. First, the quasiparticle tensors are only determined up to a gauge freedom, which shifts $\{B\}$ by terms built from $\{A\}$, $\{\tilde{A}\}$, and an arbitrary matrix on the virtual space, and which leaves the momentum eigenstate $\ket{\{B\},p}$ invariant \cite{Haegeman:2011lcd,Haegeman:2013vud,Vanderstraeten:2019voi}.\footnote{Specifically, $\ket{\{B\},p}$ is invariant under the transformation 
\begin{equation}\label{eq:B_gauge}
    \begin{tikzpicture}[diagram]
        \node[style=square] at (0,0) (B) {$\{B\}$};
        \draw (B) -- ++(-0.75,0);
        \draw (B) -- ++(0.75,0);
        \draw (B) -- ++(0,-.8);
    \end{tikzpicture}
    \;\longrightarrow\;
    \begin{tikzpicture}[diagram]
        \node[style=square] at (0,0) (B) {$\{B\}$};
        \draw (B) -- ++(-0.75,0);
        \draw (B) -- ++(0.75,0);
        \draw (B) -- ++(0,-.8);
    \end{tikzpicture}
    \;+\;
    \begin{tikzpicture}[diagram]
        \node[style=square] at (0,0) (A) {$\{A\}$};
        \node[style=circle] at (1.25,0) (X) {$X$};
        \draw (A) -- ++(-0.75,0);
        \draw (A) -- (X);
        \draw (X) -- ++(0.75,0);
        \draw (A) -- ++(0,-.8);
    \end{tikzpicture}
    \;-\; e^{-ipka}\;
    \begin{tikzpicture}[diagram]
        \node[style=circle] at (0,0) (X) {$X$};
        \node[style=square] at (1.25,0) (At) {$\{\tilde A\}$};
        \draw (X) -- ++(-0.75,0);
        \draw (X) -- (At);
        \draw (At) -- ++(0.75,0);
        \draw (At) -- ++(0,-.8);
    \end{tikzpicture}
\end{equation}
for any matrix $X$ mapping the virtual space of $\{A\}$ to that of $\{\tilde A\}$, since the added terms cancel in the telescoping sum \eqref{eq:Bsoliton} over $n$. 
} The individual terms $\ket{n, \{B\}}$ are not invariant under this transformation, however, and so neither is the wavepacket \eqref{eq:single_wavepacket}; we thus need to fix the gauge in order for \eqref{eq:single_wavepacket} to be well-defined. We will use the reflection-symmetric gauge introduced in \cite{Van_Damme_2021}. In this gauge, the energy density profile of the wavepacket is centered where the Gaussian envelope is centered \cite{Milsted:2020jmf}.

Second, note that we use the same tensors $\{B_+(p_0)\}$ for every term in \eqref{eq:single_wavepacket}, even though the wavepacket contains a spread of momenta of order $1/\sigma$ around $p_0$. This is a good approximation as long as $\{B_+(p)\}$ varies slowly over that range, which we will check by computing the momentum-space wavefunction.

We can calculate this wavefunction by projecting \eqref{eq:single_wavepacket} onto the quasiparticle states. Let $\ket{\{B_+(q)\}, q}$ be the quasiparticle eigenstate at momentum $q$, built with its own optimized tensors $\{B_+(q)\}$. We take $q$ in the Brillouin zone $-\pi/2a < q \leq \pi/2a$ and normalize these states so that
\begin{equation}\label{eq:qp_normalization}
    \braket{\{B_+(q)\}, q | \{B_+(q')\}, q'} = \frac{\pi}{a}\, \delta(q - q')\,, \qquad\text{or equivalently}\qquad \braket{\{B_+(q)\}, q | 0, \{B_+(q)\}} = 1\,.
\end{equation} 
With this normalization, $\int_{-\pi/2a}^{\pi/2a}\frac{a\,dq}{\pi}\,\ket{\{B_+(q)\}, q}\bra{\{B_+(q)\}, q}$ is the projector onto the single-soliton sector.

We thus define the momentum-space wavefunction of the wavepacket $\ket{\{B_+(p_0)\}, x_0, \sigma}$ by
\begin{equation}\label{eq:single_momentum_wf}
    \tilde{\psi}(q) = \frac{\braket{\{B_+(q)\}, q | \{B_+(p_0)\}, x_0, \sigma}}{\sqrt{\braket{\{B_+(p_0)\}, x_0, \sigma|\{B_+(p_0)\}, x_0, \sigma}}}\,.
\end{equation}
Since $\ket{\{B_+(q)\},q}$ is invariant under translation by a unit cell up to the phase $e^{2iaq}$, the overlap factorizes:
\begin{equation}\label{eq:single_momentum_wf_factor}
    \tilde{\psi}(q) = g(q)\, \hat{f}(q)\,, \qquad \hat{f}(q) = \mathcal{N}\sum_{|2na - x_0| < \ell_\text{cut}} f(n)\, e^{-2iaqn}\,, \qquad g(q) = \braket{\{B_+(q)\}, q | 0, \{B_+(p_0)\}}\,,
\end{equation}
where $\mathcal{N} = 1/\sqrt{\braket{\{B_+(p_0)\}, x_0, \sigma|\{B_+(p_0)\}, x_0, \sigma}}$. The envelope $\hat{f}(q)$ is, up to the small effect of the truncation, a Gaussian centered at $p_0$ with $|\hat{f}(q)|^2 \propto e^{-\sigma^2 (q - p_0)^2}$. The form factor $g(q)$ measures how well the fixed tensors $\{B_+(p_0)\}$ represent the soliton at momentum $q$. At $q = p_0$ we have $g(p_0) = 1$ (by \eqref{eq:qp_normalization}), and for other values of $q$ we can evaluate $g(q)$ using the transfer-matrix techniques we discussed in Section~\ref{sec:method_envs}. 

\begin{figure}
    \centering
    \includegraphics[width=.8\linewidth]{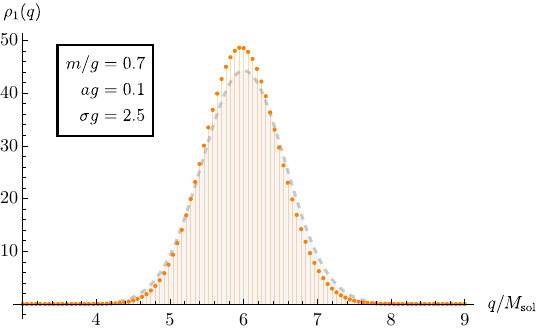}
    \caption{Probability density $\rho_1(q) = |\tilde\psi(q)|^2$ of a single positron wavepacket
(orange; see \eqref{eq:single_momentum_wf}) at $m/g = 0.7$, $ag = 0.1$,
$p_0/M_\text{sol} = 6$, and $\sigma g = 2.5$, compared with the Gaussian envelope
$|\hat{f}(q)|^2$ (dashed line). The close agreement shows that using the fixed
tensors $\{B_+(p_0)\}$ throughout the wavepacket is a good approximation.}
\label{fig:single_wavefunction}
\end{figure}

The quantity
\begin{equation}\label{eq:single_weight}
    w_1 = \int_{-\pi/2a}^{\pi/2a} \frac{a\, dq}{\pi}\, \rho_1(q)\,,\qquad \text{with }\rho_1(q) \equiv |\tilde{\psi}(q)|^2
\end{equation}
is the probability that the wavepacket lies in the single-soliton sector. In Figure~\ref{fig:single_wavefunction}, we show that $|\tilde{\psi}(q)|^2$ closely follows the Gaussian $|\hat{f}(q)|^2$, with deviations due to the form factor $g(q)$. Additionally, we find that $w_1 = 1$ to within 0.01\%, confirming that \eqref{eq:single_wavepacket} is a single-soliton state.

For a collision, we want a positron at $x_+$ with momentum $+p$ and an electron at $x_- > x_+$ with momentum $-p$, with $\ket{v_-}$ outside the pair and $\ket{v_+}$ between them. We build this state as
\begin{equation}\label{eq:two_wavepacket}
\begin{split}
    |\{B_+(+p)\}, \{B_-(-p)\},& x_+, x_-, \sigma\rangle = \sum_{\substack{|2 n_+ a-x_+|<\ell_\text{cut}\\|2 n_- a -x_-|<\ell_\text{cut}}}\;
    e^{ip(2 n_+ a)}\, e^{-(2 n_+ a - x_+)^2/2\sigma^2}\;
      e^{-ip(2 n_- a)}\, e^{-(2n_- a - x_-)^2/2\sigma^2} \\
    & \times\;
    \begin{tikzpicture}[diagram]
        \node at (-1.1,0) (dots1) {$\cdots$};
        \node[style=square] at (0,0) (A1) {$\{A_{n_+ -1}\}$};
        \node[style=square] at (1.7,0) (B1) {$\{B_+(+p)\}$};
        \node[style=square] at (3.4,0) (At1) {$\{\tilde{A}_{n_+ +1}\}$};
        \node at (4.65,0) (dotsmid) {$\cdots$};
        \node[style=square] at (5.9,0) (At2) {$\{\tilde{A}_{n_- -1}\}$};
        \node[style=square] at (7.6,0) (B2) {$\{B_-(-p)\}$};
        \node[style=square] at (9.3,0) (A2) {$\{A_{n_- +1}\}$};
        \node at (10.4,0) (dots2) {$\cdots$};
        \draw (A1) -- (B1) -- (At1);
        \draw (At2) -- (B2) -- (A2);
        \draw (At1) -- ++(0.75,0);
        \draw (At2) -- ++(-0.75,0);
        \draw (A1) -- ++(-0.75,0);
        \draw (A2) -- ++(0.75,0);
        \foreach \t in {A1,At1,At2,A2}
            \draw (\t) -- ++(0,-.8);
        \draw (B1) -- ++(0,-.8) node[below] {$n_+$};
        \draw (B2) -- ++(0,-.8) node[below] {$n_-$};
    \end{tikzpicture}\,.
\end{split}
\end{equation}
The cutoff now matters: choosing $x_- - x_+ > 2\ell_\text{cut}$ guarantees that every term has the positron to the left of the electron, so that each term has the vacuum structure $\ket{v_-}\ket{v_+}\ket{v_-}$. See Appendix~\ref{app:simulation} for how we write \eqref{eq:two_wavepacket} as a matrix product state. 

To characterize \eqref{eq:two_wavepacket} in momentum space, we would ideally project onto exact two-soliton scattering states, but we do not have a means of calculating these states. Instead, following \cite{Milsted:2020jmf, Belyansky:2023rgh}, we project onto a basis of approximate two-soliton states in which the positron and electron are confined to disjoint regions $W_L$ and $W_R$ of the lattice, separated by $n_\text{sep}$ sites:
\begin{equation}\label{eq:two_soliton_basis}
     \ket{p_1, p_2} =\sum_{n_+\in W_L} \sum_{n_-\in W_R} e^{2ia(p_1n_+ +p_2n_-)} \ket{(n_+, B_+(p_1)), (n_-, B_-(p_2))}\,,
\end{equation}
where $\ket{(n_+, B_+(p_1)), (n_-, B_-(p_2))}$ is the corresponding single term in the second line of \eqref{eq:two_wavepacket}, and $n\in W$ means that both sites of unit cell $n$ lie in $W$. For $n_\text{sep}$ larger than the soliton size, these states neglect the interaction between the solitons, which is appropriate before and long after a collision, when the wavepackets are well separated. We take $\{B_+\}$ in the left gauge and $\{B_-\}$ in the right gauge, which makes the position states orthonormal \cite{Milsted:2020jmf}. Despite the fact that \eqref{eq:two_soliton_basis} restricts the sums to the regions $W_{L/R}$, almost all the overlap with our collision states $\ket{\psi}$ comes from those regions; hence, $\int\frac{a\,dp_1}{\pi}\frac{a\,dp_2}{\pi}\,\ket{p_1,p_2}\bra{p_1,p_2}$ is a good approximation of the projector onto the two-soliton sector. In analogy with \eqref{eq:single_momentum_wf}, we define the momentum-space wavefunction of a state $\ket{\psi}$ by
\begin{equation}\label{eq:two_momentum_wf}
    \tilde{\psi}(p_1, p_2) = \frac{\braket{p_1, p_2 | \psi}}{\sqrt{\braket{\psi|\psi}}}\,,
\end{equation}
and the corresponding probability density by
\begin{equation}\label{eq:two_soliton_density}
    \rho(p_1, p_2) = |\tilde{\psi}(p_1, p_2)|^2\,.
\end{equation}
The probability that $\ket{\psi}$ lies in the two-soliton sector is then
\begin{equation}\label{eq:2solprob}
    w_2 = \int \frac{a\,dp_1}{\pi}\frac{a\,dp_2}{\pi}\, \rho(p_1, p_2)\,.
\end{equation}
For the incoming state \eqref{eq:two_wavepacket}, the overlap factorizes as in \eqref{eq:single_momentum_wf_factor}, up to corrections that are exponentially small in the separation between the two solitons, $\tilde{\psi}(p_1,p_2) \simeq g_+(p_1)\hat{f}_+(p_1)\, g_-(p_2)\hat{f}_-(p_2)$, where $\hat{f}_\pm$ are the Gaussian envelopes centered at $\pm p$ and $g_\pm$ the corresponding form factors. The natural reference distribution is thus $|\hat{f}_+(p_1)\hat{f}_-(p_2)|^2$.
\begin{figure}
    \centering  \includegraphics[width=\linewidth]{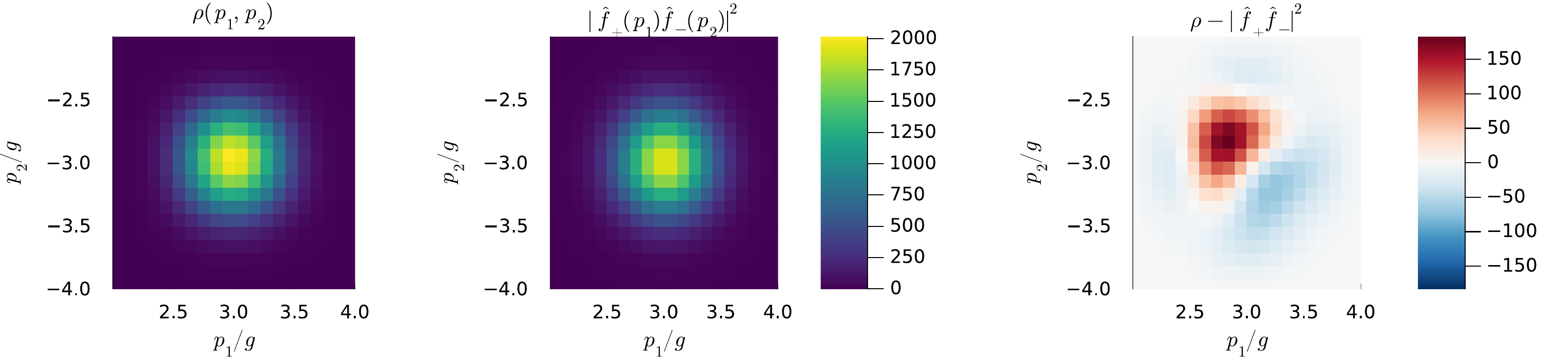}
    \caption{Momentum-space probability density $\rho(p_1,p_2)$ of the incoming two-soliton state \eqref{eq:two_wavepacket} (left), the Gaussian reference distribution $|\hat{f}_+(p_1)\hat{f}_-(p_2)|^2$ (middle), and their difference (right), for $m/g = 0.7$, $ag = 0.1$, $p = 6M_\text{sol} \simeq 3\,g$, and $\sigma g = 2.5$. Both distributions integrate to unity to within $0.01\%$. The distribution of each momentum in $\rho$ has standard deviation $\approx 0.27\,g$, slightly narrower than that of the Gaussian, $1/(\sqrt{2}\sigma) \approx 0.28\,g$, and is shifted towards smaller $|p|$ by about $0.015\,g$. The difference never exceeds $\sim 10\%$ of the peak value.}
    \label{fig:two_wavefunction}
\end{figure}
In Figure~\ref{fig:two_wavefunction}, we compare $\rho(p_1, p_2)$ to $|\hat{f}_+(p_1)\hat{f}_-(p_2)|^2$ for the incoming state of a collision at $m/g = 0.7$ and $p/M_\text{sol} = 6$. The two agree well; as for the single soliton in Fig.~\ref{fig:single_wavefunction}, the small differences are due to the form factors $g_+(p_1)g_-(p_2)$.

\section{Collision results}\label{sec:results}

We will now move to the main focus of this paper: colliding solitons on the lattice. In Section~\ref{sec:results_tdvp}, we review how the TDVP algorithm \cite{Haegeman:2011zz,Haegeman:2015ezw} allows us to time-evolve a window MPS state. Then, in Section~\ref{sec:results_elastic}, we apply this method to simulate a soliton collision at large $m/g$, where string breaking is suppressed and the collision is nearly classical; the simulation closely matches the classical prediction.

Then, in Section~\ref{sec:results_inelastic}, we dial down $m/g$ so that the flux string breaks with significant probability. We estimate this probability from our final states using two different methods, and find good agreement. Finally, in Section~\ref{sec:results_collapse}, we show how to collapse our simulation states into the string-broken and string-unbroken sectors.

\subsection{Time evolution with TDVP}\label{sec:results_tdvp}

To simulate a collision, we evolve the initial state \eqref{eq:two_wavepacket} with the lattice Hamiltonian \eqref{eq:lattice_H_fixed}, represented as a LEMPO\@. Since this state differs from the vacuum $\ket{v_-}$ only in a finite region, we represent it as a window MPS \cite{Phien:2012cgs}: outside a window of $N$ sites, the tensors are fixed to the uMPS tensors $\{A\}$ of $\ket{v_-}$, and only the $N$ tensors inside the window are allowed to change during the evolution. The window MPS is a good approximation as long as the window is large enough that the excitations never come close to its edges, so that the state outside the window remains the vacuum throughout the simulation.

Exact time evolution takes an MPS out of the manifold of MPSs with a given bond dimension. The time-dependent variational principle (TDVP) \cite{Haegeman:2011zz,Haegeman:2015ezw} instead evolves the state within this manifold, by projecting the Schr\"odinger equation onto the tangent space $T_{\psi(t)}$:
\begin{equation}\label{eq:tdvp}
    i\frac{d}{dt}\ket{\psi(t)} = \, \mathcal{P}_{T_{\psi(t)}} H \ket{\psi(t)}\,.
\end{equation}
In practice, \eqref{eq:tdvp} is integrated by sweeping through the window and evolving one local tensor at a time with an effective Hamiltonian~\cite{Haegeman:2015ezw}. The one-site version of this algorithm keeps the bond dimensions fixed. Since a collision generates entanglement, we instead use the two-site version, in which pairs of neighboring tensors are evolved together and then split by a singular value decomposition. This allows the virtual spaces to grow adaptively; we impose an upper limit $D_\text{max}$ on their dimension. We check the insensitivity of our results to $D_\text{max}$ in Appendix~\ref{app:simulation}. 

The effective Hamiltonians are built from the environments of Section~\ref{sec:method_envs}. At the edges of the window, the semi-infinite strings of vacuum tensors on either side are replaced by $\mathbb{E}_L + \alpha_L \mathbb{A}_L$ and $\mathbb{E}_R +\alpha_R \mathbb{A}_R$, using \eqref{eq:jordan_power_3}. Thus, even though only the tensors inside the window are updated, the evolution is generated by the Hamiltonian on the full infinite lattice, with the region outside the window frozen in the vacuum. This way of performing the time evolution is accurate as long as any excitations remain well inside the window; we choose window sizes so that this property holds for the full simulation time. Finally, a change in the boundary coefficients $\alpha_{L/R}$ shifts every local effective Hamiltonian by the same multiple of the identity (see Appendix~\ref{app:lempo}), so it affects only the global phase of the evolved state; thus, our method does not reliably give the global phase of the wavefunction over time. We account for our inability to keep track of the phase of the wavefunction when extracting time delays in Appendix~\ref{app:timedelay}.

Using LEMPOs in TDVP requires one modification of the two-site effective Hamiltonian. When we evolve sites $j$ and $j+1$ together, the link tensor $W^{(L)}_j$ acts on the virtual space $\mathcal{V}_j$ between them, which is contracted inside the two-site tensor. However, because the tensors are $\U(1)$-symmetric, the charge on $\mathcal{V}_j$ is fixed by the charges on $\mathcal{V}_{j-1}$ and $\mathcal{H}_j$ through \eqref{eq:gauss_symmps}. We can therefore apply $W^{(L)}_j$ to the two-site tensor by acting on the fused space $\mathcal{V}_{j-1}\otimes\mathcal{H}_j$; see \cite{Dempsey:2025wia} for details (explained there in the context of two-site DMRG).

At every time step of the evolution, we compute expectation values of observables such as the energy density and the electric field \eqref{eq:eeff_lattice}. These expectation values are calculated using a contraction like \eqref{eq:exp_On} on the full infinite lattice: inside the window, we use the evolved MPS tensors, and outside it the fixed points $l$ and $r$ of $\ket{v_-}$ account for the semi-infinite vacuum regions. To define the energy density, we use the local terms of $H_{\epsilon=0}$, so that the energy density vanishes in the vacuum. 

We discuss the convergence of our simulations in the lattice spacing, wavepacket width, and $D_\text{max}$ in Appendix~\ref{app:simulation}.

\subsection{Near-elastic soliton collisions}\label{sec:results_elastic}

We now turn to our main goal, real-time simulation of $e^+e^-$ collisions. We will start by taking $m\gg g$, so that the solitons are heavy and pair production is suppressed. Then the collision can be understood classically by treating the solitons as point particles of mass $M_\text{sol}$ and charges $\pm \langle Q_\text{sol}\rangle$.

We begin with a configuration in which the vacuum to the left is $\ket{v_-}$, a positron at position $x_+$ interpolates to $\ket{v_+}$, and then an electron at position $x_- > x_+$ interpolates back to $\ket{v_-}$. We give the positron a positive initial momentum $+p$ and the electron a negative initial momentum $-p$, so that they will eventually collide and reverse order. On the lattice, we construct such a state as the two-soliton wavepacket \eqref{eq:two_wavepacket}.

\begin{figure}[t]
    \centering
    \begin{subfigure}[t]{0.45\textwidth}%
        \centering
        \includegraphics[width=\linewidth]{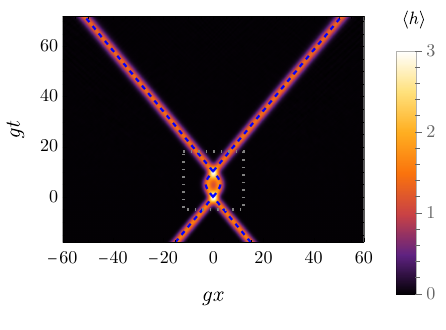}
        \caption{Energy density, tracking the particle trajectories.}
    \end{subfigure}%
    \hspace{.07\textwidth}%
    \begin{subfigure}[t]{0.45\textwidth}%
        \centering
        \includegraphics[width=\linewidth]{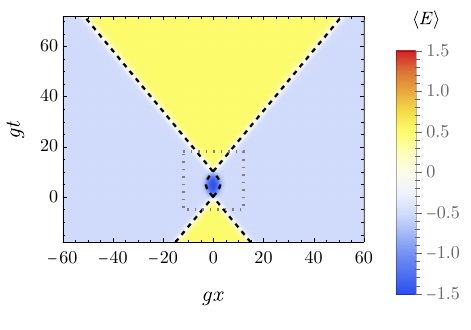}
        \caption{Electric field, showing that the solitons carry charge, and that the flux string region has an electric field of greater magnitude than that of the vacua.}
    \end{subfigure}\\[2em]
    \begin{subfigure}[t]{0.45\textwidth}%
        \centering
        \includegraphics[width=\linewidth]{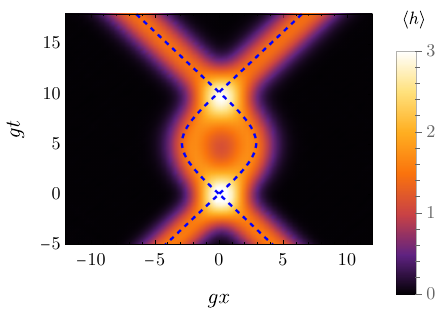}
        \caption{Energy density zoomed in on the flux string region.}
    \end{subfigure}%
    \hspace{.07\textwidth}%
    \begin{subfigure}[t]{0.45\textwidth}%
        \centering
        \includegraphics[width=\linewidth]{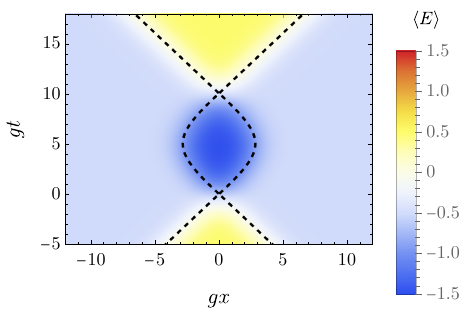}
        \caption{Electric field zoomed in on the flux string region.}
    \end{subfigure}
    \caption{An $e^+e^-$ collision that is well-described by the classical picture. The trajectory \eqref{eq:classical_trajectory} is shown in black dashed lines. The fermion mass is $m/g = 3$, the initial momentum of both particles is $|p|/g = 5$, the lattice spacing is $ag = 0.05$, and the simulation window covers 2400 sites. The dynamical soliton mass is $M_\text{sol}/g \approx 2.95$, and $\qsol \approx 0.994$.}
    \label{fig:elastic_example}
\end{figure}

\begin{figure}
    \centering
\includegraphics[width=0.6\linewidth]{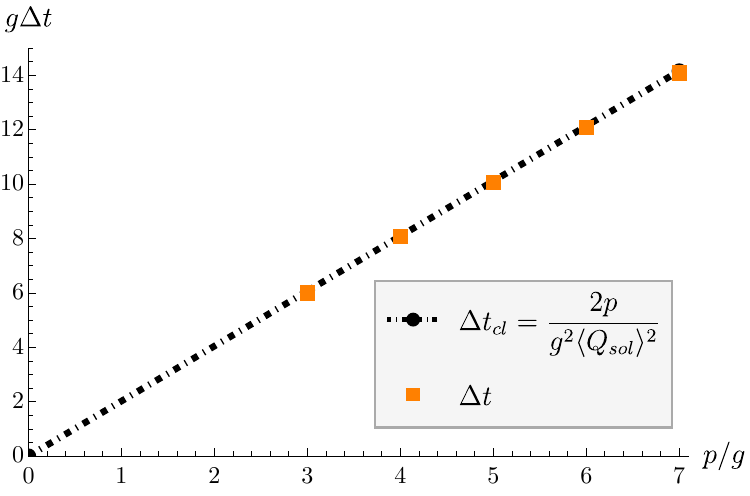}
    \caption{The classical prediction for the scattering time delay (black dotted line), and the value calculated from the quantum-mechanical phase shift as described in Appendix~\ref{app:timedelay} (orange squares), for several nearly-elastic $e^+e^-$ collisions. The initial momenta of the particles are $p/g=3,\ldots,7$ at a lattice spacing of $ag=0.05$ and fermion mass $m/g=3$. The dynamical soliton mass and charge are $M_\mathrm{sol}/g\approx 2.95$ and $\qsol \approx0.994$. }
    \label{fig:timedelay}
\end{figure}

Since the two vacua are degenerate, there is no force between the particles so long as the positron remains to the left of the electron. However, once the electron is to the left of the positron, the region between them will no longer resemble one of the degenerate vacua. Classically, we would say that the electric field in $\ket{v_-}$ is $-\qsol/2$ and the electric field between the electron and the positron is $-3\qsol/2$. This leads to a string tension of\footnote{Note that the electric field energy is given by $E^2$, and so its expectation value is $\langle E^2\rangle$, whereas here we are using $\langle E\rangle^2$. The difference between these, or relatedly the variance of $Q_\mathrm{sol}$, becomes small for $m\gg g$.}
\begin{equation}
    \sigma = \frac{1}{2}g^2\left\lbrack (-3\qsol/2)^2 - (-\qsol/2)^2\right\rbrack = g^2 \qsol^2.
\end{equation}
Conservation of energy then gives the classical trajectories,
\begin{equation}\label{eq:classical_trajectory}
    x_\pm(t) = \pm \frac{1}{\sigma}\left(\sqrt{p^2+M_\text{sol}^2} - \sqrt{(p-\sigma t)^2+M_\text{sol}^2}\right),
\end{equation}
where we have taken the first collision point to be $(x,t) = (0,0)$. The particles continue on these hyperbolic trajectories until $t = 2p/\sigma$, when they cross again and then continue to propagate freely with their initial momenta reversed.

In Figure~\ref{fig:elastic_example}, we show an example of an $e^+e^-$ collision that is well described by this classical picture. By plotting the energy density over time, we clearly see the trajectories of the positron and electron crossing and then rebounding. The detailed shape of the trajectories closely matches the prediction \eqref{eq:classical_trajectory}. Moreover, by plotting the electric field, we clearly see that it has a larger magnitude in the region between the electron and positron while the electron is to the left, giving rise to the string tension that changes the direction of both particles.

As another way of testing the classical prediction, we extract the time delay $\Delta t$ (introduced by Wigner~\cite{Wigner:1955zz}) from the post-collision state. For elastic scattering, the $2\to 2$ $S$-matrix is $S_{2\to 2} = e^{2i\delta(p)}$, and the time delay is defined by
\begin{equation} \label{eq:phaseshift_def}
    \Delta t = -i \frac{\partial(\log S_{2\to 2})}{\partial\sqrt{s}} = 2\frac{\partial
    \delta(p)}{\partial\sqrt{s}}\,.
\end{equation}
Physically, $\Delta t$ measures how much later the outgoing wavepackets emerge than they would if the solitons reflected instantaneously on contact. To extract it from our simulations, we adapt the method of \cite{Jha:2024jan} to soliton states, as explained in Appendix~\ref{app:timedelay}.

Classically, the solitons cross at $t = 0$, are pulled back by the flux string between them, and cross again at $t = 2p/\sigma$, after which they move freely with their momenta reversed. Compared with instantaneous reflection, the outgoing particles are therefore delayed by
\begin{equation}
    \Delta t_\mathrm{cl} = \frac{2p}{\sigma}\,.
\end{equation}
Figure~\ref{fig:timedelay} compares $\Delta t$ and $\Delta t_\mathrm{cl}$ for $p/g = 3,\ldots,7$ at $m/g = 3$ (where $M_\mathrm{sol}/g \approx 2.95$). The two agree closely, supporting our interpretation of collisions such as the one in Figure~\ref{fig:elastic_example} as nearly classical.

\subsection{Collisions with string breaking}\label{sec:results_inelastic}

If we collide two solitons at sufficiently high energies, it is kinematically possible for the flux string that is created to decay into additional soliton pairs. We expect the rate of string breaking to be exponentially suppressed at large $m/g$ \cite{Schwinger:1951nm}, and as we will explain in Section~\ref{sec:critical_scattering}, for $m-m_c \ll g$ we need the center-of-mass energy to be much larger than $M_\text{sol}$ to see string breaking. Thus, in order to study string breaking in this subsection, we will start by working at an intermediate mass of $m/g = 0.7$.

\begin{figure}
    \centering
    \begin{subfigure}[t]{0.45\textwidth}%
        \centering
        \includegraphics[width=\linewidth]{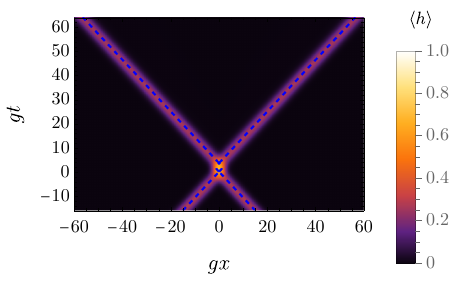}
    \end{subfigure}%
    \hspace{.07\textwidth}%
    \begin{subfigure}[t]{0.45\textwidth}%
        \centering
        \includegraphics[width=\linewidth]{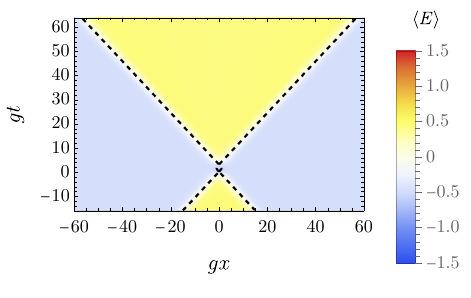}
    \end{subfigure}\\[1em]
    \begin{subfigure}[t]{0.45\textwidth}%
        \centering
        \includegraphics[width=\linewidth]{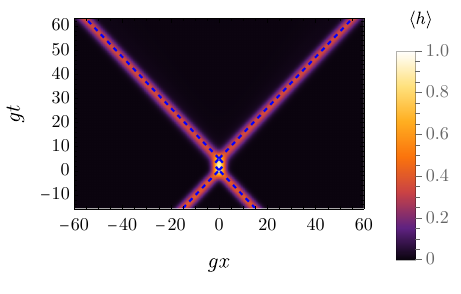}
    \end{subfigure}%
    \hspace{.07\textwidth}%
    \begin{subfigure}[t]{0.45\textwidth}%
        \centering
        \includegraphics[width=\linewidth]{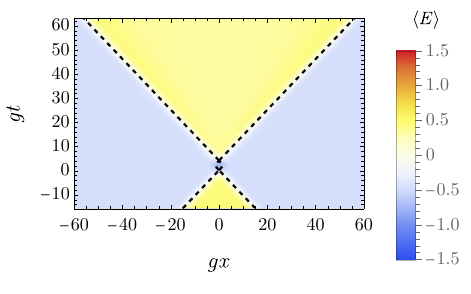}
    \end{subfigure}\\[1em]
    \begin{subfigure}[t]{0.45\textwidth}%
        \centering
        \includegraphics[width=\linewidth]{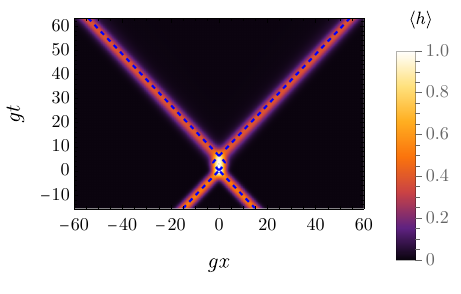}
    \end{subfigure}%
    \hspace{.07\textwidth}%
    \begin{subfigure}[t]{0.45\textwidth}%
        \centering
        \includegraphics[width=\linewidth]{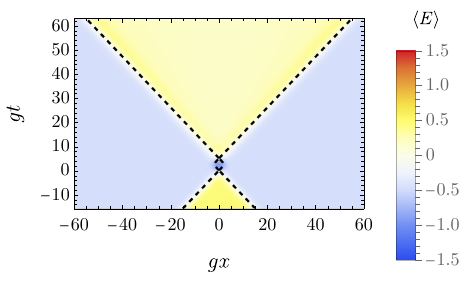}
    \end{subfigure}\\[1em]    
    \begin{subfigure}[t]{0.45\textwidth}%
        \centering
        \includegraphics[width=\linewidth]{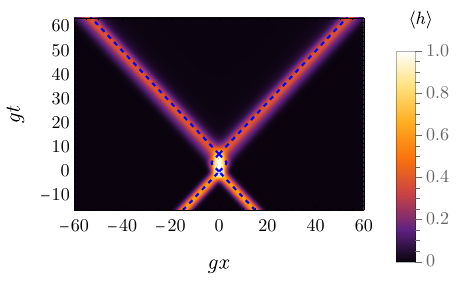}
    \end{subfigure}%
    \hspace{.07\textwidth}%
    \begin{subfigure}[t]{0.45\textwidth}%
        \centering
        \includegraphics[width=\linewidth]{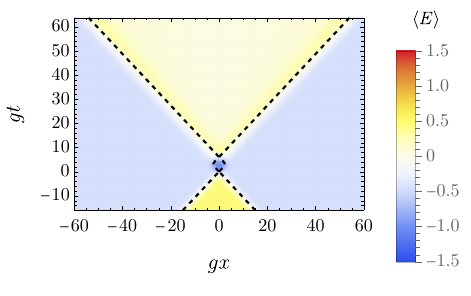}
    \end{subfigure}
    \caption{Examples of $e^+e^-$ collisions at $m/g = 0.7$, with $p/M_\text{sol} = 3$ (top), 4, 5, 6 (bottom). The black dashed line is the classical trajectory \eqref{eq:classical_trajectory}. The main change we see as we increase the scattering scale is a decrease in the magnitude of the expectation value of the electric field between the outgoing excitations in the final state.}
    \label{fig:inelastic_examples}
\end{figure}

In Figure~\ref{fig:inelastic_examples}, we plot four collisions at $m/g = 0.7$, for which the soliton mass is $M_\text{sol}/g \approx 0.5$ and the charge is $\qsol \approx 0.91$. We collide the particles at momenta $p/M_\text{sol} = 3,\, 4,\, 5,\, 6$. In each case, we plot the classical rebound trajectory \eqref{eq:classical_trajectory} in black dashed lines.

For $p/M_\text{sol} = 3$, the collision roughly resembles that of Figure~\ref{fig:elastic_example}, although the maximum extent of the string is small (and comparable to the width of the incoming wavepackets). As we increase the initial momentum, there is no obvious change in the energy density plots: up through $p/M_\text{sol} = 6$, we see bright tracks following the classical trajectory. However, in the electric field plot, we see that the magnitude of $\langle E\rangle$ in between the outgoing excitations decreases as we increase momentum. As we now argue, this is a signature of inelastic string breaking.

\begin{figure}
    \centering
    \includegraphics[width=0.6\linewidth]{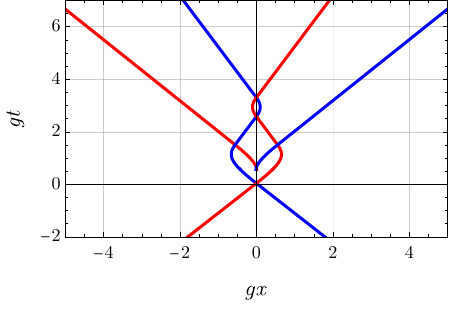}
    \caption{A schematic trajectory for a $2\to4$ collision in which the flux string breaks in its middle after reaching a critical length. Other than the pair creation, which is inserted by hand, the trajectories are determined using classical equations of motion.}
    \label{fig:classical_stringbreaking}
\end{figure}

To see why, consider which vacuum is left between the outgoing excitations in each channel. If the string does not break, it slingshots the electron and positron back out with their initial momenta reversed, as in Section~\ref{sec:results_elastic}, and the region between them is $\ket{v_+}$. If instead the string breaks, an additional soliton pair is created in its middle once it has stretched to a critical length, and the final state consists of a right-moving and a left-moving dipole. The new pair restores $\ket{v_-}$ in the center, as illustrated by the schematic trajectory for this $2\to 4$ process in Figure~\ref{fig:classical_stringbreaking}.

Quantum mechanically, the final state is a superposition of both channels. Schematically,
\begin{equation}\label{eq:psifinal}
\begin{split}
    \ket{\psi(t\to\infty)} &= \alpha_+ \left(\tikz[baseline=0cm]{
        \draw[ultra thick,dashed] (-5.5,0) -- node[below] {\footnotesize $\ket{v_-}$} (-3.5,0);
        \draw[ultra thick,dashed] (5.5,0) -- node[below] {\footnotesize $\ket{v_-}$} (3.5,0);
        \draw[ultra thick,dotted] (-3.5,0) -- node[below] {\footnotesize $\ket{v_+}$} (3.5,0);
        \draw[thick,red!50,fill=red!30] (-3.5,0) circle (.3) node[white] {$+$};
        \draw[thick,blue!50,fill=blue!30] (3.5,0) circle (.3) node[white] {$-$};
        \draw[ultra thick,-latex] (3.25,.5) -- (3.75,.5);
        \draw[ultra thick,-latex] (-3.25,.5) -- (-3.75,.5);
    }\right) \\
    &+\alpha_- \left(\tikz[baseline=0cm]{
        \draw[ultra thick,dashed] (-5.5,0) -- node[below] {\footnotesize $\ket{v_-}$} (-3.5,0);
        \draw[ultra thick,dashed] (5.5,0) -- node[below] {\footnotesize $\ket{v_-}$} (3.5,0);
        \draw[ultra thick,dotted] (-3.5,0) -- node[below] {\footnotesize $\ket{v_+}$} (-2,0);
        \draw[ultra thick,dotted] (3.5,0) -- node[below] {\footnotesize $\ket{v_+}$} (2,0);
        \draw[ultra thick,dashed] (-2,0) -- node[below] {\footnotesize $\ket{v_-}$} (2,0);
        \draw[thick,red!50,fill=red!30] (-3.5,0) circle (.3) node[white] {$+$};
        \draw[thick,blue!50,fill=blue!30] (-2,0) circle (.3) node[white] {$-$};
        \draw[thick,red!50,fill=red!30] (2,0) circle (.3) node[white] {$+$};
        \draw[thick,blue!50,fill=blue!30] (3.5,0) circle (.3) node[white] {$-$};
        \draw[ultra thick,-latex] (3.25,.5) -- (3.75,.5);
        \draw[ultra thick,-latex] (1.75,.5) -- (2.25,.5);
        \draw[ultra thick,-latex] (-1.75,.5) -- (-2.25,.5);
        \draw[ultra thick,-latex] (-3.25,.5) -- (-3.75,.5);
    }\right).
\end{split}
\end{equation}
Consistent with this, in our simulations the region between the outgoing excitations does not look locally like either of the cluster-decomposing vacua $\ket{v_\pm}$, but like a superposition of the two. The value of $\langle E\rangle$ in the central region therefore interpolates between its values in $\ket{v_+}$ and $\ket{v_-}$, weighted by $|\alpha_+|^2$ and $|\alpha_-|^2$. The decrease of $|\langle E\rangle|$ with increasing momentum thus signals a growing weight $|\alpha_-|^2$ of the string-breaking channel. (Of course, we could also include $2\to 6$ and higher production channels in this sum; we have truncated it at $2\to 4$ because, empirically, we find that $2\to 6$ and higher production happens only with small probability in the kinematic regime we study. See the discussion of Figure~\ref{fig:probability_comparison} below.) 

\begin{figure}
    \centering
    \begin{subfigure}[t]{0.5\linewidth}
    \centering
    \includegraphics[width=.98\linewidth]{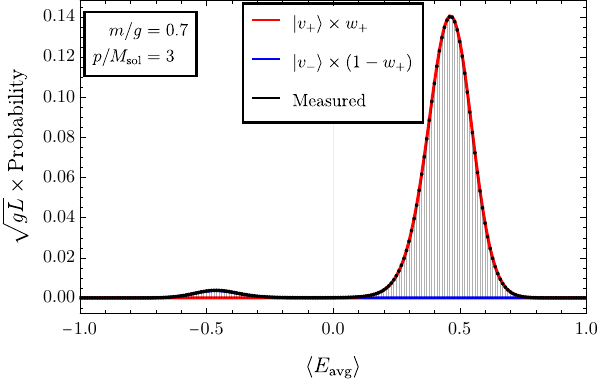}
    \caption{$p/M_\text{sol} = 3$, with a weight $w_+ \approx 0.97$.}
    \label{fig:electric_histogram_collision_pM3}
    \end{subfigure}%
    \begin{subfigure}[t]{0.5\linewidth}
    \centering
    \includegraphics[width=.98\linewidth]{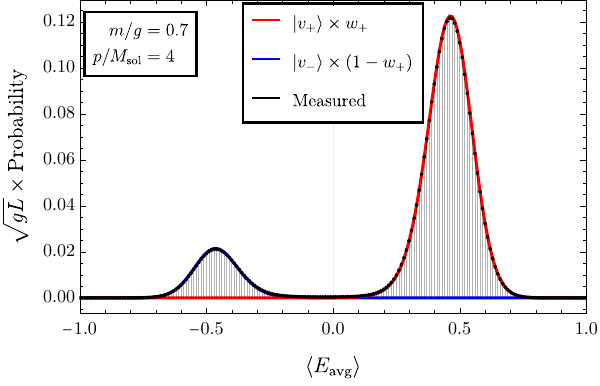}
    \caption{$p/M_\text{sol} = 4$, with a weight $w_+ \approx 0.85$.}
    \end{subfigure}\\[2em]
    \begin{subfigure}[t]{0.5\linewidth}
    \centering
    \includegraphics[width=.98\linewidth]{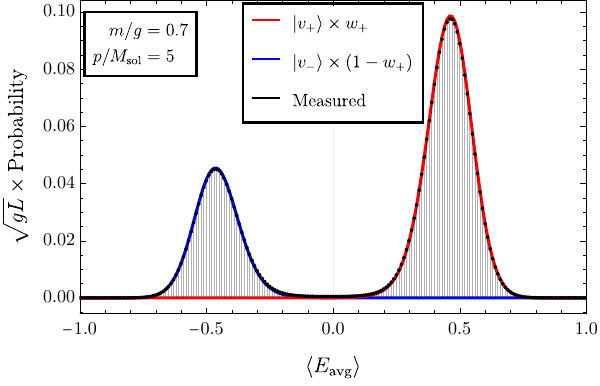}
    \caption{$p/M_\text{sol} = 5$, with a weight $w_+ \approx 0.68$.}
    \end{subfigure}%
    \begin{subfigure}[t]{0.5\linewidth}
    \centering
    \includegraphics[width=.98\linewidth]{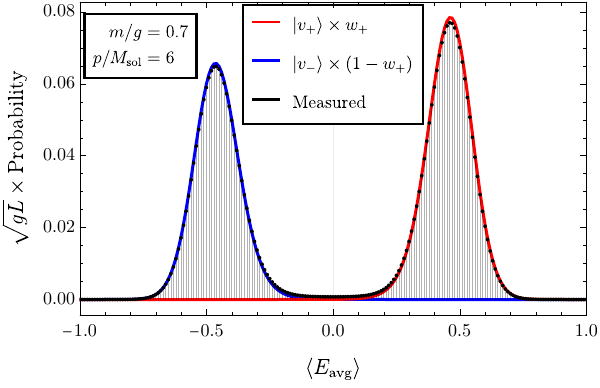}
    \caption{$p/M_\text{sol} = 6$, with a weight $w_+ \approx 0.54$.}
    \end{subfigure}
    \caption{Histograms of the electric field averaged over an interval of length $gL = 10$ in the final states of the $m/g = 0.7$ collisions shown in Figure~\ref{fig:inelastic_examples}. The histogram is well-described as a linear combination of the histograms in $\ket{v_\pm}$.}
    \label{fig:electric_histogram_collision}
\end{figure}

To support the interpretation \eqref{eq:psifinal} of the final state, we can calculate a histogram of $E_\text{avg}$ using the method described in Appendix~\ref{app:avg_e_field}. If \eqref{eq:psifinal} holds, then we should expect this histogram to be a weighted linear combination of a histogram like the one in Figure~\ref{fig:vacuum_histogram} for $\ket{v_+}$ and its mirror image for $\ket{v_-}$. And indeed, in Figure~\ref{fig:electric_histogram_collision}, we see just this. At $p/M_\text{sol} = 3$, there is a small probability ($\approx 3\%$) of ending up in $\ket{v_-}$; at $p/M_\text{sol} = 6$, that probability rises to $\approx 46\%$.

We will denote by $w_+ = |\alpha_+|^2$ the probability of finding the state $\ket{v_+}$ between the outgoing excitations. From a late-time measurement of the electric field, we can extract $w_+$ as
\begin{equation}\label{eq:wplus}
    w_+ = \frac12 + \left.\frac{\langle E(x,t)\rangle}{\qsol}\right|_{\substack{x=0\\t\to\infty}}\,.
\end{equation}
In light of \eqref{eq:psifinal}, we can interpret $w_+$ as an estimate for the probability of a $2\to 2$ event. We expect it to be an overestimate, because it also includes the probabilities of $2\to 6$, $2\to 10$, and so on.

\begin{figure}
    \centering
    \includegraphics[width=0.7\linewidth]{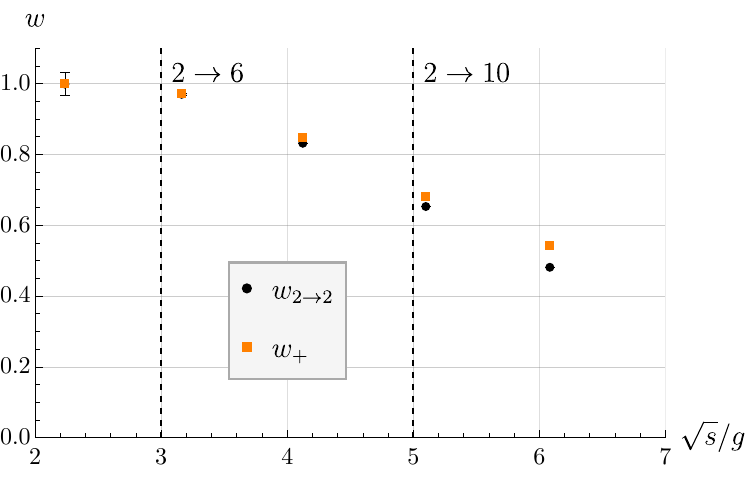}
    \caption{We compare two means of estimating the $2\rightarrow 2$ scattering probability: the proxy $w_+$ (orange squares) and via two-particle projection as in \eqref{eq:2solprob} (black circles). The quantities are given as a function of the center-of-mass energy $\sqrt{s}$ for simulations with lattice spacing $ag=0.1$ and bare mass $m/g=0.7$. Error bars for $w_{2\rightarrow2}$ are determined by varying the momentum grid spacing and the minimal separation $n_\mathrm{sep}$ between the left and right regions $W_L$ and $W_R$ as described in App.\,\ref{app:twoparticle}.}
    \label{fig:probability_comparison}
\end{figure}

To quantify by how much $w_+$ exceeds the true $2\to 2$ probability, we can estimate the latter using an independent, more direct method \cite{Milsted:2020jmf, Belyansky:2023rgh, Jha:2024jan}. We project the outgoing state onto the basis \eqref{eq:two_soliton_basis} of two-soliton states, and integrate the resulting density \eqref{eq:two_soliton_density} over a fine grid of momenta $(p_1, p_2)$ according to \eqref{eq:2solprob}.
Further details are given in Appendix~\ref{app:twoparticle}.

In Figure~\ref{fig:probability_comparison}, we plot both $w_+$ and $w_{2\to 2}$ for some collisions at $m/g = 0.7$ as a function of the center-of-mass collision energy 
\begin{equation}\label{eq:com_s}
    \sqrt{s} = 2\sqrt{M_\text{sol}^2 + p^2}.
\end{equation}
For smaller collision energies (near the six-particle threshold $\sqrt{s} = 6M_\text{sol}$), we see excellent agreement between $w_+$ and $w_{2\to 2}$, confirming our interpretation \eqref{eq:psifinal} of the outgoing state. For larger collision energies, $ w_{2\to2}$ becomes slightly lower than $w_+$. We can view this discrepancy as a rough estimate of the probability for $2\rightarrow 6$ scattering. For instance, $w_{2\to6} \approx 0.06$ for $p/M_\mathrm{sol}=6$.

\begin{figure}
    \centering
    \begin{subfigure}[t]{0.48\textwidth}
        \centering
        \includegraphics[width=\linewidth]{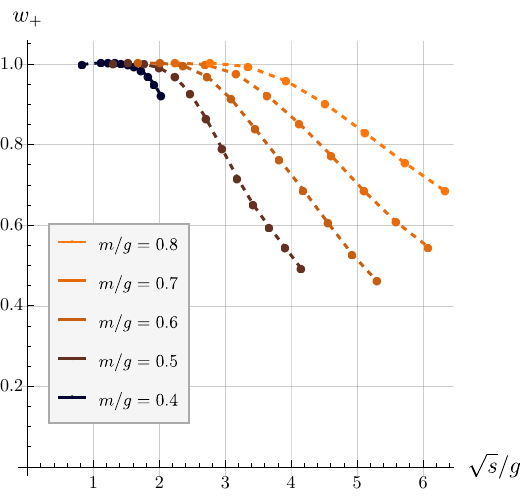}  
        \caption{}
        \label{fig:w_vs_sqrts_g}
    \end{subfigure}%
    \hspace{.02\textwidth}%
    \begin{subfigure}[t]{0.48\textwidth}
        \centering
        \includegraphics[width=\linewidth]{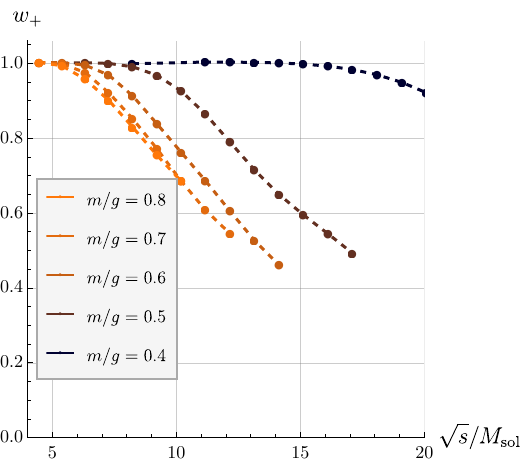}  
        \caption{}
        \label{fig:w_vs_sqrts_Msol}
    \end{subfigure}
    \caption{Estimates of the weight of $\ket{v_+}$ between the two outgoing excitations in the final state, which serves as a proxy for the probability of $2\to 2$ scattering. This probability decreases as we increase the collision energy $\sqrt{s}$, consistent with a $2\to 4$ channel opening. When we plot $w_+$ against $\sqrt{s}/M_\text{sol}$ on the right, we see that the threshold at which $w_+$ decreases significantly below 1 is much greater than the kinematic threshold $\sqrt{s}=4M_\text{sol}$, increasing as we decrease $m$ towards $m_c$. We have checked that the points are insensitive to $ag$ at a $\sim\! 1\%$ level, as we discuss in Appendix~\ref{app:simulation}.}
    \label{fig:w_vs_sqrts}
\end{figure}

Since this discrepancy remains small, we will compute $w_+$, which is computationally much easier, for a larger set of collisions at $m/g\in\{0.4,0.5,0.6,0.7,0.8\}$, and use it as an estimate of the elastic scattering probability. In Figure~\ref{fig:w_vs_sqrts}, we plot $w_+$ as a function of the collision energy $\sqrt{s}$, in units of $g$ on the left and in units of $M_\text{sol}$ on the right.

On the left, we see that the threshold value of $\sqrt{s}/g$ at which we start to see inelasticity decreases as we decrease $m/g$. This trend makes intuitive sense: lighter solitons require less energy to pair produce. On the right, however, we see that the inelasticity threshold is many times above the kinematic threshold $\sqrt{s} = 4M_\text{sol}$, and that the ratio increases as we approach the critical point.\footnote{In \cite{Belyansky:2023rgh}, the authors work at parameters near the critical point, and similarly observe the onset of inelasticity only at many times the kinematic threshold.} We explain this phenomenon in Section~\ref{sec:critical_scattering}.

\subsection{Measurement collapse}\label{sec:results_collapse}

In Figure~\ref{fig:electric_histogram_collision}, we see that in the final state of our collisions the distribution of electric field is strongly bimodal, with a lobe at $E_\text{avg}>0$ for the $\ket{v_+}$ contribution and a well-separated lobe at $E_\text{avg} < 0$ for the $\ket{v_-}$ contribution.

If we measure $E_\text{avg}$, the state will collapse into a state that looks like one of the two possibilities in \eqref{eq:psifinal}. After the collapse, we can measure expectation values of (for instance) the local electric field (outside of the region on which we measured $E_\text{avg}$), and see a state resembling one of these possibilities.
\begin{figure}
    \centering
    \includegraphics[width=.95\linewidth]{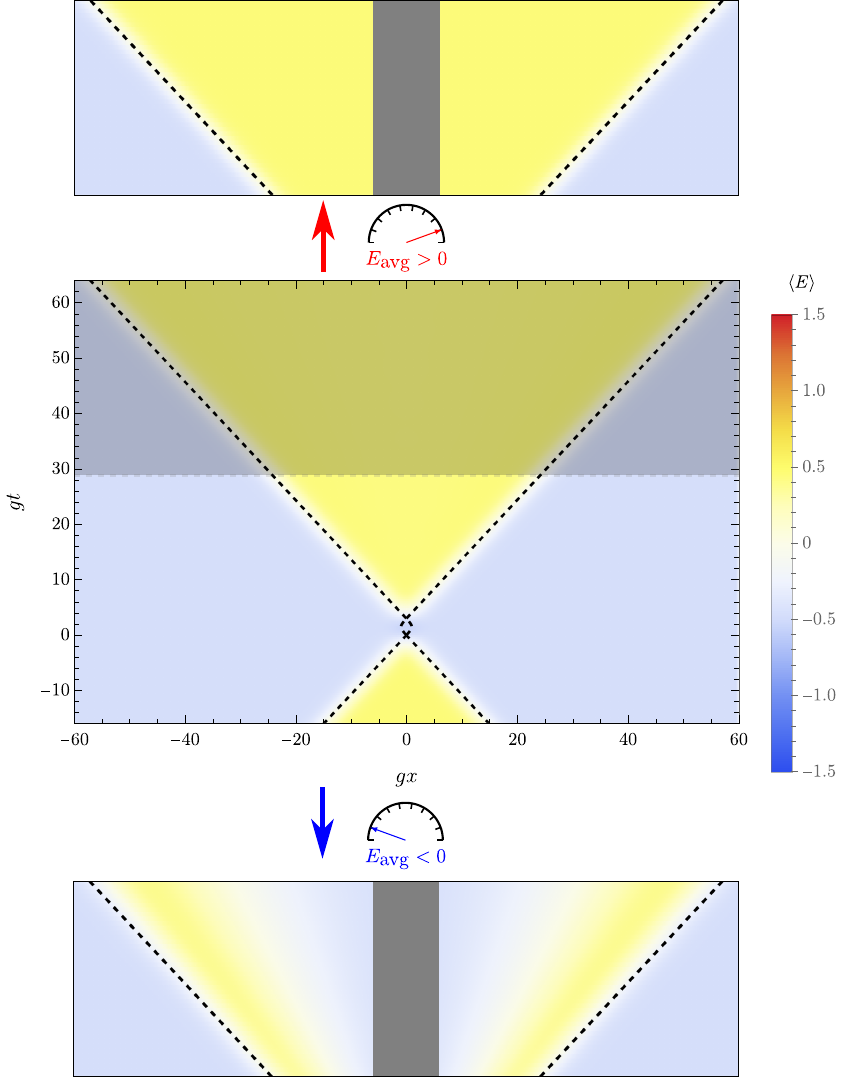}
    \caption{In the collision with $m/g = 0.7$ and $p/M_\text{sol} = 3$, for which $w_+ \approx 0.97$, we measure the average electric field in a central region and then calculate other electric field expectation values conditioned on this average being either positive or negative. In the $E_\text{avg} < 0$ case, we see a state in which the string broke, producing outgoing dipoles. This $\sim 3\%$-likelihood outcome was essentially invisible in the unprojected simulation data.}
    \label{fig:measurement_collapse}
\end{figure}
To carry this measurement out in practice, we will not actually project our state; instead, we will compute correlators of $E_\text{avg}$ with the local operators we want to measure, as described in Appendix~\ref{app:avg_e_field}\@. The result, however, is the same: we can effectively induce the measurement collapse of our simulation in a post-collision state and then measure properties of the outcome.

In Figure~\ref{fig:measurement_collapse}, we show post-collision portions of the $m/g = 0.7$, $p/M_\text{sol} = 3$ simulation (the top row of Figure~\ref{fig:inelastic_examples}) and then project onto the $E_\text{avg}>0$ and $E_\text{avg}<0$ parts of the state. When we measure $E_\text{avg} > 0$, then we project into a state in which the string did not break; since this was the case with $\sim 97\%$ probability (see Figure~\ref{fig:electric_histogram_collision_pM3}), we hardly see the change in electric field expectation values. However, if we measure $E_\text{avg} < 0$, then we project into a state in which the string broke. We then see the region of positive electric field within each outgoing dipole, consistent with the picture in the second term of \eqref{eq:psifinal}.

\section{Approaching the critical point}\label{sec:critical}

In this section, we will briefly consider the physics of solitons as we approach $m = m_c$ from above. In Section~\ref{sec:critical_charge_mass}, we derive the near-conformal charge-mass relation using integrability of the 2d Ising model, and we test our numerics against this prediction, showing good agreement. In Section~\ref{sec:critical_scattering}, we consider scattering of the light solitons in the near-conformal limit. We demonstrate analytically that, when $M_\text{sol} \ll g$, there is a separation of scales: inelastic scattering turns on only at energy scales of order $g$, and hence the ratio $\sqrt{s}/M_\text{sol}$ at which inelasticity becomes significant grows parametrically large (in particular, well above the kinematic production threshold).

\subsection{Charge-mass relation}\label{sec:critical_charge_mass}

The power laws for the onset of the soliton charge and mass as we increase $m$ above $m_c$, given in \eqref{eq:scaling}, follow straightforwardly from the scaling dimensions of operators in the 2d Ising CFT\@. The mass deformation $-(m-m_c)\bar\psi \psi$ away from the critical Schwinger model corresponds to a deformation of the 2d Ising CFT by the dimension-1 operator $\epsilon$ with coefficient $Z_m (m-m_c)$, where $Z_m$ is some non-universal factor, and so $m-m_c$ has dimension $2-\Delta_\epsilon = 1$. The dimension of $m-m_c$ establishes the first relation in \eqref{eq:scaling}. Meanwhile, $\qsol = 2\braket{v_+ | E | v_+}$ corresponds to $2Z_Q |\langle \sigma\rangle|$, where $Z_Q$ is another non-universal factor relating the electric field operator in the critical Schwinger model to the spin operator $\sigma$ in the 2d Ising CFT\@. From $\Delta_\sigma = \tfrac{1}{8}$, we find the second relation in \eqref{eq:scaling}.

These relations imply the scaling law $\qsol^8 \propto M_\text{sol}$ near the critical point. In this section, we will use results from integrability to fix the proportionality constant in terms of the electric field two-point function. To that end, we need to address the two unknown factors $Z_m$ and $Z_Q$. 

To eliminate $Z_m$, we will compare two results from integrability of the Ising model. First, the mass of the lightest particle in the ordered phase near criticality is related to the coefficient of $\epsilon$ in the deformation (which for us is $Z_m(m-m_c)$) by \cite{Delfino:2003yr,Zamolodchikov:1995xk,Fateev:1993av}
\begin{equation}
    M_\text{sol} = 2\pi Z_m (m-m_c)\,.
\end{equation}
Second, the expectation value of $\sigma$ in this region is given by \cite{Delfino:2003yr,Fonseca:2003ee}
\begin{equation}\label{eq:sigma_scaling}
    \langle \sigma\rangle = \pm 2^{5/24} (\pi/e)^{1/8} A^{3/2} \left(Z_m (m-m_c)\right)^{1/8}\,.
\end{equation}
Here, $A = 1.28243\ldots$ is the Glaisher constant. Combining these results, we find
\begin{equation}\label{eq:qm_scaling}
    \qsol = 2Z_Q \times  2^{1/12} e^{-1/8} A^{3/2} \times M_\text{sol}^{1/8}\,.
\end{equation}

\begin{figure}[t]
    \centering
    \includegraphics[width=0.8\linewidth]{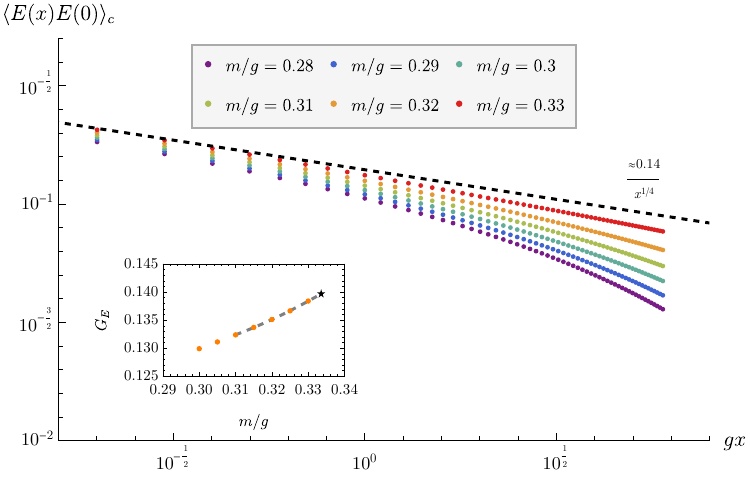}
    \caption{The connected correlator of the electric field, defined on the lattice using \eqref{eq:eeff_lattice}, as a function of physical spacing for several masses just below the critical point. In the main plot we use a lattice spacing $ag=0.1$. In the inset, we have extrapolated $\langle E(x) E(0)\rangle (gx)^{1/4}$ first to $ag\to 0$ and then to small $gx$ (i.e. much smaller than the correlation length), which lets us extract the coefficient of the near-conformal scaling at distances less than the correlation length. We finally extrapolate these $G_E$ estimates to $m_c/g \approx 0.33356$ \cite{ArguelloCruz:2024xzi,Fujii:2024reh,Okuda:2026hbg} to extract our estimate $G_E\approx 0.14$.}
    \label{fig:ee_corr}
\end{figure} 

To fix $Z_Q$, we will use two-point function normalizations. In \eqref{eq:sigma_scaling}, $\sigma$ was normalized so that, at criticality, its equal-time two-point function in the vacuum state is
\begin{equation}
    \langle \sigma(x) \sigma(0)\rangle_c = \frac{1}{|x|^{1/4}}\,,
\end{equation}
where $\langle\cdot\rangle_c$ denotes the connected correlation function. Thus, by calculating $\langle E(x) E(0)\rangle_c$ on the lattice near the critical point, and extracting the coefficient $G_E$ in
\begin{equation}\label{eq:EE_corr}
    \langle E(x) E(0)\rangle_c = \frac{G_E}{|gx|^{1/4}}\,,
\end{equation}
we may conclude that $Z_Q = g^{-1/8}\sqrt{G_E}$. Putting everything together, we obtain the proportionality
\begin{equation}
    \frac{\qsol}{(M_\text{sol}/g)^{1/8}} = (2^{13/12} e^{-1/8} A^{3/2})\times\sqrt{G_E} \approx 2.71568\times\sqrt{G_E}\,.
\end{equation}
Taking the eighth power, we find \begin{equation}\label{eq:mass_charge}
    \frac{\qsol^8}{M_\text{sol}/g} \approx 2958.2\times G_E^4\,.
\end{equation}

Now, let us test this relationship in our lattice model. First, we need to estimate $G_E$ by calculating the two-point correlation function of the electric field operator. More explicitly, we can use the identification \eqref{eq:eeff_lattice} and generalize \eqref{eq:Osandwich_limit} to allow for two of the transfer matrices in the infinite product to be modified by an insertion of $L_n$. This procedure allows us to calculate the electric field two-point function on a finite-spacing lattice at $m \neq m_c$ (because a uMPS state cannot represent the vacuum with infinite correlation length that appears at $m = m_c$ \cite{Pollmann:2009lnv,Pirvu:2012tpe}). We then extrapolate $ag\to 0$ and $m\to m_c$ to estimate the coefficient $G_E$.

In Figure~\ref{fig:ee_corr}, we carry out this procedure for masses just below the critical mass; this is the only instance in this paper for which we work in the confined phase. (It would also be possible to approach $m_c$ from above, but in practice for estimating $G_E$ we found it numerically easier to approach from below.) We see excellent consistency with the conformal power-law scaling of the correlator, and we estimate the coefficient $G_E\approx 0.14$.

\begin{figure}
    \centering
    \includegraphics[width=0.7\linewidth]{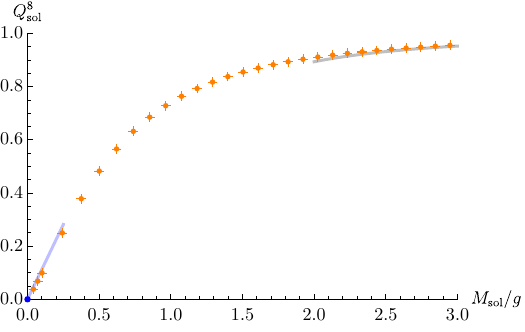}
    \caption{The relationship between $\qsol^8$ and $M_\text{sol}$. In light blue, we plot the near-critical relation \eqref{eq:mass_charge}, which uses the coefficient $G_E\approx 0.14$ estimated in Figure~\ref{fig:ee_corr}. In light gray, we plot the weak-coupling relation inferred from \eqref{eq:weak_coupling}.}
    \label{fig:charge_mass}
\end{figure}

We can substitute this estimate of $G_E$ into \eqref{eq:mass_charge} to find $\qsol^8/(M_\text{sol}/g) \approx 1.14$. In Figure~\ref{fig:charge_mass}, we combine the data of Figure~\ref{fig:eplus} (together with \eqref{eq:qsol_def}) and Figure~\ref{fig:m} to plot $\qsol^8$ as a function of $M_\text{sol}/g$. We see good agreement with the predicted linear relationship.

\subsection{Scattering light solitons}\label{sec:critical_scattering}

As discussed in Section~\ref{sec:critical_charge_mass}, for $m - m_c \ll g$, the Schwinger model at $\theta = \pi$ is described at long distances by the 2d Ising CFT deformed by the energy operator $\epsilon$. This deformation alone gives an integrable theory. Upon fermionization, this theory becomes a theory of a free Majorana fermion of mass $M_\text{sol}$ \cite{Wu:1975mw,McCoy:1977er,DiFrancesco:1997nk}, whose $2\to 2$ $S$-matrix is a pure phase, $|S(\vartheta)| = 1$; in standard conventions, it is $S(\vartheta) = -1$ (see, for instance, \cite{Berg:1978sw, Zamolodchikov:2011wd}). Here, $\vartheta$ is the difference in the rapidities of the two particles being scattered, defined by 
\begin{equation}
    s = 4M_\text{sol}^2\cosh^2(\vartheta/2) \,,
\end{equation}
where $s$ is the squared center-of-mass collision energy defined in \eqref{eq:com_s}. Since $|S(\vartheta)| = 1$, there is no particle production. Inelastic scattering, and hence string breaking, comes from corrections to this description.

To study those corrections, we write the near-critical Schwinger model as an effective field theory,
\begin{equation}\label{eq:ising_eft}
    S = S_\text{Ising  CFT} + Z_m (m - m_c) \int d^2x\, \epsilon(x) + \sum_i c_i\, g^{2-\Delta_i} \int d^2x\, \cO_i(x) + \ldots\,,
\end{equation}
where the $\cO_i$ are $C$-even, scalar, irrelevant ($\Delta_i > 2$) operators of the Ising CFT, and the ellipsis denotes terms suppressed by further powers of $(m-m_c)/g$. The structure of \eqref{eq:ising_eft} follows from the fact that only one parameter, $m$, is tuned to reach the critical point. At $m = m_c$, the only scale remaining is $g$, and the Ising CFT description emerges at distances much longer than $1/g$. Tuning $m$ close to $m_c$ makes the coefficient of $\epsilon$, the unique relevant $C$-even operator ($\sigma$ is $C$-odd), small; it does nothing special to the irrelevant couplings, which are smooth functions of $m$ near $m_c$. They approach finite, $g$-dependent limits as $m\to m_c$. 

By dimensional analysis, the operator $\cO_i$ then changes scattering amplitudes at energy $\sqrt{s}\gtrsim M_\text{sol}$ by (at most) a relative amount of order $c_i(\sqrt{s}/g)^{\Delta_i - 2}$. Deviations from free-fermion scattering are thus controlled by $\sqrt{s}/g$, not by $\sqrt{s}/M_\text{sol}$. Throughout the window
\begin{equation}\label{eq:window}
    M_\text{sol}\ll \sqrt{s}\ll g\,,
\end{equation}
scattering is nearly that of free fermions, even though most of this window lies far above the four-particle threshold $\sqrt{s} = 4M_\text{sol}$. Inelasticity can only become significant at $\sqrt{s_*}\sim g$, and since $M_\text{sol} = 2\pi Z_m (m - m_c)$,
\begin{equation}\label{eq:threshold_scaling}
    \frac{\sqrt{s_*}}{M_\text{sol}} \sim \frac{g}{M_\text{sol}} \propto \frac{g}{m - m_c}\,,
\end{equation}
which grows parametrically as $m\to m_c$. 

This separation of scales was observed in \cite{Belyansky:2023rgh}
(albeit at fixed lattice parameters rather than in a continuum limit). In near-critical collisions using a Hamiltonian lattice description of the bosonized Schwinger model, those authors found that the elastic probability stays close to one up to $\sqrt{s}/M_\text{sol}\approx 11$, and drops below $\frac12$ only at $\sqrt{s}/M_\text{sol}\approx 28$, well above the four-particle threshold. We observe a similar behavior for our collisions at $m/g = 0.4$; see Figure~\ref{fig:w_vs_sqrts_Msol}.

\section{Conclusion}\label{sec:conclusion}

In this paper, we have demonstrated the power of combining Hamiltonian lattice gauge theory \cite{Kogut:1974ag} with modern tensor network methods \cite{Banuls:2018jag,Banuls:2022vxp}. We are able to directly simulate the real-time dynamics of a non-perturbative process, an $e^+e^-$ collision and the subsequent breaking of flux strings, in the massive Schwinger model. Our simulations produce a superposition of classical outcomes, and hence encode a wealth of information about the process, which we have only just begun to explore.

Carrying out the simulations in this paper relied upon a combination of older ideas in Hamiltonian lattice gauge theory with several recent developments. We have used the Banks-Kogut-Susskind Hamiltonian \cite{Kogut:1974ag,Banks:1975gq}, modified with the mass shift prescription \cite{Dempsey:2022nys} that has been shown to hasten convergence to the continuum limit. This Hamiltonian has been studied for decades \cite{Crewther:1979ka,Hamer:1982mx,Schiller:1983sj,Hamer:1997dx,Sriganesh:1999ws,Kroger:1998se}, with the most precise results coming from the use of matrix product states \cite{Byrnes:2002gj,Banuls:2013jaa,Honda:2022edn,ArguelloCruz:2024xzi,Fujii:2024reh,Itou:2024psm}. However, the standard setup of matrix product states and operators \cite{Schollwock:2005zz,Perez-Garcia:2006nqo,Paeckel:2019yjf,Vanderstraeten:2019voi} does not naturally accommodate a local Gauss law constraint, nor the electric field energy term; much previous work has instead eliminated the electric field using the Gauss law, at the cost of long-range terms in the Hamiltonian (see, however, \cite{Buyens:2013yza,Fujii:2024reh}). LEMPOs \cite{Dempsey:2025wia}, a recently developed alternative, incorporate link degrees of freedom and the Gauss law directly, and we have employed them throughout the paper (see Section~\ref{sec:method_lempo}). This approach enables us to work on an infinite lattice, where we can sharply define soliton states. Even so, determining the energy and momentum of soliton states is subtle, as we explained in Section~\ref{sec:method}.

With all of these tools---the lattice Hamiltonian, the mass shift, tensor network algorithms, LEMPOs, and the subtle physics of solitons---we are able to prepare lattice states in which $e^+$ and $e^-$ wavepackets are directed at each other, and eventually collide. As shown in Section~\ref{sec:results}, when the solitons are heavy, our simulations match very well with the classical prediction for relativistic particles moving under a constant Coulomb force. Moreover, when the solitons are lighter and their collision energy is increased, the final states show evidence that the flux string between the solitons breaks with some probability. We can exploit the long-range entanglement between the outgoing excitations and the gauge field between them to estimate that probability, and to project onto the string-broken and string-unbroken outcomes (see Figure~\ref{fig:measurement_collapse}). 

At present, using classical computation to perform such an accurate and resolved simulation of a quantum gauge theory is only possible in $1+1$ dimensions (although see \cite{Wu:2025aly,Xu:2025ean,Zwolfer:2026yuc} for progress in higher dimensions). Eventually, quantum simulation may become the most powerful tool for this kind of analysis; there has been rapid progress in recent years in both analog \cite{Zohar:2015hwa,Gorg:2018xyc,Zache:2018jbt, Yang:2020yer,Zhou:2021kdl,Aidelsburger:2021mia,Gonzalez-Cuadra:2024xul,De:2024smi,Liu:2024lut,Halimeh:2025vvp} and digital \cite{Martinez:2016yna,Muschik:2016tws,Kokail:2018eiw,Banuls:2019bmf,deJong:2021wsd,Nguyen:2021hyk,Farrell:2023fgd,Florio:2024aix, Cochran:2024rwe,Alexandrou:2025vaj,Farrell:2024fit,Davoudi:2025rdv,John:2026gut} quantum simulators, including recent observations of string-breaking dynamics. For the moment, however, classical tensor networks outperform any of these platforms in $1+1$ dimensions, and so our simulations can be viewed as a benchmark for future work in quantum hardware.

Classical methods, when they are feasible, also offer a distinct advantage: they are not subject to the no-cloning theorem, and so we can save the quantum states we produce and study them thoroughly at little additional cost. We have begun to demonstrate the value of this capability in Section~\ref{sec:results}, by calculating time delays, electric field histograms, two-particle projections, and so on; but even more information, for instance all $2\to n$ scattering probabilities and outgoing momentum distributions, should be extractable from our simulations. We leave such analysis to future work.

Our work also suggests two natural parameter regimes for further exploration. In Section~\ref{sec:critical_scattering}, we explain the suppression of string breaking near $m = m_c$. It would be interesting to test this prediction numerically, but it is challenging since the correlation length becomes large in this regime, leading to a need for higher bond dimensions and larger lattices. Additionally, in the semiclassical regime $m\gg g$, we should be able to estimate the string-breaking probability using the Schwinger rate \cite{Schwinger:1951nm} for false vacuum decay (see \cite{Hebenstreit:2013baa,Spitz:2018eps} for related calculations). This regime is also numerically challenging because in order to have heavy fermions and long, semiclassical flux strings, we need very high incoming momenta, which in turn requires a very fine lattice spacing so that such a momentum can be resolved. We leave both questions to future work targeting the particular difficulties of these regimes. 

It would also be of great interest to carry out similar simulations in a non-abelian setting. Degenerate vacua and solitons are a generic feature of gapped QCD-like theories in $1+1$ dimensions \cite{Komargodski:2020mxz,Cordova:2024vsq,Cordova:2024goh,Cordova:2024nux}, and the toolchain we have used in this work has been demonstrated to work well for non-abelian gauge theories also (in particular, for adjoint QCD$_2$ in \cite{Dempsey:2025wia}). The rich non-invertible symmetry structure of those theories imposes nontrivial constraints on the $S$-matrix \cite{Copetti:2024rqj,Cordova:2024vsq,Cordova:2024iti,Copetti:2024dcz}, which could be tested directly in real-time simulations.

Finally, we emphasize the broader takeaway: real-time simulations of quantum field theories are made possible by the many theoretical developments detailed above, but they may also be viewed as quasi-empirical input that can aid progress in our understanding of quantum field theories, just as real-world colliders have done for the Standard Model for decades. Future non-perturbative calculations of Lorentzian observables ($S$-matrices, detector correlators, and so on), using the methods of this paper and/or rapidly developing quantum hardware, are poised to open a new window into confinement, string breaking, hadronization, and other phenomena of interest in a wide range of model theories.

\section*{Acknowledgments} 

We thank Jan Albert, Jack Donahue, Igor Klebanov, Pedro Vieira, and Raju Venugopalan for helpful discussions.   RD is supported by a Pappalardo Fellowship in Physics at MIT\@.  The work of AMG, SSP, and BTS was supported in part by the Simons Foundation Grant No.~917464 (Simons Collaboration on Confinement and QCD Strings) and by the US Department of Energy under Award No.~DE-SC0007968.

\appendix

\section{Infinite tensor network methods with LEMPOs}\label{app:lempo}
 
In this appendix, we explain how we incorporate LEMPOs into the VUMPS algorithm and the quasiparticle ansatz. Throughout, we use the notation of Section~\ref{sec:method}.
 
\subsection{VUMPS algorithm}\label{app:lempo_envs}
 
The LEMPO environments defined in \eqref{eq:E_and_A} allow us to adapt most algorithms for uMPSs to work with LEMPOs. One such algorithm is the variational uniform matrix product state algorithm (VUMPS) \cite{Zauner-Stauber:2017eqw}, which finds ground-state uMPSs by iterating a consistency condition derived from demanding that the gradient of the energy on the uMPS manifold vanishes. In defining this gradient, one has to subtract off the infinite extensive energy of the ground state, which is implemented by using the environments $\mathbb E_{L/R}$ in place of the semi-infinite contractions in the effective Hamiltonians.
 
The environments $\mathbb{E}_L$ and $\mathbb{E}_R$ of Section~\ref{sec:method_envs} are defined at the boundaries of a unit cell, i.e., at a cut through the virtual bonds just after the link tensor $W_k^{(L)}$. VUMPS, and the quasiparticle ansatz below, also need environments at cuts inside the unit cell. Since the link tensors sit between sites, we must specify which environment includes each link tensor. We place every cut just to the right of a link tensor: $\mathbb{E}_L^{(j)}$ contains everything to the left of site $j$ (including $W_{j-1}^{(L)}$), and $\mathbb{E}_R^{(j)}$ contains everything to the right of $W_j^{(L)}$. Then $\mathbb{E}_L^{(1)} = \mathbb{E}_L$, $\mathbb{E}_R^{(k)} = \mathbb{E}_R$, and the remaining environments are obtained by stepping through the unit cell,
\begin{equation}\label{eq:env_propagation}
	(\mathbb{E}_L^{(j+1)})^T = (\mathbb{E}_L^{(j)})^T\, \mathbb{T}^{(j)}\,,\qquad
	\mathbb{E}_R^{(j-1)} = \mathbb{T}^{(j)}\, \mathbb{E}_R^{(j)}\,,
\end{equation}
where $\mathbb{T}^{(j)} \equiv T(A_j, W_j, W_j^{(L)}, \bar{A}_j)$ is the LEMPO transfer matrix \eqref{eq:lempo_link_transfer_matrix} for a single site. The accumulators \eqref{eq:A_def} propagate in the same way; because of the upper-triangular form \eqref{eq:W}, the boundary vectors $\delta_i$ and $\delta_f$ pass through each step unchanged, so that $\mathbb{A}_L^{(j)} = \delta_i\otimes l^{(j)}$ and $\mathbb{A}_R^{(j)} = \delta_f \otimes r^{(j)}$, where $l^{(j)}$ and $r^{(j)}$ are the fixed points $l$ and $r$ propagated to the corresponding cuts.\footnote{Note that \eqref{eq:env_propagation} fixes the environments $\mathbb{E}^{(j)}_{L/R}$ up to the ambiguities \eqref{eq:etas} for the single left and right environments determined via \eqref{eq:E_and_A}. By contrast, solving a version of \eqref{eq:E_and_A} for each partition of the unit cell would lead to a separate freedom \eqref{eq:etas} at each $j$.}
 
VUMPS makes use of the gauge redundancy of the uMPS, under which an invertible matrix and its inverse are inserted on each virtual bond \cite{Vanderstraeten:2019voi}. In particular, one can bring the uMPS tensors $A_j$ into the left and right canonical forms $A_{L,j}$ and $A_{R,j}$, defined by the conditions
\begin{equation}\label{eq:cann_form}
	\begin{tikzpicture}[diagram]
		\node[style=square] at (0,0) (A) {$A_{L,j}$};
		\node[style=square] at (0,-1.5) (Ab) {$\bar A_{L,j}$};
		\draw (A.south) -- (Ab.north);
		\draw[rounded corners=8pt] (A.west) -- (-1,0) -- (-1,-1.5) -- (Ab.west);
		\draw (A.east) -- (.75,0) (Ab.east) -- (.75,-1.5);
	\end{tikzpicture} =
	\begin{tikzpicture}[diagram]
		\draw[rounded corners=8pt] (0,0) -- (-.8,0) -- (-.8,-1.5) -- (0,-1.5);
	\end{tikzpicture}\,,\qquad
	\begin{tikzpicture}[diagram]
		\node[style=square] at (0,0) (A) {$A_{R,j}$};
		\node[style=square] at (0,-1.5) (Ab) {$\bar A_{R,j}$};
		\draw (A.south) -- (Ab.north);
		\draw[rounded corners=8pt] (A.east) -- (1,0) -- (1,-1.5) -- (Ab.east);
		\draw (-.75,0) -- (A.west) (-.75,-1.5) -- (Ab.west);
	\end{tikzpicture} =
	\begin{tikzpicture}[diagram]
		\draw[rounded corners=8pt] (0,0) -- (.8,0) -- (.8,-1.5) -- (0,-1.5);
	\end{tikzpicture}\;.
\end{equation}
That is, the left fixed point in the left gauge is the identity, and likewise for the right fixed point in the right gauge. The bond matrix $C_j$ relating the two canonical forms defines the center tensor $A_{C,j}$ through
\begin{equation}
	\begin{tikzpicture}[diagram]
		\node[style=square] at (0,0) (A) {$A_{L,j}$};
		\node[style=circle] at (1.5,0) (C) {$C_j$};
		\draw (-.75,0) -- (A.west) (A.east) -- (C.west) (C.east) -- (2.25,0);
		\draw (A.south) -- (0,-.75);
	\end{tikzpicture} =
	\begin{tikzpicture}[diagram]
		\node[style=square] at (0,0) (A) {$A_{C,j}$};
		\draw (-.75,0) -- (A.west) (A.east) -- (.75,0);
		\draw (A.south) -- (0,-.75);
	\end{tikzpicture} =
	\begin{tikzpicture}[diagram]
		\node[style=circle] at (0,0) (C) {$C_{j-1}$};
		\node[style=square] at (1.5,0) (A) {$A_{R,j}$};
		\draw (-.75,0) -- (C.west) (C.east) -- (A.west) (A.east) -- (2.25,0);
		\draw (A.south) -- (1.5,-.75);
	\end{tikzpicture}\;.
\end{equation}
 
The dynamical input to the VUMPS algorithm comes from effective Hamiltonians $H_{A_{C,j}}$ and $H_{C_j}$, which act on the center tensors and bond tensors, respectively \cite{Zauner-Stauber:2017eqw}. We write them in terms of the environments above, with $\mathbb{E}_L^{(j)}$ evaluated using the left-canonical tensors and $\mathbb{E}_R^{(j)}$ using the right-canonical tensors. As in Section~\ref{sec:method_qp}, the semi-infinite contractions are given by \eqref{eq:jordan_power_1} as environments plus multiples of the accumulators, so we write
\begin{equation}\label{eq:HAC}
	H_{A_{C,j}}:\;\begin{tikzpicture}[diagram]
		\node[style=square] at (0.0,0) (A1) {$A_{C,j}$};
		\draw (A1) -- (0.75,0) (A1) -- (-0.75,0) (A1) -- (0,-.75);
	\end{tikzpicture}\quad \longmapsto\quad \begin{tikzpicture}[diagram]
		\draw[rounded corners,fill=white] (-1.5,-2.3) rectangle node[rotate=90] {$\mathbb{E}_L^{(j)} + \alpha_L \mathbb{A}_L^{(j)}$} (-0.5,.3);
		\draw[rounded corners,fill=white] (2.75,-2.3) rectangle node[rotate=-90] {$\mathbb{E}_R^{(j)} + \alpha_R \mathbb{A}_R^{(j)}$} (3.75,.3);
		\node[style=square] at (0.5,0) (A1) {$A_{C,j}$};
		\node[style=circle] at (0.5,-1) (W1) {$W_j$};
		\node[style=circle] at (1.75,0) (L1) {$W^{(L)}_j$};
		\draw (-0.5, 0) -- (A1);
		\draw (-0.5, -1) -- (W1);
		\draw (A1) -- (L1);
		\draw (W1) -- (L1);
		\draw (L1) -- (2.75, 0);
		\draw (L1) -- (2.75, -1);
		\draw (A1) -- (W1) -- (0.5, -3);
		\draw (-0.5, -2) -- (0, -2);
		\draw (0,-2) to[out=0, in=0] (0, -3);
		\draw (2.75, -2) -- (1, -2);
		\draw (1,-2) to[out=180, in=180] (1, -3);
	\end{tikzpicture}
\end{equation}
and
\begin{equation}\label{eq:HC}
	H_{C_j} :\begin{tikzpicture}[diagram]
		\node[style=circle] at (0.0,0) (A1) {$C_{j}$};
		\draw (A1) -- (0.75,0) (A1) -- (-0.75,0);
	\end{tikzpicture}\;\quad \longmapsto\quad \begin{tikzpicture}[diagram]
		\draw[rounded corners,fill=white] (-1,-2.3) rectangle node[rotate=90] {$\mathbb{E}_L^{(j+1)} + \alpha_L \mathbb{A}_L^{(j+1)}$} (0,.3);
		\draw[rounded corners,fill=white] (2,-2.3) rectangle node[rotate=-90] {$\mathbb{E}_R^{(j)} + \alpha_R \mathbb{A}_R^{(j)}$} (3,.3);
		\node[style=circle] at (1,0) (C) {$C_j$};
		\draw (0, 0) -- (C) -- (2, 0);
		\draw (0, -2) -- (0.5, -2) ;
		\draw (0.5,-2) to[out=0, in=0] (0.5, -3);
		\draw (1.5, -2) -- (2, -2) ;
		\draw (1.5,-2) to[out=180, in=180] (1.5, -3);
		\draw (0, -1) -- (2, -1);
	\end{tikzpicture}\,.
\end{equation}
In the VUMPS algorithm, one extracts the lowest eigenvectors $A_{C,j}'$ and $C_j'$ of the effective Hamiltonians,
\begin{equation}
        H_{A_{C,j}} A_{C,j}' = E_{A_{C,j}}A_{C,j}'\,,\qquad
        H_{C_j} C_j' = E_{C_j} C_j'\,,
    \label{eq:eigvec}
\end{equation}
which are used to iterate the uMPS tensors in the energy minimization; see \cite{Zauner-Stauber:2017eqw} for details of the update. These are the effective Hamiltonians for the single-site VUMPS algorithm; the two-site generalization, with dynamically adjusted bond dimension of the MPS tensor, is discussed in Appendix C of \cite{Dempsey:2025wia}.

Importantly, the VUMPS algorithm is insensitive to the boundary constants $\alpha_{L/R}$. Because of the upper-triangular form \eqref{eq:W}, $\delta_i$ in $\mathbb{A}_L^{(j)}$ and $\delta_f$ in $\mathbb{A}_R^{(j)}$ pass through $W_j$ unchanged, and in the canonical left and right gauges the fixed points $l^{(j)}$ and $r^{(j)}$ are identity matrices, respectively; so, just as in Section~\ref{sec:method_qp}, the accumulator terms contribute only a multiple of the identity,
\begin{equation}
    H_{A_{C,j}} = H_{A_{C,j}}\big|_{\alpha=0} + (\alpha_L+\alpha_R)\mathds 1\,,\qquad H_{C_j} = H_{C_j}\big|_{\alpha=0} + (\alpha_L+\alpha_R)\mathds 1\,,
\end{equation}
which does not change the lowest eigenvectors in \eqref{eq:eigvec}. In practice, we therefore drop the accumulator terms and use the environments $\mathbb{E}_{L/R}^{(j)}$ alone. 

The effective Hamiltonian \eqref{eq:HAC} and its two-site analog also generate the TDVP time steps of Section~\ref{sec:results_tdvp}. The only additional modification is the treatment of the link tensor inside the two-site tensor, described in Section~\ref{sec:results_tdvp}.

\subsection{Quasiparticle ansatz}
 
Having determined the vacua using VUMPS, we estimate the spectrum of excitations above them using the quasiparticle ansatz \eqref{eq:Bsoliton} \cite{Haegeman:2011lcd, Haegeman:2013xcv}, with the unit-cell tensor $\{B\}$ parametrized as in \eqref{eq:unit_cell_abbreviation} by a superposition of states in which a single $A_j$ is replaced by a new tensor $B_j$. The approximate excitation energies are then obtained by computing the matrix elements of $H_{\epsilon=0}$ between quasiparticle states and diagonalizing the resulting effective Hamiltonian on the space of tensors $\{B_j\}$ (subject to \eqref{eq:p0_ortho} for particle excitations at $p = 0$).
 
In constructing the LEMPO tensor network evaluating $\bra{\{B\}, p} H_{\epsilon=0} \ket{\{B\}, p}$, we encounter three kinds of terms:
\begin{enumerate}[1)]
	\item $B_i$ and $\bar B_i$ inserted at the same site in bra and ket,
	\item the ket $B_j$ to the right of the bra $\bar B_i$,
	\item the ket $B_j$ to the left of the bra $\bar B_i$.
\end{enumerate}
The latter two kinds of terms can be summed into environment-like vectors $\mathbb{E}^{(i)}_{B,L}(p)$ and $\mathbb{E}^{(i)}_{B,R}(p)$, giving
\begin{align}\label{eq:Mexc}
    &\bra{\{B\}, p} H_{\epsilon=0} \ket{\{B\}, p} =\frac{2\pi}{ka}\delta(0) \times \sum_{i=1}^{k}  \left(
     \begin{tikzpicture}[diagram]
        \draw[rounded corners,fill=white] (-1.5,-2.3) rectangle node {$\mathbb E_L^{(i)}$} (-0.5,.3);
        \draw[rounded corners,fill=white] (2.75,-2.3) rectangle node {$\tilde{\mathbb{E}}_R^{(i)}$} (3.75,.3);
        \node[style=square] at (0.5,0) (A1) {$B_i$};
        \node[style=circle] at (0.5,-1) (W1) {$W_i$};
        \node[style=square] at (0.5,-2) (B) {$\bar B_i$};
        \node[style=circle] at (1.75,0) (L1) {$W^{(L)}_i$};
        \draw (-0.5, 0) -- (A1) -- (L1) -- (2.75, 0);
        \draw (-0.5, -1) -- (W1) -- (L1) -- (2.75, -1);
        \draw (-0.5, -2) -- (B) -- (2.75, -2);
        \draw (A1) -- (W1) -- (B);
    \end{tikzpicture} \right. \\
    & \left.+
    \begin{tikzpicture}[diagram]
        \draw[rounded corners,fill=white] (-1.5,-2.3) rectangle node {$\mathbb E_{L}^{(i)}$} (-0.5,.3);
        \draw[rounded corners,fill=white] (2.75,-2.3) rectangle node {$\mathbb E_{B, R}^{(i)}(p)$} (3.75,.3);
        \node[style=square] at (0.5,0) (A1) {$A_i$};
        \node[style=circle] at (0.5,-1) (W1) {$W_i$};
        \node[style=square] at (0.5,-2) (B) {$\bar B_i$};
        \node[style=circle] at (1.75,0) (L1) {$W^{(L)}_i$};
        \draw (-0.5, 0) -- (A1) -- (L1) -- (2.75, 0);
        \draw (-0.5, -1) -- (W1) -- (L1) -- (2.75, -1);
        \draw (-0.5, -2) -- (B) -- (2.75, -2);
        \draw (A1) -- (W1) -- (B);
    \end{tikzpicture}
    +
    \begin{tikzpicture}[diagram]
        \draw[rounded corners,fill=white] (-1.5,-2.3) rectangle node {$\mathbb E_{B,L}^{(i)}(p)$} (-0.5,.3);
        \draw[rounded corners,fill=white] (2.75,-2.3) rectangle node {$\tilde{\mathbb{E}}_R^{(i)}$} (3.75,.3);
        \node[style=square] at (0.5,0) (A1) {$\tilde{A}_i$};
        \node[style=circle] at (0.5,-1) (W1) {$W_i$};
        \node[style=square] at (0.5,-2) (B) {$\bar B_i$};
        \node[style=circle] at (1.75,0) (L1) {$W^{(L)}_i$};
        \draw (-0.5, 0) -- (A1) -- (L1) -- (2.75, 0);
        \draw (-0.5, -1) -- (W1) -- (L1) -- (2.75, -1);
        \draw (-0.5, -2) -- (B) -- (2.75, -2);
        \draw (A1) -- (W1) -- (B);
    \end{tikzpicture}
    \right)
    \,,\notag
\end{align}
where, as in \eqref{eq:qp_energy_term_diag}, the environments to the left of the insertion are those of $\{A\}$ and the environments to the right are those of $\{\tilde{A}\}$, and the link tensors are included in the environments according to the convention of Appendix~\ref{app:lempo_envs}. The prefactor is the same divergent factor that appears in the norm $\braket{\{B\},p|\{B\},p}$, and it cancels in \eqref{eq:Esol}; see footnote~\ref{fn:normalizability}. The environments $\mathbb{E}^{(i)}_{B,L/R}(p)$ are constructed by solving the quasi-fixed point equations as detailed in \cite{Vanderstraeten:2019voi}, with the LEMPO transfer matrix and the environment conventions of Appendix~\ref{app:lempo_envs}.
 
Note that we have again dropped the accumulator terms. Restoring them, $\mathbb{E}_L^{(i)}\to\mathbb{E}_L^{(i)}+\alpha_L\mathbb{A}_L^{(i)}$ and $\tilde{\mathbb{E}}_R^{(i)}\to\tilde{\mathbb{E}}_R^{(i)}+\tilde\alpha_R\tilde{\mathbb{A}}_R^{(i)}$, adds $(\alpha_L+\tilde\alpha_R)\braket{\{B\},p|\{B\},p}$, as shown in Section~\ref{sec:method_qp}. This shifts the effective Hamiltonian on the space of $\{B_j\}$ by a multiple of the identity, so the optimal tensors $\{B_j\}$ are unaffected; only the value of the energy shifts, and this is fixed by \eqref{eq:CP_env}.

\section{Symmetries and ambiguities}\label{app:symmetries}

In this appendix, we explain how to implement the action of parity and charge conjugation on the uMPS representation of vacuum states in the spin formulation of the lattice Schwinger model. We will assume $\theta=\pi$ throughout the appendix.

\subsection{Symmetry operators}

The lattice operators $\mathcal{P}$ and $\mathcal{C}$ can be implemented by manipulating the physical and virtual spaces of uMPS tensors, as we detail below.

\subsubsection{Parity}

In terms of the spin degrees of freedom \eqref{eq:lattice_H_fixed}, a site-centered parity symmetry at $\theta=\pi$ acts as
\begin{equation}\label{eq:parity}
	\mathcal P \sigma^i_n\mathcal P^\dagger = \sigma^i_{-n}\,,\qquad \mathcal P  L_n  \mathcal P^\dagger = -  L_{-n-1} -1 \,,\qquad \mathcal P U_n\mathcal P^\dagger = U_{-n-1}^\dagger\,.
\end{equation}
Note that this map sends even sites to even sites and odd sites to odd sites. The shift in the electric variable $L_n$ is necessary to ensure that $\mathcal P \left( L_n + \frac 12 \right) \mathcal P^\dagger = - \left( L_{-n-1} + \frac 12 \right)$, which implies that the gauge kinetic term in \eqref{eq:lattice_H} for~$\theta=\pi$ remains invariant. 

To implement parity on a uMPS tensor, we twist its virtual indices in a way that corresponds to mirroring the whole uMPS while keeping the contractions fixed:
\begin{equation}\label{eq:umps_parity}
	\mathcal P:\quad\begin{tikzpicture}[diagram]
		\node[style=square] at (0,0) (A) {$A_n$};
		\draw[arrow mid] (0.35,0) -- (1,0);
		\draw[arrow mid] (-1,0) -- (-0.35,0);
		\draw[arrow mid] (0,-.8) -- (0,-0.35);
		\node at (0,-1) {$n$};
	\end{tikzpicture}\quad\longmapsto\quad
	\begin{tikzpicture}[diagram]
		\node[style=square] at (0,0) (A) {$A_{-n}$};
		\draw[arrow mid] (1,0.) to[out=180, in=0] (0,0.8) to[out=180, in=180] (-0.35,0);
		\draw[white, line width=4pt] (0,0.6) to[out=180, in=0] (-1,0.);
		\draw[arrow mid] (0.35,0) to[out=0, in=0] (0,0.6) to[out=180, in=0] (-1,0.);
		\draw[arrow mid,dashed] (1,0.2) to[out=180, in=0] (0,1) to[out=180, in=0] (-1,0.2);
		\draw[arrow mid] (0,-.8) -- (0,-0.35);
		\node at (0,-1) {$n$};
	\end{tikzpicture}\;.
\end{equation}
We have included arrows on the legs to indicate the flow of charge through the invariant tensor. In the parity operation \eqref{eq:umps_parity}, we reverse the direction of the virtual indices, which indicates that the $\U(1)$ charge assignments on the virtual spaces change sign. The dashed line indicates that we also tensor with an identity map on a one-dimensional space of charge $-1$, which implements the $-1$ shift indicated in the second equation of \eqref{eq:parity}. That is, if we decompose the virtual spaces into charge sectors
\begin{equation}\label{eq:vn_decomp}
    \mathcal{V}_n = \bigoplus_q \mathcal{V}_{n,q},
\end{equation}
then the parity transformation transforms them as
\begin{equation}\label{eq:virtual_P}
	\mathcal P:\quad {\cal V}_{n, q} \longmapsto {\cal V}_{n, q}' = {\cal V}_{-n-1, -q-1}\,.
\end{equation}

\subsubsection{Charge conjugation}
On the lattice, charge conjugation involves translation by one site. It acts on the operators of the spin model as
\begin{equation}\label{eq:CC}
	\mathcal C \sigma^\pm_n\mathcal C^\dagger = \sigma^\mp_{n+1}\,,\qquad\mathcal C \sigma^z_n\mathcal C^\dagger = -\sigma^z_{n+1}\,,\qquad \mathcal C L_n\mathcal C^\dagger = - L_{n+1}-1\,,\qquad \mathcal C U_n\mathcal C^\dagger = U_{n+1}^\dagger\,.
\end{equation}
Note that, unlike $\mathcal{P}$, this map sends even sites to odd sites and odd sites to even sites.

Charge conjugation acts on $\U(1)$ as the outer automorphism $q\mapsto -q$. Applied to a $\U(1)$-graded space $\mathcal V=\bigoplus_q\mathcal V_q$, such as \eqref{eq:vn_decomp}, it produces the conjugate space $\overline{\mathcal V}=\bigoplus_q\overline{\mathcal V}_q$, with $\overline{\mathcal V}_q\equiv\mathcal V_{-q}$ as vector spaces. We introduce the tensor $\mathsf K:\mathcal V\to\overline{\mathcal V}$, which acts as the identity on vectors but regards a vector of charge $q$ in $\mathcal V$ as a vector of charge $-q$ in $\overline{\mathcal V}$:
\begin{equation}\label{eq:C_tensor}
	\begin{tikzpicture}[diagram]
		\node[style=circle] at (0,0) {$\mathsf K$};
		\draw[arrow mid] (1,0) -- (0.35,0);
		\draw[arrow mid] (-1,0) -- (-0.35,0);
	\end{tikzpicture}:\quad \mathcal V_q\ni\ket{v}\longmapsto\ket{v}\in\overline{\mathcal V}_{-q}\,.
\end{equation}
Reversing the arrow on a leg in our diagrams likewise denotes passing from $\mathcal V$ to $\overline{\mathcal V}$. Note that $\mathsf K$ is not an invariant tensor.

Using \eqref{eq:C_tensor}, the action \eqref{eq:CC} can be implemented as
\begin{equation}\label{eq:umps_CC}
	\mathcal C:\quad\begin{tikzpicture}[diagram]
		\node[style=square] at (0,0) (A) {$A_n$};
		\draw[arrow mid] (0.35,0) -- (1,0);
		\draw[arrow mid] (-1,0) -- (-0.35,0);
		\draw[arrow mid] (0,-.8) -- (0,-0.35);
		\node at (0,-1) {$n$};
	\end{tikzpicture}\quad\longmapsto\quad
	\begin{tikzpicture}[diagram]
		\node[style=square] at (0,0) (A) {$A_n$};
		\node[style=circle] at (0,-1.1) (Cs) {$\mathsf K$};
		\draw[arrow mid] (Cs) -- (A);
		\draw[arrow mid] (Cs) -- (0,-1.9);
		\node at(0,-2.1) {$n+1$};
		\node[style=circle] at (1.1,0) (Ce) {$\mathsf K$};
		\draw[arrow mid] (A) -- (Ce);
		\draw[arrow mid] (1.9,0) -- (Ce);
		\node[style=circle] at (-1.2,0) (Cw) {$\mathsf K$};
		\draw[arrow mid] (Cw) -- (A);
		\draw[arrow mid] (Cw) -- (-1.9,0);
		\draw[arrow mid,dashed] (1.9,0.2) to[out=180, in=0] (1.1,0.5) to[out=180, in=0] (-1.1,0.5) to[out=180, in=0] (-1.9,0.2);
	\end{tikzpicture}\;,
\end{equation}
where the dashed line as in \eqref{eq:umps_parity} is a virtual isomorphism that shifts the electric field by $-1$. Note that it is necessary to act with $\mathsf{K}$ on the physical space as well, because in the one-site translation the set of charges that can appear on a site switches between $\{0,1\}$ and $\{0,-1\}$ (see \eqref{eq:Gauss}). 

Charge conjugation changes the virtual spaces of the uMPS tensors, decomposed as in \eqref{eq:vn_decomp}, to
\begin{equation}\label{eq:virtual_C}
	\mathcal C:\quad\mathcal V_{n,q}\longmapsto\mathcal V'_{n,q}=\mathcal V_{n-1,-q-1}\,.
\end{equation}

\subsubsection{$CP$ symmetry}

Finally, we can combine parity and charge conjugation to build the $CP$ symmetry that is preserved in the two symmetry-broken vacua of the Schwinger model. Composing \eqref{eq:parity} and \eqref{eq:CC}, it acts on the operators of the spin model as
\begin{equation}\label{eq:CP_ops}
\begin{aligned}
	(\mathcal C\mathcal P) \sigma^\pm_n(\mathcal C\mathcal P)^\dagger &= \sigma^\mp_{1-n}\,,&
	(\mathcal C\mathcal P) \sigma^z_n(\mathcal C\mathcal P)^\dagger &= -\sigma^z_{1-n}\,,\\
	(\mathcal C\mathcal P) L_n(\mathcal C\mathcal P)^\dagger &= L_{-n}\,,&
	(\mathcal C\mathcal P) U_n(\mathcal C\mathcal P)^\dagger &= U_{-n}\,.
\end{aligned}
\end{equation}
Unlike $\mathcal P$ and $\mathcal C$ separately, $\mathcal C\mathcal P$ involves no shift of $L_n$: the electric field $L_n+\frac12$ on each link is mapped to the electric field on the mirror-image link, consistent with each vacuum $\ket{v_\pm}$ being invariant.

This symmetry acts on the uMPS tensors by
\begin{equation}\label{eq:CP_sym}
	\mathcal C\mathcal P:\quad\begin{tikzpicture}[diagram]
		\node[style=square] at (0,0) (A) {$A_n$};
		\draw[arrow mid] (0.35,0) -- (1,0);
		\draw[arrow mid] (-1,0) -- (-0.35,0);
		\draw[arrow mid] (0,-.8) -- (0,-0.35);
		\node at (0,-1) {$n$};
	\end{tikzpicture}\quad\longmapsto\quad
	\begin{tikzpicture}[diagram]
		\node[style=square] at (0,0) (A) {$A_n$};
		\node[style=circle,opacity=0] at (-1.2,0) (Cw) {$\mathsf K$};
		\node[style=circle,opacity=0] at (1.1,0) (Ce) {$\mathsf K$};
		\draw[arrow mid] (Ce.west) to[out=180, in=0] (0,0.8) to[out=180, in=180] (-0.35,0);
		\draw[white, line width=4pt] (0,0.6) to[out=180, in=0] (Cw.east);
		\draw[arrow mid] (0.35,0) to[out=0, in=0] (0,0.6) to[out=180, in=0] (Cw.east);
		\node[style=circle] at (0,-1.1) (Cs) {$\mathsf K$};
		\draw[arrow mid] (Cs) -- (0,-1.9);
		\node at(0,-2.1) {$1-n$};
		\node[style=circle] at (1.1,0) (Ce) {$\mathsf K$};
		\draw[arrow mid] (Ce) -- (1.9,0);
		\node[style=circle] at (-1.2,0) (Cw) {$\mathsf K$};
		\draw[arrow mid] (Cs) -- (A);
		\draw[arrow mid] (-1.9,0) -- (Cw) ;
	\end{tikzpicture}\;.
\end{equation}
By combining \eqref{eq:virtual_C} with \eqref{eq:virtual_P}, we see that the virtual spaces transform as
\begin{equation}
	\mathcal C\mathcal P:\quad\mathcal V_{n,q}\longmapsto \mathcal V_{n,q}' = {\mathcal V}_{-n,q}\,.
\end{equation}

\subsection{$CP$-symmetric states}\label{app:CP_states}
When $\theta = \pi$ and $m>m_c$, charge conjugation symmetry is spontaneously broken and we have two degenerate vacua. As explained in Section~\ref{sec:method_qp}, we can fix energy and momentum ambiguities for soliton states by imposing symmetries. We will now explain the consequences of $CP$ invariance.

Each of the two degenerate vacua is left invariant by the action of $CP$\@. Thus, if we act with $CP$ on the uMPS for one of these vacua, we must recover the same state. This means that the uMPS tensors themselves must change only by a uMPS gauge transformation \cite{Pollmann_2010}: 
\begin{equation}\label{eq:CP_umps}
	\begin{tikzpicture}[diagram]
		\node[style=square] at (0,0) (A) {$A_n$};
		\node[style=circle,opacity=0] at (-1.2,0) (Cw) {$\mathsf K$};
		\node[style=circle,opacity=0] at (1.1,0) (Ce) {$\mathsf K$};
		\draw[arrow mid] (Ce.west) to[out=180, in=0] (0,0.8) to[out=180, in=180] (-0.35,0);
		\draw[white, line width=4pt] (0,0.6) to[out=180, in=0] (Cw.east);
		\draw[arrow mid] (0.35,0) to[out=0, in=0] (0,0.6) to[out=180, in=0] (Cw.east);
		\node[style=circle] at (0,-1.1) (Cs) {$\mathsf K$};
		\draw[arrow mid] (Cs) -- (0,-1.9);
		\node at(0,-2.1) {$1-n$};
		\node[style=circle] at (1.1,0) (Ce) {$\mathsf K$};
		\draw[arrow mid] (Ce) -- (1.9,0);
		\node[style=circle] at (-1.2,0) (Cw) {$\mathsf K$};
		\draw[arrow mid] (Cs) -- (A);
		\draw[arrow mid] (-1.9,0) -- (Cw) ;
	\end{tikzpicture}\quad =\quad e^{i\gamma}\;\begin{tikzpicture}[diagram]
	\node[style=square] at (0,0) (A) {$A_{1-n}$};
	\node[style=circle] at (-1.1,0) (Xi) {$X_{-n}^{-1}$};
	\node[style=circle] at (1.1,0) (X) {$X_{1-n}$};
	\draw[arrow mid] (Xi) -- (A);
	\draw[arrow mid] (A) -- (X);
	\draw[arrow mid] (X) -- (1.8,0);
	\draw[arrow mid] (-1.8,0) -- (Xi);
	\draw[arrow mid] (0,-.8) -- (0,-0.35);
	\node at (0,-1) {$1-n$};
	\end{tikzpicture}\;.
\end{equation}
Here, $X_n$ is an isomorphism on $\mathcal{V}_n$. Since $(\mathcal C\mathcal{P})^2=\mathds{1}$, $X_n$ must satisfy the consistency condition
\begin{equation}\label{eq:CP_X_consistency}
	\begin{tikzpicture}[diagram,baseline=-.5ex]
		\node[style=circle,opacity=0] at (-1.2,0) (Cw) {$\mathsf K$};
		\node[style=circle,opacity=0] at (1.2,0) (Ce) {$\mathsf K$};
		\draw[arrow mid] (Ce.west) to[out=180,in=0] (0,.8)
		to[out=180,in=180] (-.35,0);
		\draw[white,line width=4pt] (0,.6)
		to[out=180,in=0] (Cw.east);
		\draw[arrow mid] (.35,0) to[out=0,in=0] (0,.6)
		to[out=180,in=0] (Cw.east);
		\node[style=circle] at (0,0) (X) {$X_n$};
		\node[style=circle] at (-1.2,0) (Cw) {$\mathsf K$};
		\node[style=circle] at (1.2,0) (Ce) {$\mathsf K$};
		\draw[arrow mid] (-2,0) -- (Cw.west);
		\draw[arrow mid] (Ce.east) -- (2,0);
	\end{tikzpicture}
	\quad =\quad 
	e^{i\phi}\;\begin{tikzpicture}[diagram,baseline=-.5ex]
		\node[style=circle] at (0,0) (X) {$X_n$};
		\draw[arrow mid] (-.9,0) -- (X.west);
		\draw[arrow mid] (X.east) -- (.9,0);
	\end{tikzpicture}
	\,.
\end{equation}
The action of $CP$ on an MPO tensor is similar to \eqref{eq:CP_sym}; invariance under $CP$ means that this transformation only alters the tensor by an MPO gauge transformation, given by some isomorphisms\footnote{For the Schwinger model MPO \eqref{eq:schwinger_MPO}, the unique solution to \eqref{eq:mpo_sym} is
\begin{equation*}
	Y_n=\begin{pmatrix}
		0&0&0&1\\
		0&\frac{1}{2a}&0&0\\
		0&0&\frac{1}{2a}&0\\
		1&0&0&0
	\end{pmatrix}\,.
\end{equation*}} $Y_n$:
\begin{equation}\label{eq:mpo_sym}
\begin{tikzpicture}[diagram,baseline=(W)]
	\node[style=circle] at (0,0) (W) {$W_n$};
	\node[style=circle] at (0,1.) (Cn) {$\mathsf K$};
	\node[style=circle] at (0,-1.) (Cs) {$\mathsf K$};
	\node[style=circle] at (1.5,0) (Ce) {$\mathsf K$};
	\node[style=circle] at (-1.5,0) (Cw) {$\mathsf K$};
	\draw[arrow mid] (W) to[out=0,in=0] (0,0.5) to[out=180,in=0] (Cw);
	\draw[arrow mid] (Ce) to[out=180,in=0] (0,-0.5) to[out=180,in=180] (W);
	\draw[white, line width=4pt] (0,0.5) -- (0,.5) (0,-0.5) -- (0,-.5);
	\draw[arrow mid] (W) -- (Cn);
	\draw[arrow mid] (Cs) -- (W);
	\draw[arrow mid] (Ce) -- (2.2,0);
	\draw[arrow mid] (-2.2,0) -- (Cw);
	\draw[arrow mid] (0,1.7) -- (Cn);
	\draw[arrow mid] (Cs) -- (0,-1.8);
	\node at (0,-2) {$1-n$};
\end{tikzpicture}\quad=\quad
\begin{tikzpicture}[diagram,baseline=(W)]
	\node[style=circle] at (0,0) (W) {$W_{1-n}$};
	\node[style=circle] at (1,0) (Y) {$Y_{1-n}$};
	\node[style=circle] at (-1,0) (Yi) {$Y^{-1}_{-n}$};
	\draw[arrow mid] (W) -- (0,-.8);
	\draw[arrow mid] (Yi) -- (W);
	\draw[arrow mid] (W) -- (Y);
	\draw[arrow mid] (0,1) -- (W);
	\draw[arrow mid] (Y) -- (1.7,0);
	\draw[arrow mid] (-1.7,0) -- (Yi);
	\node at (0,-1) {$1-n$};
\end{tikzpicture}\,.
\end{equation}
Importantly, the map $Y_n$ interchanges $\delta_i$ and $\delta_f$:
\begin{equation}\label{eq:CP_MPO_boundaries}
	\begin{tikzpicture}[diagram,baseline=-.5ex]
		\node[style=circle] at (0,0) (Y) {$Y_n$};
		\node[style=circle] at (1.2,0) (df) {$\delta_f$};
		\draw[arrow mid] (-.7,0) -- (Y);
		\draw[arrow mid] (Y) -- (df);
	\end{tikzpicture}
	\quad = \quad
	\begin{tikzpicture}[diagram,baseline=-.5ex]
		\node[style=circle] at (0,0) (di) {$\delta_i$};
		\draw[arrow mid] (-.7,0) -- (di);
	\end{tikzpicture}
	\,,\qquad
	\begin{tikzpicture}[diagram,baseline=-.5ex]
		\node[style=circle] at (0,0) (df) {$\delta_f$};
		\node[style=circle] at (1.2,0) (Y) {$Y_n$};
		\draw[arrow mid] (df) -- (Y);
		\draw[arrow mid] (Y) -- (1.9,0);
	\end{tikzpicture}
	\quad = \quad
	\begin{tikzpicture}[diagram,baseline=-.5ex]
		\node[style=circle] at (0,0) (di) {$\delta_i$};
		\draw[arrow mid] (di) -- (.7,0);
	\end{tikzpicture}\;.
\end{equation}
The link operator $\mathcal{O}_L \equiv \left(\mathcal L_n+\frac12\right)^2$ is diagonal in the charge sectors, and $CP$ maps $\mathcal V_{n,q}$ to $\mathcal V_{-n,q}$ without changing the charge, so the link operator commutes with $CP$:
\begin{equation}
	\begin{tikzpicture}[diagram,baseline=(L)]
		\node[style=circle] at (0,0) (L) {$\mathcal{O}_L$};
		\node[style=circle] at (1.5,0) (Ce) {$\mathsf K$};
		\node[style=circle] at (-1.5,0) (Cw) {$\mathsf K$};
		\draw[arrow mid] (L) to[out=0,in=0] (0,.7) to[out=180, in=0] (Cw);
		\draw[white, line width=4pt] (1.1,0) to[out=180,in=0] (0,.5);
		\node[style=circle] at (1.5,0) (Ce) {$\mathsf K$};
		\draw[arrow mid] (Ce) to[out=180,in=0] (0,.5) to[out=180, in=180] (L);
		\draw[arrow mid] (-2.3,0) -- (Cw);
		\draw[arrow mid] (Ce) -- (2.3,0);
	\end{tikzpicture}\quad=\quad
	\begin{tikzpicture}[diagram,baseline=(L)]
		\node[style=circle] at (0,0) (L) {$\mathcal{O}_L$};
		\draw[arrow mid] (L) -- (1.,0);
		\draw[arrow mid] (-1,0) -- (L);
	\end{tikzpicture}\,.
\end{equation}
Furthermore, the properties \eqref{eq:CP_MPO_boundaries} and the structure of \eqref{eq:schwinger_link_MPO} guarantee that the corresponding link MPO will satisfy
\begin{equation}
	\begin{tikzpicture}[diagram,baseline=(L)]
		\node[style=circle,opacity=0] at (0,0) (L) {$W^{(L)}_n$};
		\node[style=circle,opacity=0] at (1.5,0) (Ce) {$\mathsf K$};
		\node[style=circle,opacity=0] at (-1.5,0) (Cw) {$\mathsf K$};
		\node[style=circle,opacity=0] at (1.5,-1.) (Ce1) {$\mathsf K$};
		\node[style=circle,opacity=0] at (-1.5,-1.) (Cw1) {$\mathsf K$};
		\draw[arrow mid] (L) to[out=0,in=0] (0,.7) to[out=180, in=0] (Cw);
		\draw[white, line width=4pt] (1.1,0) to[out=180,in=0] (0,.5);
		\draw[arrow mid] (Ce) to[out=180,in=0] (0,.5) to[out=180, in=180] (L);
		\draw[arrow mid] (-2.3,0) -- (Cw);
		\draw[arrow mid] (-2.3,-1) -- (Cw1);
		\draw[arrow mid] (Ce) -- (2.3,0);
		\draw[arrow mid] (Ce1) -- (2.3,-1);
		\draw[arrow mid] (Ce1) to[out=180, in=-135] (L);
		\draw[white, line width=4pt] (L) to[out=-45, in=0] (Cw1);
		\draw[arrow mid] (L) to[out=-45, in=0] (Cw1);
		\node[style=circle] at (0,0) (L) {$W^{(L)}_n$};
		\node[style=circle] at (1.5,0) (Ce) {$\mathsf K$};
		\node[style=circle] at (-1.5,0) (Cw) {$\mathsf K$};
		\node[style=circle] at (1.5,-1.) (Ce1) {$\mathsf K$};
		\node[style=circle] at (-1.5,-1.) (Cw1) {$\mathsf K$};
	\end{tikzpicture}\quad = \quad
\begin{tikzpicture}[diagram,baseline=(L)]
    \node[style=circle] at (0,0) (L) {$W^L_{1-n}$};
    \node[style=circle] at (-1.5,-1) (Yi) {$Y^{-1}_{1-n}$};
    \node[style=circle] at (1.5,-1) (Y) {$Y_{1-n}$};

    \draw[arrow mid] (-2.3,0) -- (L);
    \draw[arrow mid] (L) -- (2.3,0);

    \draw[arrow mid] (-2.3,-1) -- (Yi);
    \draw[arrow mid] (Yi) -- (L);
    \draw[arrow mid] (L) -- (Y);
    \draw[arrow mid] (Y) -- (2.3,-1);
\end{tikzpicture}\,.
\end{equation}

Finally, we can define $\mathbb X = X_2\otimes Y_2\otimes \bar X_2$, which acts on the full virtual space of the LEMPO transfer matrix \eqref{eq:lempo_unit_cell_transfer_matrix}. The matrix $\mathbb X$ is symmetric because the phase in \eqref{eq:CP_X_consistency} cancels between the factors of $X$ and $\bar X$. The action of $CP$ on the transfer matrix sends $\mathbb T\mapsto\mathbb C\mathbb T^T\mathbb C$, with the transpose implementing parity, and the $CP$-invariance of the vacuum implies
\begin{equation}\label{eq:CP_T}
    \mathbb C \mathbb T^T \mathbb{C} = \mathbb X^{-1} \mathbb T \mathbb X\,,
\end{equation}
where $\mathbb C$ applies $\mathsf K$ to each leg of the transfer-matrix virtual space.

Since $\mathbb{A}_L=\delta_i\otimes l$, $\mathbb{A}_R=\delta_f\otimes r$, and $\mathbb{X}$ factorizes, it follows that
\begin{equation}\label{eq:XCA}
    \mathbb{X}\mathbb{C}\mathbb{A}_L=\mathbb{A}_R\,.
\end{equation}
Indeed, $Y_2$ maps $\delta_i$ to $\delta_f$ by \eqref{eq:CP_MPO_boundaries}, and by \eqref{eq:CP_umps} the MPS part of $\mathbb{X}\mathbb{C}$ maps $l$ to a right fixed point of the MPS transfer matrix, which is proportional to $r$ because this fixed point is unique for an injective uMPS\@. We fix the overall scale of $X_2$, which \eqref{eq:CP_umps} leaves free, so that the proportionality constant is one. Neither step involves the energy density, so \eqref{eq:XCA} holds for $H$ and $H_{\epsilon=0}$ alike.

We can now fix the energy ambiguity discussed in Section~\ref{sec:method_qp}. We may choose the boundary vectors of each uMPS to be $CP$-images of each other,
\begin{equation}\label{eq:CP_BC}
    \mathbb{X}\mathbb{C} \mathbbm{v}_L= \mathbbm v_R\,;
\end{equation}
correspondingly, we fix the remaining freedom from \eqref{eq:etas} (the part not restricted by \eqref{eq:EE0}) by requiring
\begin{equation}\label{eq:CP_env}
    \mathbb{X}\mathbb{C} \mathbb{E}_L= \mathbb{E}_R\,.
\end{equation}
Applying $\mathbb X\mathbb C$ to the decomposition \eqref{eq:vDecomp} of $\mathbbm v_L$ and using \eqref{eq:XCA}, \eqref{eq:CP_BC}, and \eqref{eq:CP_env}, we find that the coefficient of $\mathbb A_R$ in $\mathbbm v_R$ is $\alpha_L$, i.e., $\alpha_L=\alpha_R$. The same argument for $\{\tilde A\}$ gives $\tilde\alpha_L=\tilde\alpha_R$.

\section{Electric field measurements}\label{app:avg_e_field}

The final state $\ket{\psi}$ of our collisions is a superposition of states for which the middle region looks locally like $\ket{v_+}$ (i.e., has $E_\text{avg} > 0$) and states for which the middle region looks locally like $\ket{v_-}$ (i.e. $E_\text{avg} < 0$). As we argued around \eqref{eq:psifinal}, in our parameter regime the former mostly correspond to elastic $2\to 2$ scattering while the latter mostly correspond to $2\to 4$ events in which the string breaks.

At $\theta = \pi$ we have $E_\text{avg} = \frac{1}{\ell}\sum L_k + \frac12$, so $E_\text{avg} > 0$ corresponds to $\sum L_k > -\ell/2$; we take $\ell$ odd so that $E_\text{avg}\neq 0$. Thus, we can approximate
\begin{equation}\label{eq:proxy_exp_val}
    \langle \mathcal{O}\rangle_\pm = \frac{\sum_{\pm(n+\ell/2)>0} \langle \mathcal{O}\rangle_{\sum L_k = n}}{\sum_{\pm(n+\ell/2)>0} \langle \mathds{1}\rangle_{\sum L_k = n}}, \qquad \text{with} \qquad\langle\mathcal{O}\rangle_{\sum L_k = n} \equiv \Braket{\psi|P_{\sum L_k = n}\mathcal{O} P_{\sum L_k = n}|\psi}.
\end{equation}
Here, $\sum L_k$ denotes the sum of the electric field operators over some set of links, and $P_{\sum L_k = n}$ projects onto states for which that sum equals $n$. The quantities $\langle \mathds{1}\rangle_{\sum L_k = n}$ are the probabilities that $E_\text{avg} = \frac{n}{\ell} + \frac12$ (which we plotted in Figures~\ref{fig:vacuum_histogram} and \ref{fig:electric_histogram_collision}). 
 
If $\sum L_k$ ranges only over a small handful of links, then we can carry out \eqref{eq:proxy_exp_val} explicitly. We would enumerate all the ways that $\sum L_k$ can give $n$, and then insert link operators projecting onto each of these possibilities. Indeed, let us define a link operator $\delta_{L_n,q}$ that acts as 1 on the sector with $\U(1)$ charge $q$ and 0 on all other sectors. Then we would have, for instance,
\begin{equation}\label{eq:link_proj_example}
    \Braket{\psi|P_{L_1 + L_2 = 1}\mathcal{O}_n P_{L_1 + L_2 = 1}|\psi} = \sum_j \begin{tikzpicture}[diagram,yscale=1.1]
			\node[style=square] at (0,0) (A1) {$A_0$};
			\node[style=square] at (0,-1.5) (B1) {$\bar{A}_0$};
			\node[style=square] at (1.5,0) (A2) {$A_1$};
			\node[style=square] at (1.5,-1.5) (B2) {$\bar{A}_1$};
			\node[style=circle] at (2.5,0) (C2) {$\delta_{L_1,j}$};
			\node[style=circle] at (2.5,-1.5) (D2) {$\delta_{L_1,j}$};
			\node[style=square] at (3.5,0) (A3) {$A_2$};
			\node[style=square] at (3.5,-1.5) (B3) {$\bar{A}_2$};
			\node[style=circle] at (4.5,0) (C3) {$\delta_{L_2,1-j}$};
			\node[style=circle] at (4.5,-1.5) (D3) {$\delta_{L_2,1-j}$};
			\node[style=square] at (5.5,0) (A4) {$A_3$};
			\node[style=square] at (5.5,-1.5) (B4) {$\bar{A}_3$};
			\node[style=square] at (7.5,0) (A5) {$A_n$};
			\node[style=square] at (7.5,-1.5) (B5) {$\bar{A}_n$};
			\node[style=circle] at (7.5,-.75) (E5) {$\mathcal{O}_n$};
			\node at (-1.,0) {$\cdots$};
			\node at (-1.,-1.5) {$\cdots$};
			\node at (6.5,0) {$\cdots$};		
			\node at (6.5,-1.5) {$\cdots$};		
			\node at (8.5,0) {$\cdots$};		
			\node at (8.5,-1.5) {$\cdots$};		
			\draw (A1) -- (A2) -- (C2) -- (A3) -- (C3) -- (A4);
			\draw (B1) -- (B2) -- (D2) -- (B3) -- (D3) -- (B4);
			\draw (A1) -- (B1);
			\draw (A2) -- (B2);
			\draw (A3) -- (B3);
			\draw (A4) -- (B4);
            \draw (A5) -- (E5) -- (B5);
			\draw (A1) -- ++(-0.75,0);
			\draw (A4) -- ++(0.75,0);
			\draw (A5) -- ++(-0.75,0);
			\draw (A5) -- ++(0.75,0);
			\draw (B1) -- ++(-0.75,0);
			\draw (B4) -- ++(0.75,0);
			\draw (B5) -- ++(-0.75,0);
			\draw (B5) -- ++(0.75,0);
		\end{tikzpicture}.
\end{equation}
The sum truncates quickly because, in practice, each MPS tensor is supported only on charge sectors $|L_n| \lesssim 5$.
 
If we want to measure the electric field summed over a larger number of links, this approach quickly becomes prohibitive, as the number of electric field configurations with a fixed average / sum grows exponentially. We can circumvent this difficulty by measuring a different expectation value, namely
\begin{equation}
\begin{split}
    \langle \mathcal{O}; \varphi\rangle &\equiv \Braket{\psi|\exp\left(i \varphi \sum L_k\right) \mathcal{O}|\psi} \\
    &= \sum_n e^{in\varphi} \langle\mathcal{O}\rangle_{\sum L_k = n},
\end{split}
\end{equation}
where the second line holds because the operators $\mathcal{O}$ we consider commute with $\sum L_k$. To evaluate $\langle \mathcal{O};\varphi\rangle$, we simply insert link operators $e^{i\varphi L_n}$ on each link we are measuring, in a contraction similar to the one in \eqref{eq:link_proj_example}. Crucially, whereas in \eqref{eq:link_proj_example} we would need to evaluate exponentially many such contractions to arrive at $\langle\mathcal{O}\rangle_{\sum L_k = n}$, by using $\langle \mathcal{O};\varphi\rangle$ we only need to evaluate it at sufficiently many values of $\varphi$ in order to read off the Fourier decomposition. We only need to evaluate as many samples as there are possible values of $\sum L_k$, and this number in turn scales linearly in the number of links we average over.

In Section~\ref{sec:results} we apply this method to ${\cal O} = L_n$, where $L_n$ is an electric field operator outside the measurement region.

\section{Simulation details}\label{app:simulation}

To perform the real-time simulations in Section~\ref{sec:results}, we first need to prepare the initial states. First, we find uMPSs for the degenerate vacuum states; we denote the uMPS tensors for the vacuum with negative electric field by $\{A_-\}$, and the tensors for the other vacuum by\footnote{The braces indicate that, because we have a two-site unit cell, we actually have a set of two tensors in each case.} $\{A_+\}$. We then fix a collision momentum $p$ and use the quasiparticle ansatz to find quasiparticle tensors $\{B_+(p)\}$ and $\{B_-(-p)\}$ that make \eqref{eq:Bsoliton} into positron and electron eigenstates, using the reflection-symmetric gauge discussed in Section~\ref{sec:method_wavepacket}. Finally, we fix a wavepacket width $\sigma$, a cutoff width $\ell_\text{cut}$, and an initial separation $|x_- - x_+|$, and build a two-soliton wavepacket state of the form \eqref{eq:two_wavepacket}.

The initial window MPS (see Section~\ref{sec:results_tdvp}) is then given exactly by
\begin{equation}\label{eq:automaton}
\begin{aligned}
    &\ket{\{B_+(+p)\}, \{B_-(-p)\}, x_+, x_-, \sigma}\\
    &\quad= \cdots A_-\,A_-\begin{pmatrix} \mathds{1} & 0 \end{pmatrix}
    M^+_{n^+_\text{min}}\cdots M^+_{n^+_\text{max}}
    \begin{pmatrix} 0 \\ \mathds{1} \end{pmatrix} A_+\cdots\\
    &\qquad\cdots A_+\begin{pmatrix} \mathds{1} & 0 \end{pmatrix}
    M^-_{n^-_\text{min}}\cdots M^-_{n^-_\text{max}}
    \begin{pmatrix} 0 \\ \mathds{1} \end{pmatrix} A_-\,A_-\cdots\,,
\end{aligned}
\end{equation}
where $\{n^\pm_\text{min},\ldots,n^\pm_\text{max}\}$ are the unit cells $n$ with $|2na - x_\pm| < \ell_\text{cut}$, and
\begin{equation}
    M^+_n = \begin{pmatrix} A_- & f_+(n)\, B_+(+p) \\ 0 & A_+ \end{pmatrix}\,,\qquad
    M^-_n = \begin{pmatrix} A_+ & f_-(n)\, B_-(-p) \\ 0 & A_- \end{pmatrix}\,,
\end{equation}
with $f_\pm(n) = e^{\pm ip(2na)}\, e^{-(2na - x_\pm)^2/2\sigma^2}$. Here each of $A_\pm$, $B_\pm$, and $M^\pm_n$ stands for the tensor of an entire two-site unit cell, with physical legs suppressed. We take $\ell_\text{cut} = 2\sigma$.

For the collisions in Figure~\ref{fig:w_vs_sqrts} that have $m/g\geq 0.5$ (which include the collisions in Figures~\ref{fig:inelastic_examples} and \ref{fig:electric_histogram_collision}), we use a lattice spacing of $ag = 0.1$, a wavepacket width of $\sigma g = 2.5$, and a virtual space dimension cutoff of $D_\text{max} = 192$. For every collision in Figure~\ref{fig:w_vs_sqrts}, we have checked that the $w_+$ value changes by less than 0.01 when we decrease the lattice spacing from $ag = 0.15$ to $ag=0.1$ while keeping the volume of the simulation window fixed. We have also verified for the highest-energy collisions (which are most susceptible to numerical difficulties) that the results are insensitive to increasing $\sigma g$ or $D_\text{max}$; see Table~\ref{tab:convergence} for an example. For the collisions at $m/g = 0.4$, the story is similar, but we use $D_\text{max} = 384$ and $\sigma g = 5$ because of the larger vacuum correlation length; convergence is demonstrated in Table~\ref{tab:convergence_mg04}.

\setlength{\tabcolsep}{1cm}
\begin{table}
\centering
\begin{tabular}{ccccc}
\toprule
$D_\text{max}$ & $ag$ & $\sigma g$ & $w_+$ & $\Delta w_+$ \\
\midrule
192 & 0.15 & 2.5 & 0.5563 & ---        \\
\textbf{216} & 0.15 & 2.5 & 0.5560 & $-0.0003$  \\
192 & \textbf{0.10} & 2.5 & 0.5478 & $-0.0085$  \\
192 & 0.15 & \textbf{5.0} & 0.5496 & $-0.0067$  \\
\bottomrule
\end{tabular}
\caption{Convergence of $w_+$ for the $m/g=0.7$, $p/M_\text{sol}=6$ soliton collision. The first row is the baseline for this test, and each subsequent row improves a single parameter.}
\label{tab:convergence}
\end{table}

\setlength{\tabcolsep}{1cm}
\begin{table}
\centering
\begin{tabular}{ccccc}
\toprule
$D_\text{max}$ & $ag$ & $\sigma g$ & $w_+$ & $\Delta w_+$ \\
\midrule
384 & 0.15 & 5.0 & 0.9218 & ---        \\
\textbf{512} & 0.15 & 5.0 & 0.9217 & $-0.0001$  \\
384 & \textbf{0.10} & 5.0 & 0.9189 & $-0.0029$  \\
384 & 0.15 & \textbf{7.5} & 0.9260 & $+0.0042$  \\
\bottomrule
\end{tabular}
\caption{Convergence of $w_+$ for the $m/g=0.4$, $p/M_\text{sol}=10$ soliton collision. The first row is the baseline for this test, and each subsequent row improves a single parameter.}
\label{tab:convergence_mg04}
\end{table}

\section{Two-soliton projections}\label{app:twoparticle}

In Section~\ref{sec:results_inelastic}, we estimate the $2\rightarrow 2$ probability by projecting initial and final states onto the two-soliton basis \eqref{eq:two_soliton_basis}. On a simulation window of $N=1200$ sites (see Section~\ref{sec:results_tdvp}), we take $W_L = \{1,\ldots,\frac{N-n_\mathrm{sep}}{2}\}$ and $W_R = \{\frac{N+n_\mathrm{sep}}{2}+1, \ldots, N\}$ with $n_\mathrm{sep}=40$, and build the basis states using the automaton construction of Appendix~\ref{app:simulation}~\cite{Milsted:2020jmf}. We evaluate the integral \eqref{eq:2solprob} numerically over $-2p \leq p_1, p_2 \leq 2p$, where $p$ is the collision momentum, on a uniform grid whose spacing is at most the momentum resolution $2\pi/(a|W_{L,R}|)$ in a square of length $p$ around the maximum, and four times as large elsewhere. We estimated the errors on the probabilities in Figure~\ref{fig:probability_comparison} by varying $n_\mathrm{sep}=20,\ldots,80$ as well as by comparing our results with an integral over a coarser grid. The resulting error bars are small, and decreasing with increasing $p/M_\mathrm{sol}$. As a check, $w_2$ for the incoming state is between $0.992$ and $1$ for all collision momenta in Figure~\ref{fig:probability_comparison} (and above $0.999$ for $p \geq 3 M_{\rm sol}$).

\begin{figure}
    \centering
    \includegraphics[width=\linewidth]{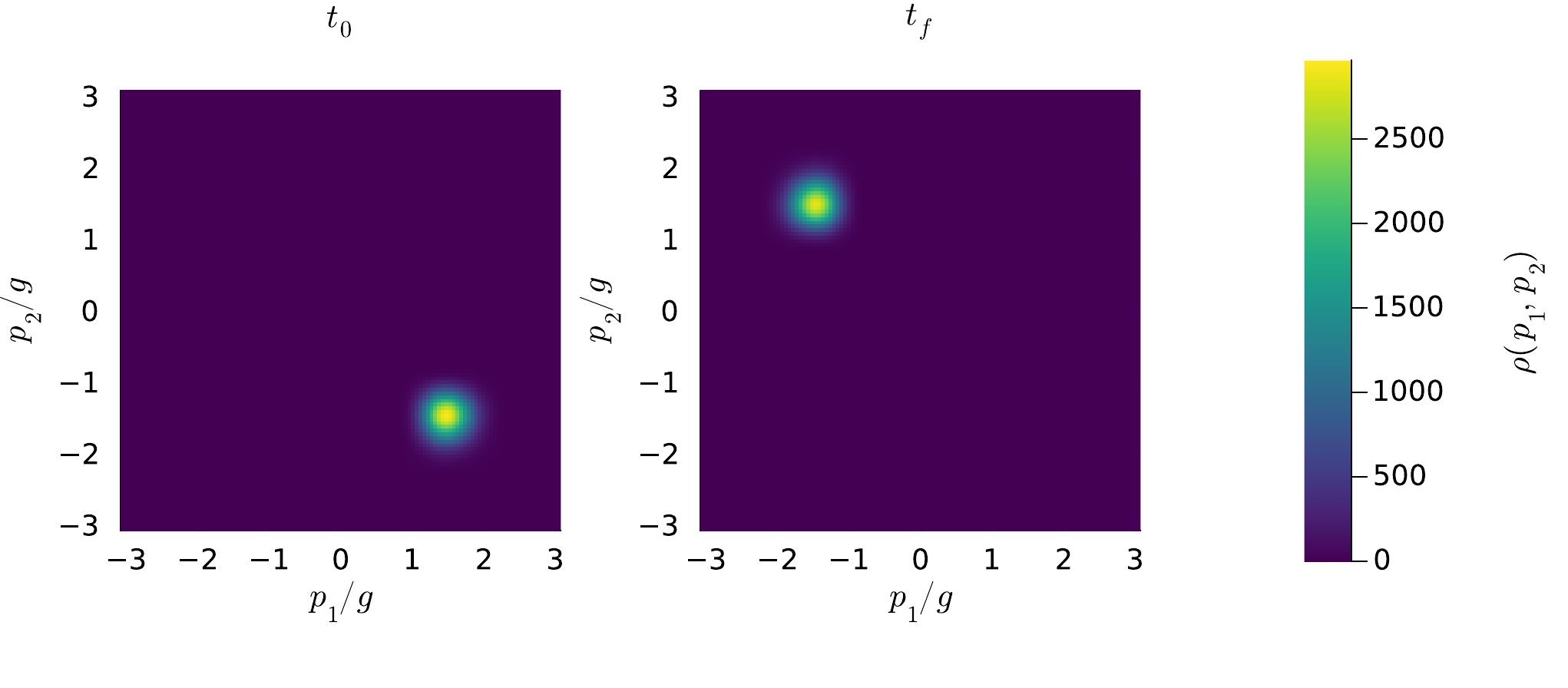}
    \caption{Two-soliton probability density $\rho(p_1, p_2)$ for the pre-collision (left, $t_0$) and post-collision (right, $t_f$) states of an $e^+e^-$ collision at $m/g=0.7$, $p/M_\mathrm{sol}=3$.}
    \label{fig:Prob_p3}
\end{figure}

\begin{figure}
    \centering
    \includegraphics[width=\linewidth]{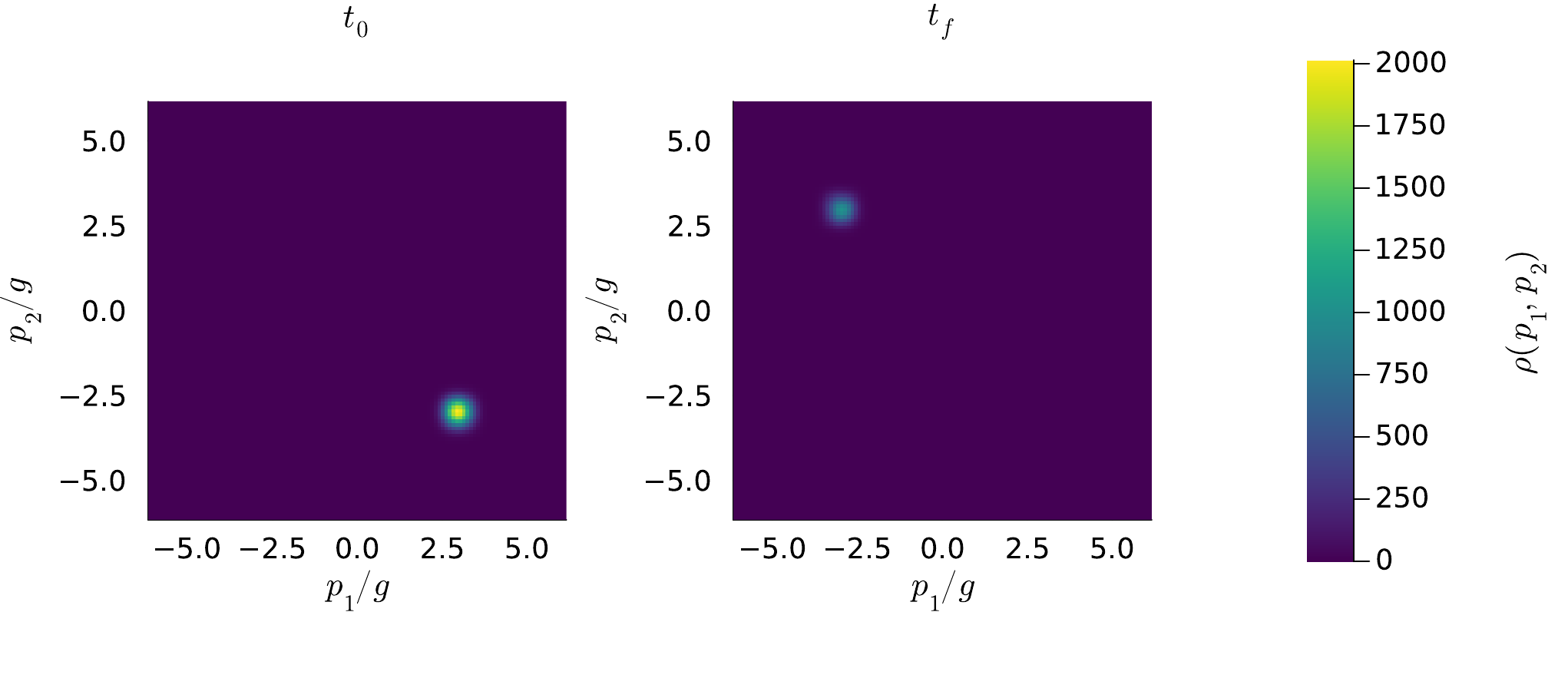}
    \caption{Two-soliton probability density $\rho(p_1, p_2)$ for the pre-collision (left, $t_0$) and post-collision (right, $t_f$) states of an $e^+e^-$ collision at $m/g=0.7$, $p/M_\mathrm{sol}=6$.}
    \label{fig:Prob_p6}
\end{figure}

In Figures~\ref{fig:Prob_p3} and \ref{fig:Prob_p6}, we plot $\rho(p_1, p_2)$ before and after scattering for the first and last collisions of Figure~\ref{fig:inelastic_examples} ($m/g = 0.7$). In both cases, the incoming state is a Gaussian around $(p_1,p_2) = (+p, -p)$ and the final state is localized around $(-p, +p)$. For $p/M_\mathrm{sol}=3$, where particle production is unlikely, the final-state density integrates to nearly 1; for $p/M_\text{sol} = 6$, its integral is substantially reduced because the state has significant overlap with sectors of higher soliton number.

\section{Time delay}\label{app:timedelay}

For elastic scattering, $S_{2\to2}$ is just a complex momentum-dependent phase $e^{i2\delta(p)}$. In the center-of-mass frame, which we always use in our collisions, the time delay is defined by $\Delta t = 2 d\delta/d\sqrt{s}$ \cite{Wigner:1955zz}. We cannot read off $\delta$ directly from our simulations, because the TDVP algorithm with window MPS states (see Section~\ref{sec:results_tdvp}) does not preserve the global phase of the evolved state. Furthermore, each quasiparticle ansatz \eqref{eq:Bsoliton} for $e^+$ and $e^-$ at different momenta carries an arbitrary phase. We therefore adapt the procedure of \cite{Jha:2024jan}, which compares projections of the incoming and outgoing states onto two-soliton states \eqref{eq:two_soliton_basis} at nearby momenta $p_\pm \equiv p \pm \Delta p$.
 
In $1+1$ dimensions, elastic scattering of equal-mass particles can only exchange their momenta. At a late time $t$, after the outgoing excitations have separated, we therefore have
\begin{equation}\label{eq:Sphase}
    \braket{-p, p|\psi(t)} = e^{i\chi(t)}\, e^{2i\delta(p)}\, e^{-2i\mathcal{E}_\text{sol}(p) t}\,\braket{p, -p|\psi(0)}\,,
\end{equation}
where $\chi(t)$ is the unknown global phase and $\mathcal{E}_\text{sol}(p)$ is the lattice soliton dispersion relation (see Figure~\ref{fig:dispersion}). We can then determine the time delay by first calculating the double ratio
\begin{equation}\label{eq:phi}
\begin{split}
     \phi(t) &= \arg \left(\frac{\braket{-p_+, p_+|\psi(t)}}{\braket{p_+, -p_+|\psi(0)}}\, \frac{\braket{p_-, -p_-|\psi(0)}}{\braket{-p_-, p_-|\psi(t)}}\right)\\
     &= 2\left(\delta(p_+)-\delta(p_-)\right) - 2\left(\mathcal{E}_\text{sol}(p_+)-\mathcal{E}_\text{sol}(p_-)\right)\,t\,.
\end{split}
\end{equation}
Then, since $\sqrt{s} = 2\mathcal{E}_\text{sol}(p)$, we can approximate the time delay as
\begin{equation}
    \Delta t = \frac{2(\delta(p_+)-\delta(p_-))}{2(\mathcal{E}_\text{sol}(p_+)-\mathcal{E}_\text{sol}(p_-))} = t + \frac{\phi(t)}{2
    (\mathcal{E}_\text{sol}(p_+)-\mathcal{E}_\text{sol}(p_-))}\,.
\end{equation}
Combining the ratios at $p_+$ and $p_-$ implements the discrete derivative while also canceling the unphysical global phase $\chi(t)$.

One might worry that \eqref{eq:Sphase} contains an additional, unphysical phase arising from the two-soliton states we project onto. With the basis \eqref{eq:two_soliton_basis}, $\ket{p,-p}$ and $\ket{-p,p}$ are independent basis states, each built from excitation tensors whose overall phase is arbitrary. A priori, the two can therefore differ by an arbitrary relative phase $e^{i\phi(p,-p)}$, which would contaminate the ratio \eqref{eq:phi}.\footnote{This issue occurs here because, unlike for particle scattering \cite{Jha:2024jan}, we are dealing with two distinct excitations (soliton and anti-soliton, corresponding to $B_+$ and $B_-$) that only make sense as asymptotic states if they are ordered with the soliton on the left and the anti-soliton on the right. Therefore, we defined the basis \eqref{eq:two_soliton_basis} with the momentum $p_1$ assigned to the soliton and $p_2$ to the anti-soliton instead of smearing two exchangeable particle excitations with momenta $p_1$ and $p_2$ across the entire lattice as is the case in \cite{Jha:2024jan}. In particular, this distinguishes the state $\ket{-p, p}$ from $\ket{p, -p}$, which for the two-particle states in \cite{Jha:2024jan} are the same, and a priori allows for an arbitrary phase between the two which could pollute the ratio \eqref{eq:phi}.}

We fix this relative phase using the discrete symmetries of the lattice theory. Acting with $C$ or $P$ on the full state $\ket{p,-p}$ places the anti-soliton to the left of the soliton, which is not an allowed asymptotic configuration in this sector, while $CP$ maps $\ket{p,-p}$ to itself. Neither yields $\ket{-p,p}$. Instead, we act on the individual excitation tensors: under $C$, $B_+(p)\to B_-(p)$ and $B_-(-p)\to B_+(-p)$; equivalently, under $P$, $B_+(p)\to B_+(-p)$ and $B_-(-p)\to B_-(p)$ (see Tab.~\ref{tab:CP}). We then build $\ket{-p,p}$ from the transformed tensors, with the soliton on the left and the same vacua as $\ket{p,-p}$. Since $C$ and $P$ act linearly on the tensors, any phase carried by $B_+(p)$ and $B_-(-p)$ appears identically in $\ket{p,-p}$ and $\ket{-p,p}$ and cancels in \eqref{eq:phi}. The relative phases $\phi(p_+,-p_+)$ and $\phi(p_-,-p_-)$ are thus fixed to zero by construction, and \eqref{eq:phi} yields a physical phase related to the time delay.

\begin{table}[h]
\centering
\renewcommand{\arraystretch}{1.3}
\setlength{\tabcolsep}{4pt}
\begin{tabular}{c|ccc}
\toprule
Symmetry & $(p, x)$ & $(E, q)$ & soliton \\
\midrule
$\mathds{1}$ & $p, x$   & $E, q$   & soliton $B_+(p)$  \\
$C$          & $p, x$   & $-E, -q$ & anti-soliton $B_-(p)$   \\
$P$          & $-p, -x$ & $-E, q$  & soliton $B_+(-p) $ \\
$CP$         & $-p, -x$ & $E, -q$  & anti-soliton $ B_-(-p)$\\
\bottomrule
\end{tabular}
\caption{Transformation properties under $C$, $P$, and $CP$, following from Appendix~\ref{app:symmetries}.}
\label{tab:CP}
\end{table}

We choose $\Delta p=10^{-3}g$ small enough so that $p_\pm$ lie well within the momentum spread of the wavepackets (i.e., $\Delta p \ll 1/\sigma$), and the error from taking the discretized derivative is small. 

\bibliographystyle{ssg}
\bibliography{bib}

\end{document}